\documentclass[journal]{IEEEtran}
\usepackage{cite}
\usepackage[cmex10]{amsmath}
\usepackage{mathrsfs,amssymb,bm,pifont,dsfont,stmaryrd,euscript}
\usepackage{graphicx}
\usepackage{epsf,url,epsfig}
\newcommand{\bb}{\mathbb}
\newcommand{\eu}{\EuScript}
\newcommand{\Scr}{\mathscr}
\newcommand{\Cal}{\mathcal}
\newtheorem{thm}{Theorem}
\newtheorem{lem}[thm]{Lemma}
\newtheorem{cor}[thm]{Corollary}

\newtheorem{rem}[thm]{Remark}

\newtheorem{fin}[thm]{Definition}
\newtheorem{exm}{Example}
\usepackage{array}
\begin{document}
\title{On Integral Probability Metrics, $\phi$-Divergences and Binary Classification}
%\title{On the Properties of Integral Probability Metrics}
%\title{On Integral Probability Metrics and $\phi$-Divergences}

\author{Bharath K. Sriperumbudur,
        Kenji Fukumizu, Arthur Gretton, Bernhard Sch\"{o}lkopf
        and Gert R. G. Lanckriet % <-this % stops a space
\thanks{Bharath K. Sriperumbudur and Gert R. G. Lanckriet are with the Department
of Electrical and Computer Engineering, University of California, San Diego,
CA, 92093, USA. E-mail: bharathsv@ucsd.edu, gert@ece.ucsd.edu.}% <-this % stops a space
\thanks{Kenji Fukumizu is with the Institute of Statistical Mathematics, 4-6-7 Minami-Azabu, Minato-ku, Tokyo 106-8569, Japan. E-mail: fukumizu@ism.ac.jp.}%
\thanks{Arthur Gretton is with the Machine Learning Department, Carnegie Mellon University, 5000 Forbes Avenue, Pittsburgh, PA 15213, USA. He is also affiliated with the Max Planck Institute for Biological Cybernetics, Spemannstra\ss e 38, 72076 T\"{u}bingen, Germany. E-mail: arthur.gretton@gmail.com.} 
\thanks{Bernhard Sch\"{o}lkopf is with the Max Planck Institute for Biological Cybernetics, Spemannstra\ss e 38, 72076 T\"{u}bingen, Germany. E-mail:  bernhard.schoelkopf@tuebingen.mpg.de.}% <-this % stops a space
%\thanks{Bharath K. Sriperumbudur wishes to acknowledge the support from the Max Planck Institute (MPI) for Biological Cybernetics, National Science Foundation (grant DMS-MSPA 0625409), the Fair Isaac Corporation and the University of California MICRO program. Part of this work was done while the author was visiting MPI.}
}%

%\markboth{IEEE Transactions on Information Theory,~Vol.~x, No.~x, August~2009}%
%{Shell \MakeLowercase{\textit{et al.}}: Bare Demo of IEEEtran.cls for Journals}
% The only time the second header will appear is for the odd numbered pages
% after the title page when using the twoside option.
% 
% If you want to put a publisher's ID mark on the page you can do it like
% this:
%\IEEEpubid{0000--0000/00\$00.00~\copyright~2007 IEEE}
% Remember, if you use this you must call \IEEEpubidadjcol in the second
% column for its text to clear the IEEEpubid mark.

% make the title area
\maketitle

\begin{abstract}
%AG2: better not to start with the thing we *don't* want to advertise! 
%$\phi$-divergences are a widely studied class of  distance measures between probabilities. 
A class of distance measures on probabilities --- the integral probability metrics (IPMs) --- is addressed: these include the Wasserstein distance, Dudley metric, and Maximum Mean Discrepancy. IPMs have thus far mostly been used in more abstract settings, for instance as theoretical tools in mass transportation problems, and in metrizing the weak topology on the set of all Borel probability measures defined on a metric space. Practical applications of IPMs are less common, with some exceptions in the kernel machines literature. 
%AG2: I say generally above, since they have been used in certain NIPS papers :)
%have been used mainly as theoretical tools in probability theory and their practical applicability has never been investigated. 
The present work contributes a number of novel properties of IPMs, which should contribute to making IPMs more widely used in practice, for instance in areas where $\phi$-divergences are currently popular. 
\par First, to understand the relation between IPMs and $\phi$-divergences, the necessary and sufficient conditions under which these classes intersect are derived: the total variation distance is shown to be the only non-trivial $\phi$-divergence that is also an IPM. This shows that IPMs are essentially different from $\phi$-divergences. Second, empirical estimates of several IPMs from finite i.i.d. samples are obtained, and their consistency and convergence rates are analyzed. These estimators are shown to be easily computable, with better rates of convergence than estimators of $\phi$-divergences.
%since the empirical estimation of $\phi$-divergences, especially the KL-divergence is well-studied, the empirical estimation of IPMs from finite i.i.d. samples is then considered and their consistency and convergence rates are analyzed. %The estimates of the Wasserstein distance and Dudley metric are obtained by solving linear programs while the empirical MMD can be simply computed in a closed form. 
%The empirical estimators of the Wasserstein distance and Dudley metric are shown to be \emph{strongly consistent}. 
Third, a novel interpretation is provided for IPMs by relating them to binary classification, where it is 
%similar to the relation between $\phi$-divergences and binary classification, IPMs are related to binary classification by 
shown that the IPM between class-conditional distributions is the negative of the optimal risk associated with a binary classifier. In addition, the smoothness of an appropriate binary classifier is proved to be inversely related to the distance between the class-conditional distributions, measured in terms of an IPM.
\end{abstract}
% IEEEtran.cls defaults to using nonbold math in the Abstract.
% This preserves the distinction between vectors and scalars. However,
% if the journal you are submitting to favors bold math in the abstract,
% then you can use LaTeX's standard command \boldmath at the very start
% of the abstract to achieve this. Many IEEE journals frown on math
% in the abstract anyway.

% Note that keywords are not normally used for peerreview papers.

% For peer review papers, you can put extra information on the cover
% page as needed:
% \ifCLASSOPTIONpeerreview
% \begin{center} \bfseries EDICS Category: 3-BBND \end{center}
% \fi
%
% For peerreview papers, this IEEEtran command inserts a page break and
% creates the second title. It will be ignored for other modes.

\IEEEpeerreviewmaketitle

\section{Introduction}\label{Sec:Introduction}
% The very first letter is a 2 line initial drop letter followed
% by the rest of the first word in caps.
% 
% form to use if the first word consists of a single letter:
% \IEEEPARstart{A}{demo} file is ....
% 
% form to use if you need the single drop letter followed by
% normal text (unknown if ever used by IEEE):
% \IEEEPARstart{A}{}demo file is ....
% 
% Some journals put the first two words in caps:
% \IEEEPARstart{T}{his demo} file is ....
% 
% Here we have the typical use of a "T" for an initial drop letter
% and "HIS" in caps to complete the first word.
%\IEEEPARstart{T}{his}
%\hfill mds
% needed in second column of first page if using \IEEEpubid
%\IEEEpubidadjcol
\IEEEPARstart{T}{he} notion of distance between probability measures has found many applications in probability theory, mathematical statistics and information theory \cite{Rachev-91,Vajda-89,Rachev-98}. Popular applications include distribution testing, %homogeneity tests (the two-sample problem), independence tests, goodness-of-fit tests, 
establishing central limit theorems, density estimation, signal detection, channel and source coding, etc. 
\par One of the widely studied and well understood families of distances/divergences between probability measures is the \emph{Ali-Silvey distance} \cite{Ali-66}, also called the \emph{Csisz\'{a}r's $\phi$-divergence} \cite{Csiszar-67}, which is defined as 
\begin{equation}\label{Eq:phidiv}
D_\phi(\bb{P},\bb{Q}):=\left\{\begin{array}{c@{\quad\quad}l}
\int_M\phi\left(\frac{d\bb{P}}{d\bb{Q}}\right)\,d\bb{Q},& \bb{P}\ll \bb{Q}\\
+\infty, & \text{otherwise}
\end{array}\right.,
\end{equation}
where $M$ is a measurable space and $\phi:[0,\infty)\rightarrow(-\infty,\infty]$ is a convex function.\footnote{Usually, the condition $\phi(1)=0$ is used in the definition of $\phi$-divergence. Here, we do not enforce this condition.} $\bb{P}\ll \bb{Q}$ denotes that $\bb{P}$ is absolutely continuous w.r.t. $\bb{Q}$. %Let $\Scr{P}$ be the set of all probability measures defined on $M$. 
Well-known distance/divergence measures %on $\Scr{P}$ 
 obtained by appropriately choosing $\phi$ include the Kullback-Liebler (KL) divergence ($\phi(t)=t\log t$), Hellinger distance ($\phi(t)=(\sqrt{t}-1)^2$), total variation distance ($\phi(t)=|t-1|$), $\chi^2$-divergence ($\phi(t)=(t-1)^2$), etc. See \cite{Vajda-89,Liese-06} and references therein for selected statistical and information theoretic applications of $\phi$-divergences.
\par In this paper, we consider %On the other hand, 
another popular family (particularly in probability theory and mathematical statistics) of distance measures: the %on $\Scr{P}$, 
 \emph{integral probability metrics} (IPMs) \cite{Muller-97}, defined as
\begin{equation}\label{Eq:ipm}
\gamma_{\eu{F}}(\bb{P},\bb{Q}):=\sup_{f\in\eu{F}}\left|\int_Mf\,d\bb{P}-\int_Mf\,d\bb{Q}\right|,
\end{equation}
where $\eu{F}$ in (\ref{Eq:ipm}) is a class of real-valued bounded measurable functions on $M$. So far, IPMs have been mainly studied as tools of theoretical interest in probability theory \cite{Rachev-85,Rachev-98},\cite[Chapter 11]{Dudley-02}, with limited applicability in practice. Therefore, in this paper, we present a number of novel properties of IPMs, %in relation to $\phi$-divergences and binary classification, 
which will serve to improve their usefulness in more applied domains. We emphasize in particular the advantages of IPMs compared to $\phi$-divergences.
%First we show that the class of IPMs are very different from that of $\phi$-divergences except in trivial cases. Therefore, if IPMs have certain advantages over $\phi$-divergences, then they can replace $\phi$-divergences in applications where they are currently being used. 
\par $\phi$-divergences, and especially the KL-divergence, are better known and more widely used in diverse fields such as neuroscience \cite{Rieke-97,Nemenman-04,Belitski-08} and distribution testing \cite{Read-88,Gyorfi-02,Inglot-90,Gretton-08b}, however they are
notoriously tough to estimate, especially in high dimensions, $d$, when $M=\bb{R}^d$, e.g., see \cite{Wang-05}. By contrast, we show that under certain conditions on $\eu{F}$, irrespective of the dimension, $d$, IPMs are very simple to estimate in a consistent manner. This property can be exploited in statistical applications where the distance between $\bb{P}$ and $\bb{Q}$ is to be estimated from finite data. %Therefore, in all these applications, IPMs can be find wide applicability similar to $\phi$-divergences. 
Further, we show that IPMs are naturally related to binary classification, which gives these distances a clear and natural interpretation. Specifically, we show that (a) the smoothness of a binary classifier is inversely related to the distance between the class-conditional distributions, measured in terms of IPM, and (b) the IPM between the class-conditional distributions is the negative of the optimal risk associated with an appropriate binary classifier. We will go into more detail regarding these contributions in Section~\ref{Subsec:contributions}. First, we provide some examples of IPMs and their applications.
\subsection{Examples and Applications of IPMs}
The definition of IPMs in (\ref{Eq:ipm}) is motivated from the notion of \emph{weak convergence} of probability measures on metric spaces \cite[Section 9.3, Lemma 9.3.2]{Dudley-02}. In probability theory, IPMs are used in proving central limit theorems using Stein's method \cite{Stein-72,Barbour-05}. 
 They are also the fundamental quantities that appear in empirical process theory \cite{Vaart-96}, where $\bb{Q}$ is replaced by the \emph{empirical distribution} of $\bb{P}$. 
\par Various popular distance measures in probability theory and statistics can be obtained by appropriately choosing $\eu{F}$. Suppose $(M,\rho)$ is a metric space with $\eu{A}$ being the Borel $\sigma$-algebra induced by the metric topology. Let $\Scr{P}$ be the set of all Borel probability measures on $\eu{A}$. \vspace{2mm}
\subsubsection{Dudley metric}
Choose $\eu{F}=\{f:\Vert f\Vert_{BL}\le 1\}$ in (\ref{Eq:ipm}), where $\Vert f\Vert_{BL}:=\Vert f\Vert_\infty +\Vert f \Vert_L$, $\Vert f\Vert_\infty:=\sup\{|f(x)|:x\in M\}$ and $\Vert f\Vert_L:=\sup\{|f(x)-f(y)|/\rho(x,y):x\ne y\,\,\text{in}\,\, M\}$. $\Vert f\Vert_L$ is called the Lipschitz semi-norm of a real-valued function $f$ on $M$ \cite[Chapter 19, Definition 2.2]{Shorack-00}. The Dudley metric is popularly used in the context of proving the convergence of probability measures with respect to the weak topology \cite[Chapter 11]{Dudley-02}.\vspace{2mm}
\subsubsection{Kantorovich metric and Wasserstein distance}
Choosing $\eu{F}=\{f:\Vert f\Vert_L\le 1\}$ in (\ref{Eq:ipm}) yields the \emph{Kantorovich metric}. The famous Kantorovich-Rubinstein theorem \cite[Theorem 11.8.2]{Dudley-02} shows that when $M$ is separable, the Kantorovich metric is the dual representation of the so called \emph{Wasserstein distance} defined as 
\begin{equation}\label{Eq:primal-wasserstein}
W_1(\bb{P},\bb{Q}):=\inf_{\mu\in \eu{L}(\bb{P},\bb{Q})}\int\rho(x,y)\,d\mu(x,y),
\end{equation}
where $\bb{P},\bb{Q}\in\Scr{P}_1:=\{\bb{P}:\int \rho(x,y)\,d\bb{P}(x)<\infty,\,\forall\,y\in M\}$ and $\eu{L}(\bb{P},\bb{Q})$ is the set of all measures on $M\times M$ with marginals $\bb{P}$ and $\bb{Q}$. Due to this duality, in this paper, we refer to the Kantorovich metric as the Wasserstein distance and denote it as $W$ when $M$ is separable. The Wasserstein distance has found applications in information theory \cite{Gray-75}, mathematical statistics \cite{Zolotarev-83, Rachev-84}, \emph{mass transportation problems} \cite{Rachev-85} and is also called as the \emph{earth mover's distance} in engineering applications \cite{Levina-01}.%see \cite{Levina-01} and references therein for a list of engineering applications where the Wasserstein distance is used). %We refer the interested reader to \cite[Chapter 5]{Vajda-89} for the generalizations of $W_1$. 
\vspace{2mm}
\subsubsection{Total variation distance and Kolmogorov distance}
$\gamma_{\eu{F}}$ is the \emph{total variation metric} when $\eu{F}=\{f:\Vert f\Vert_\infty\le 1\}$ while it is the \emph{Kolmogorov distance} when $\eu{F}=\{\mathds{1}_{(-\infty,t]}\,:\,t\in\mathbb{R}^d\}$. Note that the classical central limit theorem and the Berry-Ess\'{e}en theorem in $\bb{R}^d$ use the Kolmogorov distance. The Kolmogorov distance also appears in hypothesis testing as the Kolmogorov-Smirnov statistic \cite{Shorack-00}. \vspace{2mm}
\subsubsection{Maximum mean discrepancy} $\gamma_\eu{F}$ is called the \emph{maximum mean discrepancy (MMD)} \cite{Gretton-06, Sriperumbudur-08} when $\eu{F}=\{f:\Vert f\Vert_\eu{H}\le 1\}$. Here, $\eu{H}$ represents a reproducing kernel Hilbert space (RKHS) \cite{Aronszajn-50,Saitoh-88} with $k$ as its reproducing kernel (r.k.).\footnote{A function $k:M\times M \rightarrow\bb{R},\,\,(x,y)\mapsto k(x,y)$ is a \emph{reproducing kernel} of the Hilbert space $\eu{H}$ if and only if the following hold: \emph{(i)} $\forall\,y\in M,\,\,k(\cdot,y)\in\eu{H}$ and \emph{(ii)} $\forall\,y\in M,\,\forall\,f\in\eu{H},\,\,\langle f,k(\cdot,y)\rangle_\eu{H}=f(y)$. $\eu{H}$ is called a reproducing kernel Hilbert space.} MMD is used in statistical applications including homogeneity testing \cite{Gretton-06}, independence testing \cite{Gretton-08a}, and testing for conditional independence \cite{Fukumizu-08a}.
\subsection{Contributions}\label{Subsec:contributions}
Some of the previously mentioned IPMs, e.g., the Kantorovich distance and Dudley metric, are mainly  tools of theoretical interest in probability theory. That said, their application in practice is generally less well established. The Dudley metric has been used only in the context of metrizing the weak topology on $\Scr{P}$ \cite[Chapter 11]{Dudley-02}. The Kantorovich distance is more widespread, although it is better known in its primal form in (\ref{Eq:primal-wasserstein}) as the Wasserstein distance than as an IPM \cite{Rachev-85,Rachev-98}. The goal of this work is to present a number of favourable statistical and implementational properties of IPMs, and to specifically compare IPMs and $\phi$-divergences. Our hope is to broaden the applicability of IPMs, and to encourage their wider adoption in data analysis and statistics. The contributions of this paper are three-fold, and explained in detail below.\vspace{2mm}
\subsubsection{IPMs and $\phi$-divergences} Since $\phi$-divergences are well studied and understood, the first question we are interested in is whether IPMs have any relation to $\phi$-divergences. In particular, we would like to know whether any of the IPMs can be realized as a $\phi$-divergence, so that 
%Since the properties of $\phi$-divergences are widely studied (see \cite{Vajda-89,Liese-06} and references therein), in this work, the first question we are interested in is whether any of the IPMs can be realized as a $\phi$-divergence so that
%we first investigate the relation between IPMs and $\phi$-divergences, the motivation being that if some of the IPMs can be realized as $\phi$-divergences, then 
the properties of $\phi$-divergences will carry over to those IPMs. %AG: this is a very poor motivation, given that the TV distance is the only one for which this occurs. A better motivation would be to demonstrate that  IPMs are generally not phi-divergences.
%BK: I dont understand ur point. first of all why wud one do a study to compared IPMs and phi-div. the outcome of the study is that ipms are not phi-div. but i feel the outcome is not the motivation. what say?
%AG2: I'm happy to go with current wording.
In Section~\ref{Sec:intersection}, we first show that $\gamma_\eu{F}$ is closely related to the variational form of $D_\phi$ \cite{Keziou-03,Nguyen-09,Broniatowski-09} and is ``trivially" a $\phi$-divergence if $\eu{F}$ is chosen to be the set of all real-valued measurable functions on $M$ (see Theorem~\ref{thm:phi-mmd}). Next, we generalize this result by determining the necessary and sufficient conditions on $\eu{F}$ and $\phi$ for which $\gamma_{\eu{F}}(\bb{P},\bb{Q})=D_\phi(\bb{P},\bb{Q}),\,\forall\,\bb{P},\bb{Q}\in\mathscr{P}_0\subset\mathscr{P}$, where $\mathscr{P}_0$ is some subset of $\mathscr{P}$. This leads to our first contribution, answering the question, ``Given a set of distance/divergence measures, $\{\gamma_{\eu{F}}:\eu{F}\}$ (indexed by $\eu{F}$) and $\{D_\phi:\phi\}$ (indexed by $\phi$) defined on $\mathscr{P}$, is there a set of distance measures that is common to both these families?" %in we consider the question of whether the class of IPMs and the class of $\phi$-divergences are different or not.  So, in Section~\ref{Sec:intersection}, we consider the problem of deriving necessary and sufficient conditions on $\eu{F}$ and $\phi$ so that $\forall\,\bb{P},\bb{Q}\in\mathscr{P}_0\subset\mathscr{P},\,\gamma_{\eu{F}}(\bb{P},\bb{Q})=D_\phi(\bb{P},\bb{Q})$, where $\mathscr{P}_0$ is some subset of $\mathscr{P}$. 
%This is equivalent to finding conditions on $\eu{F}$ and $\phi$ so that $\gamma_\eu{F}$ is a $\phi$-divergence, which is addressed in Section~\ref{Sec:intersection}, wherein 
We show that the classes $\{\gamma_{\eu{F}}:\eu{F}\}$ and $\{D_\phi:\phi\}$ of distance measures intersect \emph{non-trivially only} at the \emph{total variation distance}, which in turn indicates that these classes are essentially different and therefore the properties of $\phi$-divergences will not carry over to IPMs.\vspace{2mm} %This study also answers the open question that has recently been posed by Reid and Williamson~\cite[pp. 56]{Reid-09}, where they questioned, ``Whether there exist $\eu{F}$ such that $\gamma_\eu{F}$ is not a metric but equals $D_\phi$ for some $\phi\ne t\mapsto |t-1|$?"\footnote{Note that $D_\phi$ is a metric for $\phi(t)=|t-1|$, which is the total variation distance. If $\gamma_\eu{F}=D_\phi$ for $\phi(t)=|t-1|$, then $\gamma_\eu{F}$ is a metric on $\Scr{P}$ and as mentioned before, $\eu{F}=\{f:\Vert f\Vert_\infty\le 1\}$ provides the total variation distance.}
%, while they
%We also show that $\{\gamma_{\eu{F}}\}_{\eu{F}}$ and $\{D_\phi\}_\phi$ 
%\emph{intersect trivially} by choosing $\eu{F}$ to be the set of all real-valued measurable functions on $M$; $\phi$ to be $\phi(t)=0$ for $t=1$ and $\phi(t)=+\infty$ for $t\ne 1$, so that $D_\phi(\bb{P},\bb{Q})=0$ when $\bb{P}=\bb{Q}$ and $+\infty$ otherwise. 
%Since these results show that $\gamma_\eu{F}$ is a $\phi$-divergence for certain $\eu{F}$ and $\phi$, we relate $\gamma_\eu{F}$ to the problem of binary classification and statistical information~\cite{DeGroot-62,DeGroot-70}, which we elaborate below.
\subsubsection{Estimation of IPMs} \label{subsubsec:estimation}
Many statistical inference applications such as distribution testing involve the estimation of distance between probability measures $\bb{P}$ and $\bb{Q}$ based on finite samples drawn i.i.d. from each. We first consider the properties of finite sample estimates of the $\phi$-divergence, which is a well-studied problem (especially for the KL-divergence; see \cite{Wang-05,Nguyen-08} and references therein).
%Though IPMs like the Wasserstein distance and Dudley metric are far-reaching as theoretical tools, they have a definite drawback: explicit calculation is difficult for most concrete examples. The same issue also arises with MMD and $\phi$-divergences where the exact computation is not straightforward for certain distributions. %%AG: what do you mean here? Empirical estimates of the MMD for RKHS kernels are very easy to compute! And when computing the population expressions, you only need to convolve the distribution with a kernel, which could be a little tricky, but is not as hard as you make it sound.
%BK: thats not correct right. what if i give two triangular distributions and the kernel is Gaussian. in this though u can do convolution, u cannot get a closed form solution. u have to use either numerical integration methods or use the empirical estimator of MMD with large samples.
% Therefore, given two probability measures $\bb{P}$ and $\bb{Q}$, one approach to compute the distance (say the Wasserstein distance) between them is to estimate it based on finite samples drawn i.i.d. from $\bb{P}$ and $\bb{Q}$ and hope that the estimate converges to the true distance between $\bb{P}$ and $\bb{Q}$ given a large number of samples. This situation also arises in statistical inference applications where $\bb{P}$ and $\bb{Q}$ are known only through finite samples drawn i.i.d. from them and one would like to estimate the distance between $\bb{P}$ and $\bb{Q}$. 
 Wang \emph{et al.} \cite{Wang-05} used a data-dependent space partitioning scheme and showed that the non-parametric estimator of KL-divergence is \emph{strongly consistent}. However, the rate of convergence of this estimator can be arbitrarily slow depending on the distributions. In addition, for increasing dimensionality of the data (in $\bb{R}^d$), the method is increasingly difficult to implement. On the other hand, by exploiting the variational representation of $\phi$-divergences, Nguyen \emph{et al.} \cite{Nguyen-08} provide a consistent estimate of a lower bound %AG: why the quotes? It is or it isn't.
%BK: i removed the quotes...it is a lower bound.
of the KL-divergence by solving a convex program. Although this approach is efficient and the dimensionality of the data is not an issue, the estimator provides a lower bound and not the KL-divergence itself. Given the disadvantages associated with the estimation of $\phi$-divergences, it is of interest to compare with the convergence behaviour of finite sample estimates of IPMs.
%AG: ``therefore'' is a poor transition here. ``By contrast, we show that IPMs have fast convergence....'' would be better
%BK: I think I have rectified this.
\par To this end, as our second and ``main" contribution, in Section~\ref{Sec:consistency}, we consider the non-parametric estimation of some IPMs, in particular the Wasserstein distance, Dudley metric and MMD based on finite samples drawn i.i.d. from $\bb{P}$ and $\bb{Q}$. The estimates of the Wasserstein distance and Dudley metric are obtained by solving linear programs while an estimator of MMD is computed in closed form (see Section~\ref{subsec:empirical}). One of the advantages with these estimators is that they are quite simple to implement and are not affected by the dimensionality of the data, unlike $\phi$-divergences. Next, in Section~\ref{subsec:consistency}, we show that these estimators are strongly consistent and provide their rates of convergence, using concentration inequalities and tools from empirical process theory \cite{Vaart-96}. In Section~\ref{subsec:simulation}, we describe simulation results that demonstrate the practical viability of these estimators. The results show that it is simpler and more efficient to use IPMs instead of $\phi$-divergences in many statistical inference applications. 
\par Since the total variation distance is also an IPM, in Section~\ref{subsec:tv}, we discuss its empirical estimation and show that the empirical estimator is not strongly consistent. Because of this, we provide new lower bounds for the total variation distance in terms of the Wasserstein distance, Dudley metric, and MMD, which can be consistently estimated. These bounds also translate as lower bounds on the KL-divergence through Pinsker's inequality \cite{Fedotov-03}. 
\par Our study shows that estimating IPMs (especially the Wasserstein distance, Dudley metric and MMD) is much simpler than estimating $\phi$-divergences, and that the estimators are strongly consistent while exhibiting good rates of convergence. In addition, IPMs also account for the properties of the underlying space $M$ (the metric property is determined by $\rho$ in the case of Wasserstein and Dudley metrics, while the similarity property is determined by the kernel $k$ \cite{Scholkopf-02} in the case of MMD) while computing the distance between $\bb{P}$ and $\bb{Q}$, which is not the case with $\phi$-divergences. This property is useful when $\bb{P}$ and $\bb{Q}$ have disjoint support.\footnote{When $\bb{P}$ and $\bb{Q}$ have disjoint support, $D_\phi(\bb{P},\bb{Q})=+\infty$ irrespective of the properties of $M$, while $\gamma_\eu{F}(\bb{P},\bb{Q})$ varies with the properties of $M$. Therefore, in such cases, $\gamma_\eu{F}(\bb{P},\bb{Q})$ provides a better notion of distance between $\bb{P}$ and $\bb{Q}$.} With these advantages, we believe that IPMs can find many applications in information theory, image processing, machine learning, neuroscience and other areas.\vspace{2mm}
\subsubsection{Interpretability of IPMs: Relation to Binary Classification} Finally, as our third contribution, we provide a nice interpretation for IPMs by showing they naturally appear in binary classification. Many previous works \cite{Osterreicher-93,Buja-05,Liese-06,Nguyen-09} relate $\phi$-divergences (between $\bb{P}$ and $\bb{Q}$) to binary classification (where $\bb{P}$ and $\bb{Q}$ are the class conditional distributions) as the negative of the optimal risk associated with a loss function (see \cite[Section 1.3]{Reid-09} for a detailed list of references). %As an example, Nguyen \emph{et al.} \cite{Nguyen-05,Nguyen-09} have shown that for each \emph{margin-based surrogate} loss function (that acts as a surrogate to $0-1$ loss), there exists exactly one corresponding $\phi$-divergence such that the optimal risk is equal to the negative of the $\phi$-divergence. 
In Section~\ref{Sec:surrogate}, we present a series of results that relate IPMs to binary classification. First, in Section~\ref{Sec:Lrisk_interpret}, we provide a result (similar to that for $\phi$-divergences), which shows $\gamma_\eu{F}(\bb{P},\bb{Q})$ is the negative of the optimal risk associated with a binary classifier that separates the class conditional distributions, $\bb{P}$ and $\bb{Q}$, where the classification rule is restricted to $\eu{F}$. Therefore, the Dudley metric, Wasserstein distance, total variation distance and MMD can be understood as the negative of the optimal risk associated with a classifier for which the classification rule is restricted to $\{f:\Vert f\Vert_{BL}\le 1\}$, $\{f:\Vert f\Vert_L\le 1\}$, $\{f:\Vert f\Vert_\infty\le 1\}$ and $\{f:\Vert f\Vert_\eu{H}\le 1\}$ respectively. 
%Since we show in Section~\ref{Sec:intersection} that integral probability metrics and $\phi$-divergences are different, in Section~\ref{Sec:surrogate}, we also relate integral probability metrics to binary classification and show that the Dudley metric, Wasserstein distance, total variation distance and maximum mean discrepancy can be seen as the negative of the optimal risk associated with a particular classification rule. 
Next, in Sections~\ref{Sec:Lipschitz} and \ref{Sec:mmd}, we present a second result that relates the empirical estimators studied in Section~\ref{Sec:consistency} to the binary classification setting, by relating the empirical estimators of the Wasserstein distance and Dudley metric to the \emph{margins} of the Lipschitz \cite{Luxburg-04} and bounded Lipschitz classifiers, respectively; and MMD to the Parzen window classifier \cite{Scholkopf-02,ShaweTaylor-04} (see \emph{kernel classification rule}~\cite[Chapter 10]{Devroye-96}). The significance of this result is that the smoothness of the classifier is inversely related to the empirical estimate of the IPM between class conditionals $\bb{P}$ and $\bb{Q}$. Although this is intuitively clear, our result provides a theoretical justification.% and support vector machine (SVM) \cite{Cortes-95}. %This connection of MMD to the Parzen window classifier also shows how to derive a Parzen window classifier within the framework of empirical risk minimization and regularization \cite{Evgeniou-00,Scholkopf-02}. %In a related work, recently, \cite{Reid-09} has connected MMD to the Fisher discriminant analysis \cite{Devroye-96} and support vector machines \cite{Cortes-95}. 
\\
\par Before proceeding with our main presentation, we introduce the notation we will use throughout the paper. Certain proofs and supplementary results are presented in a collection of appendices, and referenced as needed.
\subsection{Notation}
For a measurable function $f$ and a signed measure $\bb{P}$, $\bb{P}f:=\int f\,d\bb{P}$ denotes the expectation of $f$ under $\bb{P}$. $\llbracket A\rrbracket$ represents the indicator function for set $A$. Given an i.i.d. sample $X_1,\ldots,X_n$ drawn from $\bb{P}$, $\bb{P}_n:=\frac{1}{n}\sum^n_{i=1}\delta_{X_i}$ represents the empirical distribution, where $\delta_x$ represents the Dirac measure at $x$. We use $\bb{P}_nf$ to represent the empirical expectation $\frac{1}{n}\sum^n_{i=1}f(X_i)$. We define: 
$\text{Lip}(M,\rho):=\{f:M\rightarrow\bb{R}\,\vert\,\Vert f\Vert_L<\infty\}$,
$BL(M,\rho):=\{f:M\rightarrow\bb{R}\,\vert\,\Vert f\Vert_{BL}<\infty\}$, and
\begin{equation}
%\vspace{-2mm}
\begin{array}{c@{\quad}l}
%0,& t=1\\
%+\infty,& t\ne 1
%\end{array}
\eu{F}_W:=\{f:\Vert f\Vert_L\le 1\}, & W(\bb{P},\bb{Q}):=\gamma_{\eu{F}_W}(\bb{P},\bb{Q}),\\ 
\eu{F}_\beta:=\{f:\Vert f\Vert_{BL}\le 1\}, & \beta(\bb{P},\bb{Q}):=\gamma_{\eu{F}_\beta}(\bb{P},\bb{Q}),\\
\eu{F}_k:=\{f:\Vert f\Vert_\eu{H}\le 1\}, & 
\gamma_k(\bb{P},\bb{Q}):=\gamma_{\eu{F}_k}(\bb{P},\bb{Q}),\\
\eu{F}_{TV}:=\{f:\Vert f\Vert_\infty\le 1\}, & TV(\bb{P},\bb{Q}):=\gamma_{\eu{F}_{TV}}(\bb{P},\bb{Q}).
\end{array}\nonumber
\end{equation}
\section{IPMs and $\phi$-divergences}\label{Sec:intersection}
In this section, we consider $\{\gamma_\eu{F}:\eu{F}\}$ and $\{D_\phi:\phi\}$, which are classes of IPMs and $\phi$-divergences on $\mathscr{P}$ indexed by $\eu{F}$ and $\phi$, respectively. We derive conditions on $\eu{F}$ and $\phi$ such that $\forall\,\bb{P},\,\bb{Q}\in\mathscr{P}_0\subset\mathscr{P},\,\gamma_{\eu{F}}(\bb{P},\bb{Q})=D_\phi(\bb{P},\bb{Q})$ for some chosen $\mathscr{P}_0$. This shows the degree of overlap between the class of IPMs and the class of $\phi$-divergences. %To this end, we introduce some notation where we define $\bb{P}f:=\int_Mf\,d\bb{P}$. Denote $\llbracket A\rrbracket$ to be the indicator function for set $A$. 
\par Consider the variational form of $D_\phi$ \cite{Keziou-03,Nguyen-08,Broniatowski-09} given by %AG: function class? Bounded? Measurable?
%BK: i dont understand by what u mean here?
\setlength{\arraycolsep}{0.0em}
\begin{eqnarray}\label{Eq:dual-phi-div}
D_\phi(\bb{P},\bb{Q})&{}={}&\sup_{f:M\rightarrow\bb{R}}\left[\int_Mf\,d\bb{P}-\int_M\phi^*(f)\,d\bb{Q}\right]\nonumber\\
&{}={}&\sup_{f:M\rightarrow\bb{R}}(\bb{P}f-\bb{Q}\phi^*(f)),
\end{eqnarray} 
where $\phi^*(t)=\sup\{tu-\phi(u):u\in\mathbb{R}\}$ is the \emph{convex conjugate} of $\phi$. Suppose $\eu{F}$ is such that $f\in\eu{F}\Rightarrow -f\in\eu{F}$. Then, 
\begin{equation}\label{Eq:ipm-rewrite}
\gamma_\eu{F}(\bb{P},\bb{Q})=\sup_{f\in\eu{F}}|\bb{P}f-\bb{Q}f|=\sup_{f\in\eu{F}}(\bb{P}f-\bb{Q}f).
\end{equation}
Recently, Reid and Williamson \cite[Section 8.2]{Reid-09}  considered the generalization of $D_\phi$ by modifying its variational form as \begin{equation}\label{Eq:generalization-d-phi}
D_{\phi,\eu{F}}(\bb{P},\bb{Q}):=\sup_{f\in\eu{F}}(\bb{P}f-\bb{Q}\phi^*(f)).
\end{equation}
Let $\eu{F}_\star$ be the set of all real-valued measurable functions on $M$ and let $\phi_\star$ be the convex function defined as in (\ref{Eq:phi-mmd}). It is easy to show that $\phi^*_\star(u)=u$. Comparing $\gamma_\eu{F}$ in (\ref{Eq:ipm-rewrite}) to $D_\phi$ in (\ref{Eq:dual-phi-div}) through $D_{\phi,\eu{F}}$ in (\ref{Eq:generalization-d-phi}), we see that $\gamma_\eu{F}=D_{\phi_\star,\eu{F}}$ and $D_\phi=D_{\phi,\eu{F}_\star}$. This means $\gamma_\eu{F}$ is obtained by fixing $\phi$ to $\phi_\star$ in $D_{\phi,\eu{F}}$ with $\eu{F}$ as the variable and $D_\phi$ is obtained by fixing $\eu{F}$ to $\eu{F}_\star$ in $D_{\phi,\eu{F}}$ with $\phi$ as the variable. This provides a nice relation between $\gamma_\eu{F}$ and $D_\phi$, leading to the following simple result which shows that $\gamma_{\eu{F}_\star}$ is ``trivially" a $\phi$-divergence. \vspace{2mm}%If $\eu{F}$ is such that $f\in\eu{F}\Rightarrow -f\in\eu{F}$, then $\gamma_\eu{F}=D_{\phi_\star,\eu{F}}$, where $\phi_\star$ is defined in (\ref{Eq:phi-mmd}). Interested reader can refer to \cite[Section 8.2]{Reid-09} for further generalizations of $D_{\phi,\eu{F}}$.\vspace{2mm}
%To start with, the following theorem shows that $\gamma_{\eu{F}}$ is ``trivially" a $\phi$-divergence.
\begin{thm}[$\gamma_{\eu{F}_\star}$ is a $\phi$-divergence]\label{thm:phi-mmd}
Let $\eu{F}_\star$ be the set of all real-valued measurable functions on $M$ and 
let
\begin{equation}\label{Eq:phi-mmd}
%\phi_\star(t)=\left\{\begin{array}{c@{\quad\quad}l}
%0,& t=1\\
%+\infty,& t\ne 1
%\end{array}\right..
\phi_\star(t)=0\llbracket t=1\rrbracket+\infty\llbracket t\ne 1\rrbracket.
\end{equation} Then
\begin{equation}\label{Eq:gamma-phi}
%\gamma_{\eu{F}_\star}(\bb{P},\bb{Q})=D_{\phi_\star}(\bb{P},\bb{Q})=\left\{\begin{array}{c@{\quad\quad}l}
%0,& \bb{P}=\bb{Q}\\
%+\infty,& \bb{P}\ne \bb{Q}
%\end{array}\right..
\gamma_{\eu{F}_\star}(\bb{P},\bb{Q})=D_{\phi_\star}(\bb{P},\bb{Q})=0\llbracket\bb{P}=\bb{Q}\rrbracket+\infty\llbracket\bb{P}\ne\bb{Q}\rrbracket.
\end{equation}
Conversely, $\gamma_\eu{F}(\bb{P},\bb{Q})=D_\phi(\bb{P},\bb{Q})=0\llbracket\bb{P}=\bb{Q}\rrbracket+\infty\llbracket\bb{P}\ne\bb{Q}\rrbracket$ implies $\eu{F}=\eu{F}_\star$ and $\phi=\phi_\star$.\vspace{2mm}
\end{thm}
\begin{IEEEproof}
(\ref{Eq:gamma-phi}) simply follows by using $\eu{F}_\star$ and $\phi_\star$ in $\gamma_\eu{F}$ and $D_\phi$ or by using $\phi^*_\star(u)=u$ in (\ref{Eq:dual-phi-div}). For the converse, note that $D_\phi(\bb{P},\bb{Q})=0\llbracket\bb{P}=\bb{Q}\rrbracket+\infty\llbracket\bb{P}\ne\bb{Q}\rrbracket$ implies $\phi(1)=0$ and $\int \phi(d\bb{P}/d\bb{Q})\,d\bb{Q}=\infty,\,\forall\,\bb{P}\ne\bb{Q}$, which means $\phi(x)=\infty,\,\forall\,x\ne 1$ and so $\phi=\phi_\star$. Consider $\gamma_\eu{F}(\bb{P},\bb{Q})=\gamma_{\eu{F}_\star}(\bb{P},\bb{Q})=\sup\{\bb{P}f-\bb{Q}f\,:\,f\in\eu{F}_\star\},\,\forall\,\bb{P},\bb{Q}\in\Scr{P}$. Suppose $\eu{F}\subsetneq\eu{F}_\star$. Then it is easy to see that $\gamma_\eu{F}(\bb{P},\bb{Q})<\gamma_{\eu{F}_\star}(\bb{P},\bb{Q})$ for some $\bb{P},\bb{Q}\in\Scr{P}$, 
%This means for some $\bb{P}\ne\bb{Q}$, we can have $\gamma_\eu{F}(\bb{P},\bb{Q})<\gamma_{\eu{F}_\star}(\bb{P},\bb{Q})$, 
which leads to a contradiction. Therefore, $\eu{F}=\eu{F}_\star$.\vspace{2mm}
\end{IEEEproof}
%The above result shows that MMD is a $\phi$-divergence when $\eu{F}$ is the set of all real-valued functions on $M$. 
From (\ref{Eq:gamma-phi}), it is clear that $\gamma_{\eu{F}_\star}(\bb{P},\bb{Q})$ is the strongest way to measure the distance between probability measures, and is not a very useful metric in practice.\footnote{Unless $\bb{P}$ and $\bb{Q}$ are exactly the same, $\gamma_{\eu{F}_\star}(\bb{P},\bb{Q})=+\infty$ and therefore is a trivial and useless metric in practice.} 
%AG:the next paragraph is clumsy, and skirts the point: you can rather say upfront ``we show that only TV is both an IPM and a phi-divergence.
%BK: I mention ur point.
We therefore consider a more restricted function class than $\eu{F}_\star$ resulting in a variety of more interesting IPMs, including the Dudley metric, Wasserstein metric, total variation distance, etc. Now, the question is for what other, more restricted function classes $\eu{F}$ does there exist a $\phi$ such that $\gamma_\eu{F}$ is a $\phi$-divergence? We answer this in the following theorem, where we show that the total-variation distance is the only ``non-trivial" IPM that is also a $\phi$-divergence. We first introduce some notation. Let us define $\Scr{P}_\lambda$ as the set of all probability measures, $\bb{P}$ that are absolutely continuous with respect to some $\sigma$-finite measure, $\lambda$. For $\bb{P}\in\Scr{P}_\lambda$, let $p=\frac{d\bb{P}}{d\lambda}$ be the Radon-Nikodym derivative of $\bb{P}$ with respect to $\lambda$. Let $\Phi$ be the class of all convex functions $\phi:[0,\infty)\rightarrow(-\infty,\infty]$ continuous at $0$ and finite on $(0,\infty)$.\vspace{2mm}%\footnote{The recent submission by Reid and Williamson~\cite{Reid-09} mentions that the class of $\phi$-divergences $\{D_\phi\}_{\phi}$ on $M$ and the class of integral probability metrics $\{\gamma_\eu{F}\}_{\eu{F}\subset[-1,1]^M}$ on $M$ intersect at $D_{t\mapsto |t-1|}(\bb{P},\bb{Q})=:TV(\bb{P},\bb{Q})$, where $TV$ is the total variation distance. Here, we have proved a much stronger result which shows that there does not exist an $\eu{F}$ such that $\gamma_\eu{F}(\cdot,\cdot)$ is not a metric on $\mathscr{P}$ but equals $D_\phi(\cdot,\cdot)$ for some $\phi\ne t\mapsto |t-1|$. This is posed as one of the open problems of their work~\cite[pp. 56]{Reid-09}.}
\begin{thm}[Necessary and sufficient conditions]\label{thm:tv-mmd}
%Let $\eu{F}_\star$ be the set of all real-valued measurable functions on $M$ and let $\eu{F}_0\subsetneq\eu{F}_\star$. Let $\mathscr{P}_\lambda=\{\bb{P}:\bb{P}\ll \lambda,\,p=\frac{d\bb{P}}{d\lambda}\}$, where $\lambda$ is some $\sigma$-finite measure. Then 
Let $\eu{F}\subset\eu{F}_\star$ and $\phi\in\Phi$. Then for any $\bb{P},\bb{Q}\in\mathscr{P}_\lambda$, $\gamma_{\eu{F}}(\bb{P},\bb{Q})=D_{\phi}(\bb{P},\bb{Q})$ if and only if any one of the following hold:
\begin{itemize}
\item[\emph{(i)}] $\eu{F}=\{f:\Vert f\Vert_\infty\le \frac{\beta-\alpha}{2}\},$\\ $\phi(u)=\alpha(u-1)\llbracket 0\le u\le 1\rrbracket+\beta(u-1)\llbracket u\ge 1\rrbracket$ for some $\alpha<\beta<\infty$.\vspace{2mm}
\item[\emph{(ii)}] $\eu{F}=\{f:f=c,\,c\in\mathbb{R}\},$\\$\phi(u)=\alpha(u-1)\llbracket u\ge 0\rrbracket,\,\alpha\in\mathbb{R}$.\vspace{2mm}
\end{itemize}
\end{thm}
The proof idea is as follows. First note that $\gamma_\eu{F}$ in (\ref{Eq:ipm}) is a pseudometric\footnote{Given a set $M$, a \emph{metric} for $M$ is a function $\rho:M\times M\rightarrow\mathbb{R}_+$ such that \emph{(i)} $\forall\,x,\,\rho(x,x)=0$, \emph{(ii)} $\forall\,x,y,\,\rho(x,y)=\rho(y,x)$, \emph{(iii)} $\forall\,x,y,z,\,\rho(x,z)\le \rho(x,y)+\rho(y,z)$, and \emph{(iv)} $\rho(x,y)=0\Rightarrow x=y$. A pseudometric only satisfies \emph{(i)-(iii)} of the properties of a metric. Unlike a metric space $(M,\rho)$, points in a pseudometric space need not be distinguishable: one may have $\rho(x,y)=0$ for $x\ne y$.} on $\mathscr{P}_\lambda$ for any $\eu{F}$. Since we want to prove $\gamma_\eu{F}=D_\phi$, this suggests that we first study the conditions on $\phi$ for which $D_\phi$ is a pseudometric. This is answered by Lemma~\ref{lem:phipseudo}, which is a simple modification of a result in \cite[Theorem 2]{Khosravifard-07}.\vspace{2mm}
\begin{lem}\label{lem:phipseudo}
For $\phi\in\Phi$, $D_\phi$ is a pseudometric on $\mathscr{P}_\lambda$ if and only if $\phi$ is of the form 
\begin{equation}\label{Eq:discern-triangle}
\phi(u)=\alpha(u-1)\llbracket 0\le u \le 1\rrbracket+\beta(u-1)\llbracket u\ge 1\rrbracket,
\end{equation}
for some $\beta\ge\alpha$.\vspace{2mm}
\end{lem}
\begin{IEEEproof}
See Appendix~\ref{appendix-lem-phipseudo}.\vspace{2mm}
\end{IEEEproof}
The proof of Lemma~\ref{lem:phipseudo} uses the following result from \cite{Khosravifard-07}, which is quite easy to prove. 
\vspace{2mm}
\begin{lem}[\cite{Khosravifard-07}]\label{lem:tv}
For $\phi$ in (\ref{Eq:discern-triangle}), \begin{equation}\label{Eq:prop-tv}
D_\phi(\bb{P},\bb{Q})=\frac{\beta-\alpha}{2}\int_M |p-q|\,d\lambda,
\end{equation} for any $\bb{P},\bb{Q}\in\Scr{P}_\lambda$, where $p$ and $q$ are the Radon-Nikodym derivatives of $\bb{P}$ and $\bb{Q}$ with respect to $\lambda$.\vspace{2mm}
\end{lem}
Lemma \ref{lem:tv} shows that $D_\phi(\bb{P},\bb{Q})$ in (\ref{Eq:prop-tv}) associated with $\phi$ in (\ref{Eq:discern-triangle}) is proportional to the total variation distance between $\bb{P}$ and $\bb{Q}$. Note that the total variation distance between $\bb{P}$ and $\bb{Q}$ can be written as $\int_M|p-q|\,d\lambda$, where $p$ and $q$ are defined as in Lemma~\ref{lem:tv}.
\begin{IEEEproof}[Proof of Theorem~\ref{thm:tv-mmd}] 
%\begin{itemize}
$(\Leftarrow)\,\,$ Suppose \emph{(i)} holds. Then for any $\bb{P},\bb{Q}\in\mathscr{P}_\lambda$, we have 
\setlength{\arraycolsep}{0.0em}
\begin{eqnarray}
\gamma_{\eu{F}}(\bb{P},\bb{Q})&{}={}&\sup\left\{|\bb{P}f-\bb{Q}f|:\Vert f\Vert_\infty\le \frac{\beta-\alpha}{2}\right\}\nonumber\\
&{}={}&\frac{\beta-\alpha}{2}\sup\{|\bb{P}f-\bb{Q}f|:\Vert f\Vert_\infty\le 1\}\nonumber\\
&{}={}&\frac{\beta-\alpha}{2}\int_M |p-q|\,d\lambda\stackrel{(a)}{=}D_{\phi}(\bb{P},\bb{Q}),\nonumber
\end{eqnarray}
where $(a)$ follows from Lemma~\ref{lem:tv}.
\par Suppose \emph{(ii)} holds. Then $\gamma_{\eu{F}}(\bb{P},\bb{Q})=0$ and $D_{\phi}(\bb{P},\bb{Q})=\alpha\int_Mq\phi(p/q)\,d\lambda=\alpha\int_M(p-q)\,d\lambda=0$.\vspace{2mm}\\
$(\Rightarrow)\,\,$ Suppose $\gamma_\eu{F}(\bb{P},\bb{Q})=D_\phi(\bb{P},\bb{Q})$ for any $\bb{P},\bb{Q}\in\mathscr{P}_\lambda$. Since $\gamma_\eu{F}$ is a pseudometric on $\mathscr{P}_\lambda$ (irrespective of $\eu{F}$), $D_\phi$ is a pseudometric on $\mathscr{P}_\lambda$. Therefore, by Lemma~\ref{lem:phipseudo}, $\phi(u)=\alpha(u-1)\llbracket 0\le u\le 1\rrbracket +\beta(u-1)\llbracket u\ge 1\rrbracket$ for some $\beta\ge\alpha$. Now, let us consider two cases.\vspace{2mm}\\
\noindent \emph{Case 1:} $\beta>\alpha$\\
By Lemma~\ref{lem:tv}, $D_\phi(\bb{P},\bb{Q})=\frac{\beta-\alpha}{2}\int_M|p-q|\,d\lambda$. Since $\gamma_\eu{F}(\bb{P},\bb{Q})=D_\phi(\bb{P},\bb{Q})$ for all $\bb{P},\bb{Q}\in\mathscr{P}_\lambda$, we have $\gamma_\eu{F}(\bb{P},\bb{Q})=\frac{\beta-\alpha}{2}\int_M|p-q|\,d\lambda=\frac{\beta-\alpha}{2}\sup\{|\bb{P}f-\bb{Q}f|:\Vert f\Vert_\infty\le 1\}=\sup\{|\bb{P}f-\bb{Q}f|:\Vert f\Vert_\infty\le \frac{\beta-\alpha}{2}\}$ and therefore $\eu{F}=\{f:\Vert f\Vert_\infty\le\frac{\beta-\alpha}{2}\}$.\vspace{2mm}\\
\noindent \emph{Case 2:} $\beta=\alpha\\\phi(u)=\alpha(u-1),u\ge 0,\,\alpha<\infty$. Now, $D_\phi(\bb{P},\bb{Q})=\int_M q\phi(p/q)\,d\lambda=\alpha\int_M (p-q)=0$ for all $\bb{P},\bb{Q}\in\mathscr{P}_\lambda$. Therefore, $\gamma_\eu{F}(\bb{P},\bb{Q})=\sup_{f\in\eu{F}}|\bb{P}f-\bb{Q}f|=0$ for all $\bb{P},\bb{Q}\in\mathscr{P}_\lambda$, which means $\forall\,\bb{P},\bb{Q}\in\mathscr{P}_\lambda$, $\forall\,f\in\eu{F}$, $\bb{P}f=\bb{Q}f$. This, in turn, means $f$ is a constant on $M$, i.e., $\eu{F}=\{f:f=c,\,c\in\mathbb{R}\}$.\vspace{2mm}%\qed
%\end{itemize}
\end{IEEEproof}
Note that in Theorem~\ref{thm:tv-mmd}, the cases \emph{(i)} and \emph{(ii)} are disjoint as $\alpha<\beta$ in case \emph{(i)} and $\alpha=\beta$ in case \emph{(ii)}. Case \emph{(i)} shows that the family of $\phi$-divergences and the family of IPMs intersect only at the total variation distance, which follows from Lemma~\ref{lem:tv}. %\footnote{The intersection happens only at the variational distance as it is the only case for which $D_\phi$ is a metric. However, some functions of $D_\phi$ are shown to be metrics on $\mathscr{P}_0$, for example, the square root of variational distance, the square root of Hellinger's distance, the square root of Jensen-Shannon divergence etc., are shown to be metrics on $\mathscr{P}_0$~\cite{Endres-03, Fuglede-03, Hein-05}. Similarly, \cite[Theorem 1]{Osterreicher-03} has shown that certain powers of $D_\phi$ are a metric on $\mathscr{P}_0$. So, one can ask for conditions on $\eu{F}$ so that $\gamma_\eu{F}$ equals some function of $D_\phi$. This is one of the open questions of this work.} 
Case \emph{(ii)} is trivial as the distance between any two probability measures is zero. %Note that choosing a trivial kernel, $k(x,y)=C,\,C>0$ induces an RKHS $\eu{H}\subsetneq\eu{F}$ (where $\eu{F}$ is defined in \emph{(ii)}) which trivially coincides with the class of $\phi$-divergence. 
This result shows that IPMs and $\phi$-divergences are essentially different. %Note that we have restricted $\eu{F}$ to a proper subset of $\eu{F}_\star$. Otherwise, $\gamma_\eu{F}$ trivially coincides with the family of $\phi$-divergences through Theorem~\ref{thm:phi-mmd}. 
Theorem~\ref{thm:tv-mmd} also addresses the open question posed by Reid and Williamson~\cite[pp. 56]{Reid-09} of ``whether there exist $\eu{F}$ such that $\gamma_\eu{F}$ is not a metric but equals $D_\phi$ for some $\phi\ne t\mapsto |t-1|$?" This is answered affirmatively by case \emph{(ii)} in Theorem~\ref{thm:tv-mmd} as $\gamma_\eu{F}$ with $\eu{F}=\{f\,:\,f=c,\,c\in\bb{R}\}$ is a pseudometric (not a metric) on $\Scr{P}_\lambda$ but equals $D_\phi$ for $\phi(u)=\alpha(u-1)\llbracket u\ge 0\rrbracket\ne u\mapsto |u-1|$. 
\section{Non-parametric Estimation of IPMs}%: Consistency and Convergence Rate Analysis}
\label{Sec:consistency}
As mentioned in Section~\ref{subsubsec:estimation}, the estimation of distance between $\bb{P}$ and $\bb{Q}$ is an important problem in statistical inference applications like distribution testing, where $\bb{P}$ and $\bb{Q}$ are known only through random i.i.d. samples. Another instance where an estimate of the distance between $\bb{P}$ and $\bb{Q}$ is useful is as follows. 
% In the previous section, we showed that IPMs and $\phi$-divergences are essentially different classes of distance measures on $\Scr{P}$. 
Suppose one wishes to compute the Wasserstein distance or Dudley metric between $\bb{P}$ and $\bb{Q}$. This is not straightforward as the explicit calculation, i.e., in closed form, is difficult for most concrete examples.\footnote{The explicit form for the Wasserstein distance in (\ref{Eq:primal-wasserstein}) is known for $(M,\rho(x,y))=(\bb{R},|x-y|)$ \cite{Vallander-73,Vajda-89}, which is given as $W_1(\bb{P},\bb{Q})=\int_{(0,1)}|F^{-1}_\bb{P}(u)-F^{-1}_\bb{Q}(u)|\,du=\int_\bb{R}|F_\bb{P}(x)-F_\bb{Q}(x)|\,dx$, where $F_\bb{P}(x)=\bb{P}((-\infty,x])$ and $F^{-1}_\bb{P}(u)=\inf\{x\in\bb{R}|F_\bb{P}(x)\ge u\}$, $0<u<1$. It is easy to show that this explicit form can be extended to $(\bb{R}^d,\Vert \cdot\Vert_1)$. However, the exact computation (in closed form) of $W_1(\bb{P},\bb{Q})$ is not straightforward %AG: empirically, or in closed form? %BK: closed form-clarified it
 for all $\bb{P}$ and $\bb{Q}$. See Section~\ref{subsec:simulation} for some examples where $W_1(\bb{P},\bb{Q})$ can be computed exactly. Note that since $\bb{R}^d$ is separable, by the Kantorovich-Rubinstein theorem, $W(\bb{P},\bb{Q})=W_1(\bb{P},\bb{Q}),\,\forall\,\bb{P},\bb{Q}$.%Therefore, statistical applications (e.g. hypothesis tests) involving the Wasserstein distance are restricted to $\bb{R}$ \cite{Barrio-99}. However, by appealing to the dual representation, we show that the Wasserstein distance can be simply estimated by solving a linear program and the estimator is strongly consistent. Therefore, the Wasserstein distance can be used in statistical applications for any separable $M$.
\label{footnote:vallander}} Similar is the case with MMD and $\phi$-divergences for certain distributions, where the one approach to compute the distance between $\bb{P}$ and $\bb{Q}$ is to draw random i.i.d. samples from each, and estimate the distance based on these samples. We need the estimator to be such that the estimate converges to the true distance with large sample sizes.  
\par To this end, the non-parametric estimation of $\phi$-divergences, especially the KL-divergence is well studied (see \cite{Wang-05,Wang-06,Nguyen-08} and references therein). As mentioned before, the drawback with $\phi$-divergences is that they are difficult to estimate in high dimensions and the rate of convergence of the estimator can be arbitrarily slow depending on the distributions \cite{Wang-05}. Since IPMs and $\phi$-divergences are essentially different classes of distance measures on $\Scr{P}$, in Section~\ref{subsec:empirical}, we consider the non-parametric estimation of IPMs, especially the Wasserstein distance, Dudley metric and MMD. We show that the Wasserstein and Dudley metrics can be estimated by solving linear programs (see Theorems \ref{thm:Lipschitz} and \ref{thm:Dudley}) whereas an estimator for MMD can be obtained in closed form (\cite{Gretton-06}; see Theorem \ref{thm:parzen} below). These results are significant because to our knowledge, statistical applications (e.g. hypothesis tests) involving the Wasserstein distance in (\ref{Eq:primal-wasserstein}) are restricted only to $\bb{R}$ \cite{Barrio-99} as the closed form expression for the Wasserstein distance is known only for $\bb{R}$ (see footnote~\ref{footnote:vallander}). 
%In our case, the results in Section~\ref{subsec:empirical} show that the estimation of above mentioned IPMs is possible without any difficulty even in $\bb{R}^d$ ($d>1$) and therefore can be used in testing applications.
%In Sections \ref{Sec:Lipschitz}-\ref{Sec:mmd}, we have considered the empirical estimators of the Wasserstein distance, Dudley metric and MMD, while relating them to the binary classification problem. Ideally, given $\bb{P}$ and $\bb{Q}$, for example say, two normal distributions on $\bb{R}$, one would like to compute these distances directly without resorting to estimating them from finite samples. However, this is not straightforward in the case of Wasserstein and Dudley metrics whereas in the case of MMD, \cite{Gretton-06,Sriperumbudur-08} have shown that $\gamma_\eu{F}(\bb{P},\bb{Q})=\Vert \bb{P}k-\bb{Q}k\Vert_\eu{H}$, where $\eu{F}=\{f:\Vert f\Vert_\eu{H}\le 1\}$, $\eu{H}$ is an RKHS with a measurable and bounded r.k., $k$. Even in the case of MMD, one might have to resort to numerical integration techniques to compute it depending on $\bb{P}$ and $\bb{Q}$. Instead as we have shown before, estimating these distances from samples drawn i.i.d. from $\bb{P}$ and $\bb{Q}$ is very simple and straightforward (see Eqs.~(\ref{Eq:lp}), (\ref{Eq:lp-dudley}) and (\ref{Eq:gamma-opt})). Therefore, if these estimators are statistically consistent, then a pretty good estimate of these distances (in the population sense) can be obtained by sampling arbitrarily large number of i.i.d. samples from $\bb{P}$ and $\bb{Q}$. 
\par In Section~\ref{subsec:consistency}, we present the consistency and convergence rate analysis of these estimators. To this end, in Theorem~\ref{Thm:consistency}, we present a general result on the statistical consistency of the estimators of IPMs by using tools from empirical process theory~\cite{Vaart-96}. As a special case, in Corollary~\ref{cor:wasserstein-dudley}, we show that the estimators of Wasserstein distance and Dudley metric are \emph{strongly consistent}, i.e., suppose $\{\theta_l\}$ is a sequence of estimators of $\theta$, then $\theta_l$ is strongly consistent if $\theta_l$ converges a.s. to $\theta$ as $l\rightarrow \infty$. Then, in Theorem~\ref{Thm:rate}, we provide a probabilistic bound on the deviation between $\gamma_\eu{F}$ and its estimate for any $\eu{F}$ in terms of the Rademacher complexity (see Definition~\ref{def:Rademacher}), which is then used to derive the rates of convergence for the estimators of Wasserstein distance, Dudley metric and MMD in Corollary~\ref{cor:rate}. Using the Borel-Cantelli lemma, we then show that the estimator of MMD is also strongly consistent. In Section~\ref{subsec:simulation}, we present simulation results to demonstrate the performance of these estimators. Overall, the results in this section show that IPMs (especially the Wasserstein distance, Dudley metric and MMD) are easier to estimate than the KL-divergence and the IPM estimators exhibit better convergence behavior \cite{Wang-05,Nguyen-08}.
\par Since the total variation distance is also an IPM, we discuss its empirical estimation and consistency in Section~\ref{subsec:tv}. By citing earlier work \cite{Devroye-90}, we show that the empirical estimator of the total variation distance is not consistent. Since the total variation distance cannot be estimated consistently, in Theorem~\ref{thm:TVbound}, we provide two lower bounds on the total variation distance, one involving the Wasserstein distance and Dudley metric and the other involving MMD. These bounds can be estimated consistently based on the results in Section~\ref{subsec:consistency} and, moreover, they translate to lower bounds on the KL-divergence through Pinsker's inequality (see \cite{Fedotov-03} and references therein for more lower bounds on the KL-divergence in terms of the total variation distance).
%\footnote{Using similar techniques, recently \cite{Nguyen-08} has studied the statistical consistency of an estimator of $\phi$-divergence, where the estimator is constructed by replacing $\bb{P}$, $\bb{Q}$ and $\eu{F}_\star$ with $\bb{P}_m$, $\bb{Q}_n$ and $\eu{F}\subsetneq\eu{F}_\star$ respectively in (\ref{Eq:dual-phi-div}), which is the variational form of $\phi$-divergence.} 
\subsection{Non-parametric estimation of Wasserstein distance, Dudley metric and MMD}\label{subsec:empirical}
Let $\{X^{(1)}_1,X^{(1)}_2,\ldots,X^{(1)}_m\}$ and $\{X^{(2)}_1,X^{(2)}_2,\ldots,X^{(2)}_n\}$ be i.i.d. samples drawn randomly from $\bb{P}$ and $\bb{Q}$ respectively. The empirical estimate of $\gamma_\eu{F}(\bb{P},\bb{Q})$ is given by
\begin{equation}\label{Eq:ipm-empirical}
\gamma_\eu{F}(\bb{P}_m,\bb{Q}_n)=\sup_{f\in\eu{F}}\left|\sum^N_{i=1}\widetilde{Y}_if(X_i)\right|,
\end{equation}
where $\bb{P}_m$ and $\bb{Q}_n$ represent the empirical distributions of $\bb{P}$ and $\bb{Q}$, $N=m+n$ and 
\begin{equation}
\widetilde{Y}_i=\left\{\begin{array}{c@{\quad\quad}l}
\frac{1}{m},& X_i=X^{(1)}_.\\
-\frac{1}{n},& X_i=X^{(2)}_.
\end{array}\right..
\end{equation}
The computation of $\gamma_\eu{F}(\bb{P}_m,\bb{Q}_n)$ in (\ref{Eq:ipm-empirical}) is not straightforward for any arbitrary $\eu{F}$. In the following, we restrict ourselves to $\eu{F}_W:=\{f:\Vert f\Vert_L\le 1\}$, $\eu{F}_\beta:=\{f:\Vert f\Vert_{BL}\le 1\}$ and $\eu{F}_k:=\{f:\Vert f\Vert_\eu{H}\le 1\}$ and compute (\ref{Eq:ipm-empirical}). 
%AG: starting to get too much notation below:
%BK: i defined these in the "notation" section. but still for easyness, i wrote here too. may be we can remove it.
Let us denote $W:=\gamma_{\eu{F}_W}$, $\beta:=\gamma_{\eu{F}_\beta}$ and $\gamma_k:=\gamma_{\eu{F}_k}$. %Now, we have the following results. 
%AG: don't need this:
%BK: removed
%Later in Sections \ref{Sec:Lipschitz} and \ref{Sec:mmd}, we show how these empirical estimates appear in binary classification.
\vspace{2mm}
\begin{thm}[Estimator of Wasserstein distance]\label{thm:Lipschitz}
For all $\alpha\in[0,1]$, the following function solves (\ref{Eq:ipm-empirical}) for $\eu{F}=\eu{F}_W$:
\begin{eqnarray}\label{Eq:h-alpha}
\setlength{\arraycolsep}{0.0em}
f_\alpha(x)&{}:={}&\alpha\min_{i=1,\ldots,N}(a^\star_i+\rho(x,X_i))\nonumber\\
&&\quad+(1-\alpha)\max_{i=1,\ldots,N}(a^\star_i-\rho(x,X_i)),
\end{eqnarray}
where \begin{equation}
W(\bb{P}_m,\bb{Q}_n)=\sum^N_{i=1}\widetilde{Y}_ia^\star_i,
\end{equation} and $\{a^\star_i\}^N_{i=1}$ solve the following linear program,
\setlength{\arraycolsep}{0.0em}
\begin{eqnarray}\label{Eq:lp}
\max_{a_1,\ldots,a_N} &{}{}& \,\,\,\sum^N_{i=1}\widetilde{Y}_ia_i\nonumber\\
\text{s.t.} &{}{}& \,\,\,-\rho(X_i,X_j)\le a_i-a_j\le \rho(X_i,X_j),\forall\,i,j.
\end{eqnarray}
%In addition, if $\{a^\star_i\}^N_{i=1}$ satisfies $\min_{i\in I^+}a^\star_i\ge 0$, $\max_{i\in I^-}a^\star_i\le 0$, $\min_{i\in I^+}a^\star_i+\min_{i\in I^-}a^\star_i=0$ and $\max_{i\in I^+}a^\star_i+\max_{i\in I^-}a^\star_i=0$, then $\text{sign}(f_{\frac{1}{2}}(x))$ is a $1$-NN classifier.\vspace{1mm}
\end{thm}
\begin{IEEEproof}
Consider $W(\bb{P}_m,\bb{Q}_n)=\sup\{\sum^N_{i=1}\widetilde{Y}_if(X_i):\Vert f\Vert_L\le 1\}$. Note that
\begin{equation}
1\ge\Vert f\Vert_L=\sup_{x\ne x'}\frac{|f(x)-f(x')|}{\rho(x,x')}\ge \max_{X_i\ne X_j}\frac{|f(X_i)-f(X_j)|}{\rho(X_i,X_j)},\nonumber
\end{equation}
which means
\setlength{\arraycolsep}{0.0em}
\begin{eqnarray}\label{Eq:wass-temp}
W(\bb{P}_m,\bb{Q}_n)\le \sup&{}{}&\,\,\,\sum^N_{i=1}\widetilde{Y}_if(X_i)\nonumber\\
\text{s.t.}&{}{}&\,\,\, \max_{X_i\ne X_j}\frac{|f(X_i)-f(X_j)|}{\rho(X_i,X_j)}\le 1.
\end{eqnarray}
The right hand side of (\ref{Eq:wass-temp}) can be equivalently written as 
\setlength{\arraycolsep}{0.0em}
\begin{eqnarray}
\sup&{}{}&\,\,\,\sum^N_{i=1}\widetilde{Y}_if(X_i)\nonumber\\
\text{s.t.}&{}{}&\,\,\,-\rho(X_i,X_j)\le f(X_i)-f(X_j)\le \rho(X_i,X_j),\forall\,i,j.\nonumber
\end{eqnarray} 
Let $a_i:=f(X_i)$. Therefore, we have $W(\bb{P}_m,\bb{Q}_n)\le \sum^N_{i=1}\widetilde{Y}_ia^\star_i$, where $\{a^\star_i\}^N_{i=1}$ solve the linear program in (\ref{Eq:lp}). 
%\begin{eqnarray}\label{Eq:lp-ineq}
%\max_{a_1,\ldots,a_n} && \sum^N_{i=1}\widetilde{Y}_ia_i\nonumber\\
%\text{s.t.} && -\rho(X_i,X_j)\le a_i-a_j\le\rho(X_i,X_j),\,\forall\,i,j=1\ldots,n.
%\end{eqnarray}
%In addition, it is easy to see that
Note that the objective in (\ref{Eq:lp}) is linear in $\{a_i\}^N_{i=1}$ with linear inequality constraints and therefore by Theorem~\ref{Thm:rockafeller} (see Appendix~\ref{appendix-supp}), %For readability, it makes no sense to have Theorem 22 in the appendix, when the entropy definition needed to parse it is given here in the text.
the optimum lies on the boundary of the constraint set, which means 
%$\{(a_1,\ldots,a_n):a_i-a_j=\rho(X_i,X_j),\,\forall\,i,j=1\ldots,n\}$. So, we 
%have 
$\max_{X_i\ne X_j}\frac{|a^\star_i-a^\star_j|}{\rho(X_i,X_j)}=1$. Therefore, by Lemma~\ref{lem:lip} (see Appendix~\ref{appendix-supp}), $f$ on $\{X_1,\ldots,X_N\}$ can be extended to a function $f_\alpha$ (on $M$) defined in (\ref{Eq:h-alpha}) where $f_\alpha(X_i)=f(X_i)=a^\star_i$ and $\Vert f_\alpha\Vert_L=\Vert f\Vert_L=1$, which means $f_\alpha$ is a maximizer of (\ref{Eq:ipm-empirical}) and $W(\bb{P}_m,\bb{Q}_n)=\sum^N_{i=1}\widetilde{Y}_ia^\star_i$.\vspace{2mm}
\end{IEEEproof}
%\begin{rem}\label{rem:wasserstein}
%The conditions on $\{a^\star_i\}^N_{i=1}$ in Theorem~\ref{thm:Lipschitz} are satisfied if $m=n$ and $\rho(X_i,X_j)=2c,\,\forall\,i,j$ for some $c>0$ as $a^\star_i=c,\,\forall\,i\in I^+$ and $a^\star_i=-c,\,\forall\,i\in I^-$. It is not clear whether this is also necessary for the conditions in Theorem~\ref{thm:Lipschitz} to hold.\vspace{1mm}%However, such a condition on $\{X_i\}^N_{i=1}$ is too restrictive. 
%\end{rem}
\begin{thm}[Estimator of Dudley metric]\label{thm:Dudley}
For all $\alpha\in[0,1]$, the following function solves (\ref{Eq:ipm-empirical}) for $\eu{F}=\eu{F}_\beta$: 
\begin{equation}\label{Eq:g-alpha-dudley}
g_\alpha(x):=\max\left(-\max_{i=1,\ldots,N}|a^\star_i|,\min\left(h_\alpha(x),\max_{i=1,\ldots,N}|a^\star_i|\right)\right)
\end{equation}
where 
\setlength{\arraycolsep}{0.0em}
\begin{eqnarray}\label{Eq:h-alpha-dudley}
h_\alpha(x)&{}:={}&\alpha\min_{i=1,\ldots,N}(a^\star_i+L^\star\rho(x,X_i))\nonumber\\
&{}{}&\quad+(1-\alpha)\max_{i=1,\ldots,N}(a^\star_i-L^\star\rho(x,X_i)),
\end{eqnarray}
\begin{equation}\label{Eq:dudley-estimate}
\beta(\bb{P}_m,\bb{Q}_n)=\sum^N_{i=1}\widetilde{Y}_ia^\star_i,
\end{equation}
\begin{equation}
L^\star=\max_{X_i\ne X_j}\frac{|a^\star_i-a^\star_j|}{\rho(X_i,X_j)},
\end{equation}
and $\{a^\star_i\}^N_{i=1}$ solve the following linear program,
\setlength{\arraycolsep}{0.0em}
\begin{eqnarray}\label{Eq:lp-dudley}
\max_{a_1,\ldots,a_N,b,c} &{}{}&\,\,\, \sum^N_{i=1}\widetilde{Y}_ia_i\nonumber\\
\text{s.t.} &{}{}& \,\,\,-b\,\rho(X_i,X_j)\le a_i-a_j\le b\,\rho(X_i,X_j),\,\forall\,i,j\nonumber\\
&{}{}&\,\,\, -c\le a_i\le c,\,\forall\,i\nonumber\\
&{}{}&\,\,\, b+c\le 1.
\end{eqnarray}
%In addition, if $\{a^\star_i\}^N_{i=1}$ satisfies $\min_{i\in I^+}a^\star_i\ge 0$, $\max_{i\in I^-}a^\star_i\le 0$, $\min_{i\in I^+}a^\star_i+\min_{i\in I^-}a^\star_i=0$ and $\max_{i\in I^+}a^\star_i+\max_{i\in I^-}a^\star_i=0$, then $\text{sign}(g_{\frac{1}{2}}(x))$ is a $1$-NN classifier.\vspace{1mm}
\end{thm}
\begin{IEEEproof}
The proof is similar to that of Theorem~\ref{thm:Lipschitz}. Note that 
\setlength{\arraycolsep}{0.0em}
\begin{eqnarray}
1&{}\ge{}&\Vert f\Vert_L+\Vert f\Vert_\infty=\sup_{x\ne y}\frac{|f(x)-f(y)|}{\rho(x,y)}+\sup_{x\in M}|f(x)|\nonumber\\
&{}{}&\quad \ge \max_{X_i\ne X_j}\frac{|f(X_i)-f(X_j)|}{\rho(X_i,X_j)}+\max_{i}|f(X_i)|,\nonumber
\end{eqnarray}
which means
\setlength{\arraycolsep}{0.0em}
\begin{eqnarray}
\beta(\bb{P}_m,\bb{Q}_n)&{}\le{}&\sup\Big\{\sum^N_{i=1}\widetilde{Y}_if(X_i):\max_{i}|f(X_i)|\nonumber\\
&{}{}&\quad+\max_{X_i\ne X_j}\frac{|f(X_i)-f(X_j)|}{\rho(X_i,X_j)}\le 1\Big\}.
\end{eqnarray}
Let $a_i:=f(X_i)$. Therefore, $\beta(\bb{P}_m,\bb{Q}_n)\le \sum^N_{i=1}\widetilde{Y}_ia^\star_i$, where $\{a^\star_i\}^N_{i=1}$ solve
\setlength{\arraycolsep}{0.0em}
\begin{eqnarray}\label{Eq:lp-dudley-1}
\max_{a_1,\ldots,a_N} &{}{}&\,\,\, \sum^N_{i=1}\widetilde{Y}_ia_i\nonumber\\
\text{s.t.} &{}{}& \,\,\,\max_{X_i\ne X_j}\frac{|a_i-a_j|}{\rho(X_i,X_j)}+\max_i|a_i|\le 1.
\end{eqnarray}
Introducing variables $b$ and $c$ such that $\max_{X_i\ne X_j}\frac{|a_i-a_j|}{\rho(X_i,X_j)}\le b$ and $\max_i|a_i|\le c$ reduces the program in (\ref{Eq:lp-dudley-1}) to (\ref{Eq:lp-dudley}). In addition, it is easy to see that the optimum occurs at the boundary of the constraint set and therefore $\max_{X_i\ne X_j}\frac{|a_i-a_j|}{\rho(X_i,X_j)}+\max_i|a_i|=1$. Hence, by Lemma~\ref{lem:lip-bounded} (see Appendix~\ref{appendix-supp}), $g_\alpha$ in (\ref{Eq:g-alpha-dudley}) extends $f$ defined on $\{X_1,\ldots,X_n\}$ to $M$, i.e., $g_\alpha(X_i)=f(X_i)$ and $\Vert g_\alpha\Vert_{BL}=\Vert f\Vert_{BL}=1$. Note that $h_\alpha$ in (\ref{Eq:h-alpha-dudley}) is the Lipschitz extension of $g$ to $M$ (by Lemma~\ref{lem:lip}). Therefore, $g_\alpha$ is a solution to (\ref{Eq:ipm-empirical}) and (\ref{Eq:dudley-estimate}) holds.\vspace{2mm}
%\par To show that $\text{sign}(g_{\frac{1}{2}}(x))$ is a $1$-NN classifier, first note that $\text{sign}(h_{\frac{1}{2}}(x))$ is a $1$-NN classifier under the conditions on $\{a^\star_i\}^N_{i=1}$ mentioned in Theorem~\ref{thm:Dudley}, whose proof is the same as that of Theorem~\ref{thm:Lipschitz}. Therefore, it is easy to check that $\text{sign}(g_{\frac{1}{2}}(x))$ is a $1$-NN classifier.\vspace{1mm}
\end{IEEEproof}
%As mentioned in Remark~\ref{rem:wasserstein}, the same sufficient condition ensures that the conditions on $\{a^\star_i\}^N_{i=1}$ in Theorem~\ref{thm:Dudley} are satisfied. We now relate $\beta$ to the bounded Lipschitz classifier.\vspace{1mm}
\begin{thm}[Estimator of MMD \cite{Gretton-06}]\label{thm:parzen}
For $\eu{F}=\eu{F}_k$, the following function is the unique solution to (\ref{Eq:ipm-empirical}):
\begin{equation}\label{Eq:optimum-g}
f=\frac{1}{\Vert\sum^N_{i=1}\widetilde{Y}_ik(\cdot,X_i)\Vert_\eu{H}}\sum^N_{i=1}\widetilde{Y}_ik(\cdot,X_i),
\end{equation}
and \begin{equation}\label{Eq:gamma-opt}
\gamma_k(\bb{P}_m,\bb{Q}_n)=\left\Vert\sum^N_{i=1}\widetilde{Y}_ik(\cdot,X_i)\right\Vert_\eu{H}=\sqrt{\sum^N_{i,j=1}\widetilde{Y}_i\widetilde{Y}_jk(X_i,X_j)}.
\end{equation}
%In addition, $\text{sign}(g(x))$ is the Parzen window classifer, given as
%\begin{equation}\label{Eq:Parzen}
%y=\left\{\begin{array}{c@{\quad}l}
%+1,& \frac{1}{m}\sum_{i\in I^+}k(x,X_i)>\frac{1}{n}\sum_{i\in I^-}k(x,X_i)\\
%-1,& \frac{1}{n}\sum_{i\in I^+}k(x,X_i)\le \frac{1}{m}\sum_{i\in I^-}k(x,X_i)\\
%\end{array}\right.,
%\end{equation}
%where $y:=\text{sign}(g(x))$.\vspace{1mm}
\end{thm}
\begin{IEEEproof}
Consider $\gamma_k(\bb{P}_m,\bb{Q}_n):=\sup\{\sum^N_{i=1}\widetilde{Y}_if(X_i)\,:\,\Vert f\Vert_\eu{H}\le 1\}$, which can be written as
\begin{equation}\label{Eq:rewrite-empirical}
\gamma_k(\bb{P}_m,\bb{Q}_n)=\sup_{\Vert f\Vert_\eu{H}\le 1}\left\langle f,\sum^N_{i=1}\widetilde{Y}_ik(\cdot,X_i)\right\rangle_\eu{H},
\end{equation}
where we have used the reproducing property of $\eu{H}$, i.e., $\forall\,f\in\eu{H},\,\forall\,x\in M,\,f(x)=\langle f,k(\cdot,x)\rangle_\eu{H}$. The result follows from using the Cauchy-Schwartz inequality. 
%It is easy to see that the objective function in (\ref{Eq:rewrite-empirical}) is linear in $f$ and the constraint set is convex in $f$ and therefore by Theorem~\ref{Thm:rockafeller} (see the Appendix), the optimum occurs on the boundary of the constraint set, i.e., $\{f\in\eu{H}:\Vert f\Vert_\eu{H}=1\}$.
%The Lagrangian function associated with (\ref{Eq:rewrite-empirical}) is given by
%\begin{equation}\label{Eq:lagrangian}
%J(f,\lambda)=\langle f,\sum^N_{i=1}\widetilde{Y_i}k(\cdot,X_i)\rangle_\eu{H}-\lambda(\langle f,f\rangle^{1/2}_\eu{H}-1),
%\end{equation}
%where $\lambda\ge 0$. Minimizing $J$ with respect to $f$ gives 
%\begin{equation}
%f=\frac{\Vert f\Vert_\eu{H}}{\lambda}\sum^N_{i=1}\widetilde{Y}_ik(\cdot,X_i)=\frac{1}{\lambda}\sum^N_{i=1}\widetilde{Y}_ik(\cdot,X_i),\nonumber
%\end{equation}
%which implies $\lambda=\Vert\sum^N_{i=1}\widetilde{Y}_ik(\cdot,X_i)\Vert_\eu{H}$, therefore resulting in $f$ as in (\ref{Eq:optimum-g}). Using this $f$ in (\ref{Eq:rewrite-empirical}) with $k(X_i,X_j)=\langle k(\cdot,X_i),k(\cdot,X_j)\rangle_\eu{H}$ yields (\ref{Eq:gamma-opt}).
\vspace{2mm}
\end{IEEEproof}
One important observation to be made is that estimators in Theorems \ref{thm:Lipschitz}--\ref{thm:parzen} depend on $\{X_i\}^N_{i=1}$ through $\rho$ or $k$. This means, once $\{\rho(X_i,X_j)\}^N_{i,j=1}$ or $\{k(X_i,X_j)\}^N_{i,j=1}$ is known, the complexity of the corresponding estimators is independent of the dimension $d$ when $M=\bb{R}^d$, unlike in the estimation of KL-divergence. In addition, because these estimators depend on $M$ %AG2: correct? ``On M'' was missing.
only through $\rho$ or $k$, the domain $M$ is immaterial as long as $\rho$ or $k$ is defined on $M$. Therefore, these estimators extend to arbitrary domains unlike the KL-divergence, where the domain is usually chosen to be $\bb{R}^d$. 
%The above result related to MMD also appears in \cite{Gretton-06}. We presented here for completeness. %AG: first, I would not re-derive this result here when it has been proved earlier. Second, why do you use this complex proof when Cauchy-Schwarz suffices? 
%BK: now i used cauchy-swartz. with the proof being short, its ok i believe. i just wanted to give it, as we are using this technqiue at many places.
\subsection{Consistency and rate of convergence}\label{subsec:consistency}
In Section~\ref{subsec:empirical}, we presented the empirical estimators of $W,\,\beta$ and $\gamma_k$. For these estimators to be reliable, we need them to converge to the population values as $m,n\rightarrow\infty$. Even if this holds, we would like to have a fast rate of convergence such that in practice, fewer samples are sufficient to obtain reliable estimates. We address these issues in this section.
\par Before we start presenting the results, we briefly introduce some terminology and notation from empirical process theory. For any $r\ge 1$ and probability measure $\bb{Q}$, define the $L_r$ norm $\Vert f\Vert_{\bb{Q},r}:=(\int |f|^r\,d\bb{Q})^{1/r}$ and let $L_r(\bb{Q})$ denote the metric space induced by this norm. The \emph{covering number} $\mathcal{N}(\varepsilon,\eu{F},L_r(\bb{Q}))$ is the minimal number of $L_r(\bb{Q})$ balls of radius $\varepsilon$ needed to cover $\eu{F}$. $\mathcal{H}(\varepsilon,\eu{F},L_r(\bb{Q})):=\log\mathcal{N}(\varepsilon,\eu{F},L_r(\bb{Q}))$ is called the \emph{entropy} of $\eu{F}$ using the $L_r(\bb{Q})$ metric. Define the minimal envelope function: $F(x):=\sup_{f\in \eu{F}}|f(x)|$.
\par We now present a general result on the strong consistency of $\gamma_\eu{F}(\bb{P}_m,\bb{Q}_n)$, which simply follows from Theorem~\ref{Thm:vandegeer} (see Appendix~\ref{appendix-supp}).\vspace{2mm}
\begin{thm}\label{Thm:consistency}
Suppose the following conditions hold:
\begin{itemize}
\item[\emph{(i)}] $\int F\,d\bb{P}<\infty$. 
\item[\emph{(ii)}] $\int F\,d\bb{Q}<\infty$. 
\item[\emph{(iii)}] $\forall\varepsilon>0$, $\frac{1}{m}\mathcal{H}(\varepsilon,\eu{F},L_1(\bb{P}_m))\stackrel{\bb{P}}{\longrightarrow} 0$ as $m\rightarrow\infty$.
\item[\emph{(iv)}] $\forall\varepsilon>0$, $\frac{1}{n}\mathcal{H}(\varepsilon,\eu{F},L_1(\bb{Q}_n))\stackrel{\bb{Q}}{\longrightarrow} 0$ as $n\rightarrow\infty$.
\end{itemize}
Then, $|\gamma_\eu{F}(\bb{P}_m,\bb{Q}_n)-\gamma_\eu{F}(\bb{P},\bb{Q})|\stackrel{a.s.}{\longrightarrow} 0$ as $m,n\rightarrow\infty$.\vspace{2mm}
\end{thm}
\begin{IEEEproof}
See Appendix~\ref{appendix-strong}.
\vspace{2mm}
\end{IEEEproof}
The following corollary to Theorem~\ref{Thm:consistency} shows that $W(\bb{P}_m,\bb{Q}_n)$ and $\beta(\bb{P}_m,\bb{Q}_n)$ are strongly consistent.\vspace{2mm}
\begin{cor}[Consistency of $W$ and $\beta$]\label{cor:wasserstein-dudley}
Let $(M,\rho)$ be a totally bounded metric space. Then, as $m,n\rightarrow\infty$,
\begin{itemize}
\item[\emph{(i)}] $|W(\bb{P}_m,\bb{Q}_n)-W(\bb{P},\bb{Q})|\stackrel{a.s.}{\longrightarrow} 0$.
\item[\emph{(ii)}] $|\beta(\bb{P}_m,\bb{Q}_n)-\beta(\bb{P},\bb{Q})|\stackrel{a.s.}{\longrightarrow} 0$.
\end{itemize}
\vspace{2mm}
\end{cor}
\begin{IEEEproof}
For any $f\in \eu{F}_W$, 
\setlength{\arraycolsep}{0.0em}
\begin{eqnarray}
\lefteqn{f(x)\le\sup_{x\in M}|f(x)|\le\sup_{x,y}|f(x)-f(y)|\le}\hspace{5mm}\nonumber\\
&&\Vert f\Vert_L\sup_{x,y}\rho(x,y)\le \Vert f\Vert_L \text{diam}(M)\le \text{diam}(M)<\infty,\nonumber
\end{eqnarray} 
where $\text{diam}(M)$ represents the diameter of $M$. Therefore, $\forall\,x\in M,\,F(x)\le \text{diam}(M)<\infty$, which satisfies \emph{(i)} and \emph{(ii)} in Theorem~\ref{Thm:consistency}. Kolmogorov and Tihomirov \cite{Kolmogorov-61} have shown that 
\begin{equation}\label{Eq:covering}
\mathcal{H}(\varepsilon,\eu{F}_W,\Vert\cdot\Vert_\infty)\le \mathcal{N}(\frac{\varepsilon}{4},M,\rho)\log\left(2\left\lceil\frac{2\text{diam}(M)}{\varepsilon}\right\rceil+1\right).
\end{equation}
Since $\mathcal{H}(\varepsilon,\eu{F}_W,L_1(\bb{P}_m))\le\mathcal{H}(\varepsilon,\eu{F}_W,\Vert\cdot\Vert_\infty)$, the conditions \emph{(iii)} and \emph{(iv)} in Theorem~\ref{Thm:consistency} are satisfied and therefore, $|W(\bb{P}_m,\bb{Q}_n)-W(\bb{P},\bb{Q})|\stackrel{a.s.}{\longrightarrow} 0$ as $m,n\rightarrow\infty$. Since $\eu{F}_\beta\subset\eu{F}_W$, the envelope function associated with $\eu{F}_\beta$ is upper bounded by the envelope function associated with $\eu{F}_W$ and $\mathcal{H}(\varepsilon,\eu{F}_\beta,\Vert\cdot\Vert_\infty)\le \mathcal{H}(\varepsilon,\eu{F}_W,\Vert\cdot\Vert_\infty)$. Therefore, the result for $\beta$ follows.\vspace{2mm}
\end{IEEEproof}
Similar to Corollary~\ref{cor:wasserstein-dudley}, a strong consistency result for $\gamma_k$ can be provided by estimating the entropy number of $\eu{F}_k$. See Cucker and Zhou \cite[Chapter 5]{Cucker-07} for the estimates of entropy numbers for various $\eu{H}$. However, in the following, we adopt a different approach to prove the strong consistency of $\gamma_k$. To this end, we first provide a general result on the rate of convergence of $\gamma_\eu{F}(\bb{P}_m,\bb{Q}_n)$ and then, as a special case, obtain the rates of convergence of the estimators of $W$, $\beta$ and $\gamma_k$. Using this result, we then prove the strong consistency of $\gamma_k$. We start with the following definition.\vspace{2mm}
\begin{fin}[Rademacher complexity]\label{def:Rademacher}
Let $\eu{F}$ be a class of functions on $M$ and $\{\sigma_i\}^m_{i=1}$ be independent Rademacher random variables, i.e., $\text{Pr}(\sigma_i=+1)=\text{Pr}(\sigma_i=-1)=\frac{1}{2}$. The Rademacher process is defined as $\{\frac{1}{m}\sum^m_{i=1}\sigma_if(x_i)\,:\,f\in\eu{F}\}$ for some $\{x_i\}^m_{i=1}\subset M$. The Rademacher complexity over $\eu{F}$ is defined as 
\begin{equation}
R_m(\eu{F};\{x_i\}^m_{i=1}):=\bb{E}\sup_{f\in\eu{F}}\left|\frac{1}{m}\sum^m_{i=1}\sigma_if(x_i)\right|.
\end{equation}
%where $\bb{E}_\rho$ represents the expectation with respect to $\{\rho_i\}^n_{i=1}$.
\end{fin}
We now present a general result that provides a probabilistic bound on the deviation of $\gamma_\eu{F}(\bb{P}_m,\bb{Q}_n)$ from $\gamma_\eu{F}(\bb{P},\bb{Q})$.
This generalizes \cite[Theorem 4]{Gretton-06}, the main difference being that we now consider
 function classes other than RKHSs, and thus express the bound in terms of the Rademacher complexities 
(see the proof for further discussion).
\vspace{2mm}
\begin{thm}\label{Thm:rate}
For any $\eu{F}$ such that $\nu:=\sup_{x\in M} F(x)<\infty$, with probability at least $1-\delta$, the following holds:
\begin{eqnarray}\label{Eq:consistency}
\lefteqn{\left|\gamma_\eu{F}(\bb{P}_m,\bb{Q}_n)-\gamma_\eu{F}(\bb{P},\bb{Q})\right|\le
\sqrt{18\nu^2\log\frac{4}{\delta}}\left(\frac{1}{\sqrt{m}}+\frac{1}{\sqrt{n}}\right)}\hspace{2.2cm}\nonumber\\
&&+2R_{m}(\eu{F};\{X^{(1)}_i\})+2R_{n}(\eu{F};\{X^{(2)}_i\}).
\end{eqnarray} 
\end{thm}
\begin{IEEEproof}
See Appendix~\ref{appendix-rate}.
\vspace{2mm}%\qed
\end{IEEEproof}
Theorem~\ref{Thm:rate} holds for any $\eu{F}$ for which $\nu$ is finite. However, to obtain the rate of convergence for $\gamma_\eu{F}(\bb{P}_m,\bb{Q}_n)$, one requires an estimate of $R_m(\eu{F};\{X^{(1)}_i\}^m_{i=1})$ and $R_n(\eu{F};\{X^{(2)}_i\}^n_{i=1})$. Note that if $R_m(\eu{F};\{X^{(1)}_i\}^m_{i=1})\stackrel{\bb{P}}{\longrightarrow} 0$ as $m\rightarrow \infty$ and $R_n(\eu{F};\{X^{(2)}_i\}^n_{i=1})\stackrel{\bb{Q}}{\longrightarrow} 0$ as $n\rightarrow \infty$, then \begin{equation}
|\gamma_\eu{F}(\bb{P}_m,\bb{Q}_n)-\gamma_\eu{F}(\bb{P},\bb{Q})|\stackrel{\bb{P},\bb{Q}}{\longrightarrow}0\,\,\,\text{as}\,\,\,m,n\rightarrow\infty.\nonumber
\end{equation}
Also note that if $R_m(\eu{F};\{X^{(1)}_i\}^m_{i=1})=O_\bb{P}(r_m)$ and $R_n(\eu{F};\{X^{(2)}_i\}^n_{i=1})=O_\bb{Q}(r_n)$, then from (\ref{Eq:consistency}), %(\ref{Eq:p}) and (\ref{Eq:q}), 
\begin{equation}
|\gamma_\eu{F}(\bb{P}_m,\bb{Q}_n)-\gamma_\eu{F}(\bb{P},\bb{Q})|=O_{\bb{P},\bb{Q}}(r_m\vee\, m^{-1/2}+r_n\vee\, n^{-1/2}),\nonumber
\end{equation} %as $\gamma_\eu{F}(\bb{P}_m,\bb{P})=O_\bb{P}(r_m\,\vee\,m^{-1/2})$ and $\gamma_\eu{F}(\bb{Q}_n,\bb{Q})=O_\bb{Q}(r_n\,\vee\,n^{-1/2})$, 
where $a\,\vee\, b:=\max(a,b)$. The following corollary to Theorem~\ref{Thm:rate} provides the rate of convergence for $W$, $\beta$ and $\gamma_k$. Note that Corollary \ref{cor:rate}\emph{(ii)} was proved in \cite{Gretton-06},\cite[Appendix A.2]{Gretton-07} by a more direct argument, where the fact that $\eu{F}_k$ is an RKHS was used at an earlier stage of the proof to simplify the reasoning. We include the result here for completeness.
% (for a fixed $k$), by choosing a specific $\eu{F}$.
\vspace{2mm}
%Before we prove this result, we use it to prove the consistency of estimators of the Wasserstein distance, Dudley metric and MMD. First, we provide a bound on the Rademacher complexity in (\ref{Eq:consistency}) for various $\eu{F}$. 
%need the following result due to \cite{Dudley-87}, called the \emph{classical entropy bound} that provides a bound on $R_n(\eu{F})$ in terms of the covering number of $\eu{F}$.
%\begin{lem}[Classical entropy bound~\cite{Dudley-87}]
%For every class $\eu{F}$ of functions, there exists a universal constant, $C$ such that 
%\begin{equation}\label{Eq:entropybound}
%R_{n^+}(\eu{F})\le \frac{C}{\sqrt{n^+}}\int^\infty_0\sqrt{\log\Cal{N}(\eu{F},\epsilon,L^2(\bb{P}_m)}\,d\epsilon.
%\end{equation}
%\end{lem}
%The following result shows that the Wasserstein distance, Dudley metric and MMD are strongly consistent.
\begin{cor}[Rates of convergence for $W$, $\beta$ and $\gamma_k$]\label{cor:rate}
%Let $\eu{F}_W=\{f:\Vert f\Vert_L\le 1\}$, $\eu{F}_\beta=\{f:\Vert f\Vert_{BL}\le 1\}$ and $\eu{F}_k=\{f:\Vert f\Vert_\eu{H}\le 1\}$. Then, the following hold:
\emph{(i)} Let $M$ be a bounded subset of $(\bb{R}^d,\Vert\cdot\Vert_s)$ for some $1\le s\le\infty$. Then, 
\begin{equation}
|W(\bb{P}_m,\bb{Q}_n)-W(\bb{P},\bb{Q})|=O_{\bb{P},\bb{Q}}(r_m+r_n)\nonumber
\end{equation} and 
\begin{equation}
|\beta(\bb{P}_m,\bb{Q}_n)-\beta(\bb{P},\bb{Q})|=O_{\bb{P},\bb{Q}}(r_m+r_n),\nonumber
\end{equation}
where 
\begin{equation}\label{Eq:rates}
r_m=\left\{\begin{array}{c@{\quad\quad}l}
m^{-1/2}\log m,& d=1\\
m^{-1/(d+1)},& d\ge 2
\end{array}\right..
\end{equation}
In addition if $M$ is a bounded, convex subset of $(\bb{R}^d,\Vert\cdot\Vert_s)$ with non-empty interior, then
\begin{equation}\label{Eq:rates-1}
r_m=\left\{\begin{array}{c@{\quad\quad}l}
m^{-1/2},& d=1\\
m^{-1/2}\log m,& d=2\\
m^{-1/d},& d>2
\end{array}\right..
\end{equation}
%for every $\epsilon>0$,
%\begin{equation}\label{Eq:rademacher-wasserstein}
%R_{n^+}(\eu{F}_W)\le 2\epsilon+\frac{4\sqrt{2}}{\sqrt{n^+}}\int^{4\text{diam}(M)}_{\frac{\epsilon}{4}}\sqrt{\Cal{N}(M,\frac{u}{4},\rho)\log\left(2\left\lceil\frac{2\text{diam}(M)}{u}\right\rceil+1\right)}\,du,
%\end{equation}
%where $\Cal{N}(M,\tau,\rho)$ is the $\tau$-covering number of $(M,\rho)$.
%\item[(ii)] $R_{n^+}(\eu{F}_\beta)\le R_{n^+}(\eu{F}_W)$. 
\emph{(ii)} Let $M$ be a measurable space. Suppose $k$ is measurable and $\sup_{x\in M}k(x,x)\le C<\infty$. Then, \begin{equation}
|\gamma_k(\bb{P}_m,\bb{Q}_n)-\gamma_k(\bb{P},\bb{Q})|=O_{\bb{P},\bb{Q}}(m^{-1/2}+n^{-1/2}). \nonumber
\end{equation}
In addition, \begin{equation}
|\gamma_k(\bb{P}_m,\bb{Q}_n)-\gamma_k(\bb{P},\bb{Q})|\stackrel{a.s.}{\longrightarrow} 0\,\,\, \text{as}\,\,\, m,n\rightarrow \infty,\nonumber\end{equation}
i.e., the estimator of MMD is strongly consistent.
%$R_{n^+}(\eu{F}_k)\le\frac{\sqrt{C}}{\sqrt{n^+}}$.
\vspace{2mm}
\end{cor}
\begin{IEEEproof}
\emph{(i)} Define $R^1_m(\eu{F}):=R_m(\eu{F};\{X^{(1)}_i\}^m_{i=1})$. The generalized entropy bound \cite[Theorem 16]{Luxburg-04} gives that for every $\varepsilon>0$,
\begin{equation}\label{Eq:rademacher-generalized}
R^1_{m}(\eu{F})\le  2\varepsilon+\frac{4\sqrt{2}}{\sqrt{m}}\int^{\infty}_{\varepsilon/4}\sqrt{\mathcal{H}(\tau,\eu{F},L_2(\bb{P}_m))}\,d\tau.
\end{equation}
Let $\eu{F}=\eu{F}_W$. Since $M$ is a bounded subset of $\bb{R}^d$, it is totally bounded and therefore the entropy number in (\ref{Eq:rademacher-generalized}) can be bounded through (\ref{Eq:covering}) by noting that
\begin{equation}\label{Eq:covering-1}
\mathcal{H}(\tau,\eu{F}_W,L_2(\bb{P}_m))\le\mathcal{H}(\tau,\eu{F}_W,\Vert\cdot\Vert_\infty)\le \frac{C_1}{\tau^{d+1}}+\frac{C_2}{\tau^d},
\end{equation}
where we have used the fact that $\mathcal{N}(\varepsilon,M,\Vert\cdot\Vert_s)=O(\varepsilon^{-d}),\,1\le s\le\infty$ and $\log(\lceil x \rceil +1)\le x+1$.\footnote{Note that for any $x\in M\subset\bb{R}^d$, $\Vert x\Vert_\infty\le\cdots\le\Vert x\Vert_s\le\cdots\le\Vert x\Vert_2\le\Vert x\Vert_1\le \sqrt{d}\Vert x\Vert_2$. Therefore, $\forall\,s\ge 2,\,\Cal{N}(\varepsilon,M,\Vert\cdot\Vert_s)\le \Cal{N}(\varepsilon,M,\Vert \cdot\Vert_2)$ and $\forall\,1\le s\le 2,\,\Cal{N}(\varepsilon,M,\Vert\cdot\Vert_s)\le \Cal{N}(\varepsilon,M,\sqrt{d}\Vert\cdot\Vert_2)=\Cal{N}(\varepsilon/\sqrt{d},M,\Vert\cdot\Vert_2)$. Use $\Cal{N}(\varepsilon,M,\Vert\cdot\Vert_2)=O(\varepsilon^{-d})$ \cite[Lemma 2.5]{vandeGeer-00}.} The constants $C_1$ and $C_2$ depend only on the properties of $M$ and are independent of $\tau$. Substituting (\ref{Eq:covering-1}) in (\ref{Eq:rademacher-generalized}), we have
\setlength{\arraycolsep}{0.0em}
\begin{eqnarray}\label{Eq:rademacher-generalized-1}
R^1_{m}(\eu{F}_W)&{}\le{}&  \inf_{\varepsilon>0}\left[2\varepsilon+\frac{4\sqrt{2}}{\sqrt{m}}\int^{\infty}_{\varepsilon/4}\sqrt{\mathcal{H}(\tau,\eu{F}_W,L_2(\bb{P}_m))}\,d\tau\right]\nonumber\\
&{}\le{}&\inf_{\varepsilon>0}\left[2\varepsilon+\frac{4\sqrt{2}}{\sqrt{m}}\int^{4R}_{\varepsilon/4}\left(\frac{\sqrt{C_1}}{\tau^{(d+1)/2}}+\frac{\sqrt{C_2}}{\tau^{d/2}}\right)\,d\tau\right],\nonumber
\end{eqnarray}
where $R:=\text{diam}(M)$. Note the change in upper limits of the integral from $\infty$ to $4R$. This is because $M$ is totally bounded and $\mathcal{H}(\tau,\eu{F}_W,\Vert\cdot\Vert_\infty)$ depends on $\mathcal{N}(\tau/4,M,\Vert\cdot\Vert_s)$. The rates in (\ref{Eq:rates}) are simply obtained by solving the right hand side of the above inequality. As mentioned in the paragraph preceding the statement of Corollary~\ref{cor:rate}, we have $r_m\vee m^{-1/2}=r_m$ and so the result for $W(\bb{P}_m,\bb{Q}_n)$ follows.
\par Suppose $M$ is convex. Then $M$ is connected. It is easy to see that $M$ is also centered, i.e., for all subsets $A\subset M$ with $\text{diam}(A)\le 2r$ there exists a point $x\in M$ such that $\Vert x-a\Vert_s\le r$ for all $a\in A$. Since $M$ is connected and centered, we have from \cite{Kolmogorov-61} that 
\setlength{\arraycolsep}{0.0em}
\begin{eqnarray}\label{Eq:entropy-convex}
\lefteqn{\mathcal{H}(\tau,\eu{F}_W,L_2(\bb{P}_m))\le\mathcal{H}(\tau,\eu{F}_W,\Vert\cdot\Vert_\infty)\le}\hspace{4mm}\nonumber\\ &&\mathcal{N}(\frac{\tau}{2},M,\Vert\cdot\Vert_s)\log 2+\log\left(2\left\lceil \frac{2\,\text{diam}(M)}{\tau}\right\rceil+1\right)\nonumber\\
&&\quad\le C_3\tau^{-d}+C_4\tau^{-1}+C_5,
\end{eqnarray} where we used the fact that
$\mathcal{N}(\varepsilon,M,\Vert\cdot\Vert_s)=O(\varepsilon^{-d})$. $C_3$, $C_4$ and $C_5$ are constants that depend only on the properties of $M$ and are independent of $\tau$. Substituting (\ref{Eq:entropy-convex}) in (\ref{Eq:rademacher-generalized}), we have,
\begin{equation}\label{Eq:rademacher-generalized-R}
R^1_{m}(\eu{F}_W)\le  \inf_{\varepsilon>0}\left[2\varepsilon+\frac{4\sqrt{2}}{\sqrt{m}}\int^{2R}_{\varepsilon/4}\frac{\sqrt{C_3}}{\tau^{d/2}}\,d\tau\right]+O(m^{-1/2}).\nonumber
\end{equation}
Again note the change in upper limits of the integral from $\infty$ to $2R$. This is because $\Cal{H}(\tau,\eu{F}_W,\Vert\cdot\Vert_\infty)$ depends on $\Cal{N}(\tau/2,M,\Vert\cdot\Vert_s)$. The rates in (\ref{Eq:rates-1}) are obtained by solving the right hand side of the above inequality. Since $r_m\vee m^{-1/2}=r_m$, the result for $W(\bb{P}_m,\bb{Q}_n)$ follows.
%Since (\ref{Eq:rademacher-generalized-R}) holds for all $\varepsilon>0$, the bound can be tightened by taking the infimum of the right hand side of (\ref{Eq:rademacher-generalized-R}) over $\varepsilon>0$. Now, for different values of $d$, the rates in (\ref{Eq:rates}) can be simply obtained by computing the integral in (\ref{Eq:rademacher-generalized-R}) and finding the tightest bound by taking infimum over $\varepsilon$.
\par Since $\eu{F}_{\beta}\subset\eu{F}_{W}$, we have $R^1_m(\eu{F}_{\beta})\le R^1_m(\eu{F}_{W})$ and therefore, the result for $\beta(\bb{P}_m,\bb{Q}_n)$ follows. The rates in (\ref{Eq:rates-1}) can also be directly obtained for $\beta$ by using the entropy number of $\eu{F}_{\beta}$, i.e., $\mathcal{H}(\varepsilon,\eu{F}_{\beta},\Vert\cdot\Vert_\infty)=O(\varepsilon^{-d})$ \cite[Theorem 2.7.1]{Vaart-96} in (\ref{Eq:rademacher-generalized}).\vspace{2mm}\\
%\emph{(i)} simply follows from \cite[Theorem 18]{Luxburg-04}. For \emph{(ii)}, note that 
%\begin{equation}
%R_{n^+}(\eu{F}_\beta)=\bb{E}_\sigma\sup_{f\in\eu{F}_\beta}\left|\frac{1}{n^+}\sum^{n^+}_{i=1}\sigma_if(X_i)\right|\le \bb{E}_\sigma\sup_{f\in\eu{F}_W}\left|\frac{1}{n^+}\sum^{n^+}_{i=1}\sigma_if(X_i)\right|=R_{n^+}(\eu{F}_W).\nonumber
%\end{equation}
\emph{(ii)} By \cite[Lemma 22]{Bartlett-02}, $R^1_{m}(\eu{F}_k)\le\frac{\sqrt{C}}{\sqrt{m}}$ and $R_n(\eu{F}_k;\{X^{(2)}_i\}^n_{i=1})\le\frac{\sqrt{C}}{\sqrt{n}}$.
%can be bounded as
%\begin{eqnarray}\label{Eq:rademacher-rkhs}
%\lefteqn{\bb{E}_\sigma\sup_{f\in\eu{F}_k}\Big|\sum^{m}_{i=1}\frac{\sigma_if(X^{(1)}_i)}{m}\Big|\stackrel{(a)}{=}\bb{E}_\sigma\sup_{f\in\eu{F}_k}\Big|\langle f,\sum^{m}_{i=1}\frac{\sigma_ik(\cdot,X^{(1)}_i)}{m}\rangle_\eu{H}\Big|}\nonumber\\
%&&\stackrel{(b)}{=}\bb{E}_\sigma\Big\Vert\sum^{m}_{i=1}\frac{\sigma_ik(\cdot,X^{(1)}_i)}{m}\Big\Vert_\eu{H}
%=\bb{E}_\sigma\sqrt{\sum^{m}_{i,j=1}\frac{\sigma_i\sigma_jk(X^{(1)}_i,X^{(1)}_j)}{m^2}}\nonumber\\
%&&\stackrel{(c)}{\le}\sqrt{\bb{E}_\sigma\sum^{m}_{i,j=1}\frac{\sigma_i\sigma_jk(X^{(1)}_i,X^{(1)}_j)}{m^2}}
%=\frac{1}{m}\sqrt{\sum^{m}_{i=1}k(X^{(1)}_i,X^{(1)}_i)}\nonumber\\
%&&\le\frac{\sqrt{C}}{\sqrt{m}},
%\end{eqnarray}
%where $(a)$ follows from the reproducing property of $\eu{H}$, $(b)$ follows from the Cauchy-Schwartz inequality and $(c)$ follows from the Jensen's inequality. 
%Performing similar analysis for $R_n(\eu{F}_k;\{X^{(2)}_i\}^n_{i=1})$ yields $R_n(\eu{F}_k;\{X^{(2)}_i\}^n_{i=1})=O_\bb{Q}(n^{-1/2})$. 
Substituting these %along with (\ref{Eq:rademacher-rkhs}) 
in (\ref{Eq:consistency}) yields the result. 
%By using (\ref{Eq:rademacher-rkhs}) in (\ref{Eq:p}), it is easy to see that $\gamma_k(\bb{P}_m,\bb{P})=O_\bb{P}(m^{-1/2})$. 
In addition, by the Borel-Cantelli lemma, the strong consistency of $\gamma_k(\bb{P}_m,\bb{Q}_n)$ follows. %$\gamma_k(\bb{P}_m,\bb{P})\stackrel{a.s.}{\longrightarrow} 0$ as $m\rightarrow\infty$. Performing similar analysis for $\gamma_k(\bb{Q}_n,\bb{Q})$ and adding (\ref{Eq:p}) and (\ref{Eq:q}) yields the result.
\vspace{2mm}
\end{IEEEproof}
\begin{figure*}[t]
  \centering
  \begin{tabular}{cccccc}\hspace{-5mm}
    \begin{minipage}{6cm}
      \center{\epsfxsize=6cm
      \epsffile{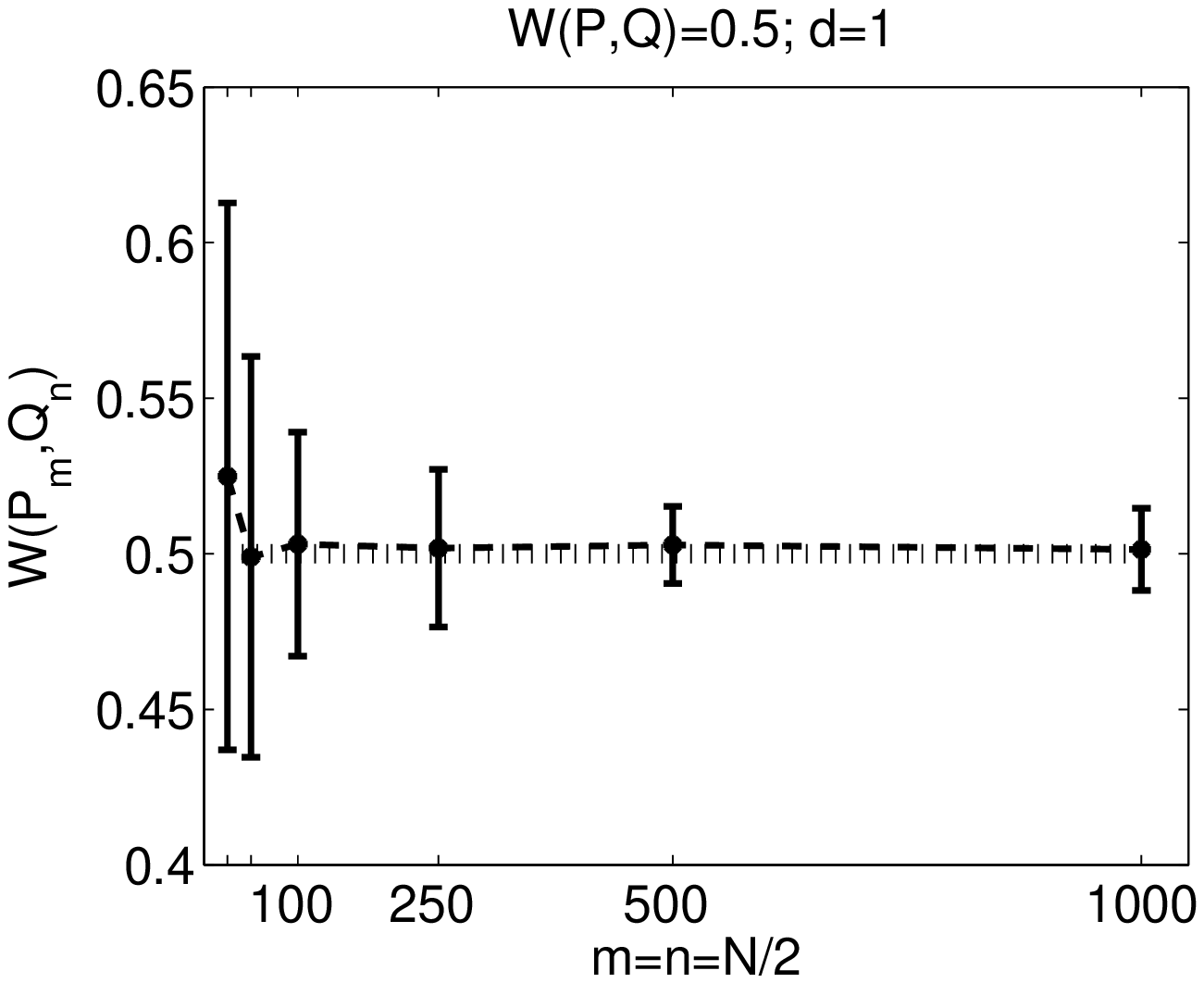}}\vspace{-2.5mm}
      {\small \center{(a)}}
    \end{minipage}\hspace{-3mm}
    \begin{minipage}{6cm}
      \center{\epsfxsize=6cm
      \epsffile{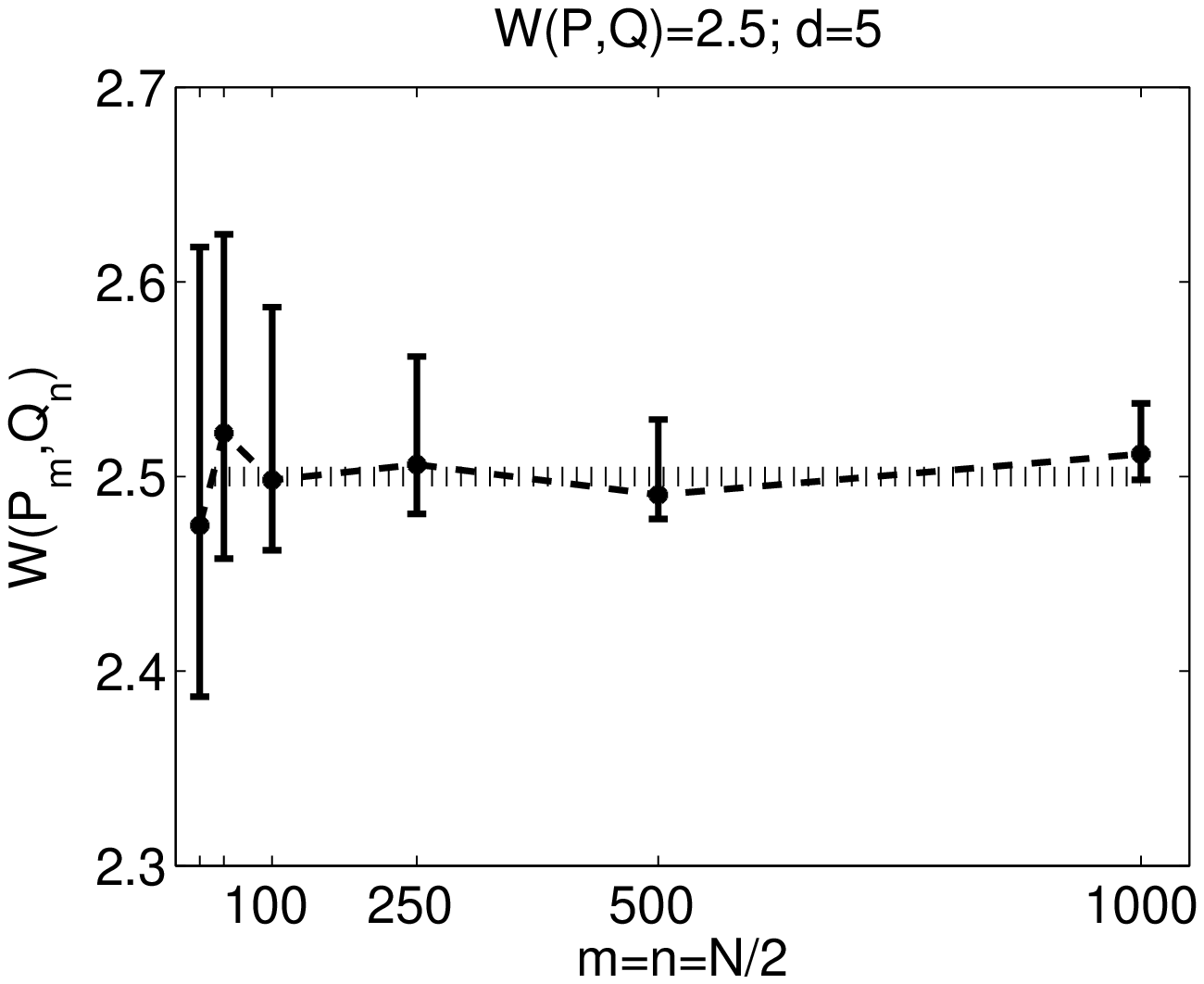}}\vspace{-2.5mm}
      {\small \center{(b)}}
    \end{minipage}\hspace{-3mm}
    \begin{minipage}{6cm}
      \center{\epsfxsize=6cm
      \epsffile{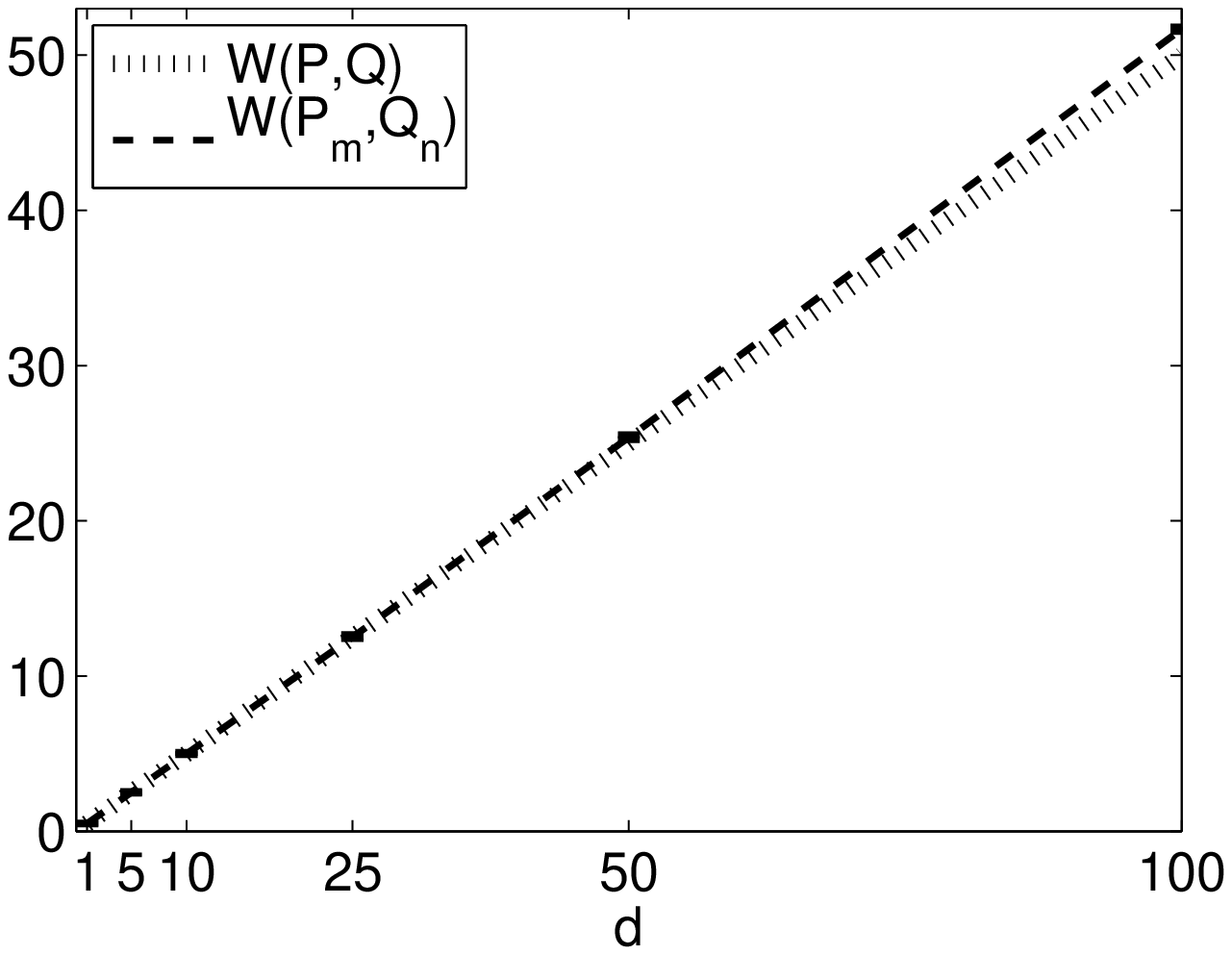}}\vspace{-2.5mm}
      {\small \center{(c)}}
    \end{minipage}
       \vspace{3.5mm}
  \end{tabular}
\begin{tabular}{cccccc}\hspace{-5mm}
    \begin{minipage}{6cm}
      \center{\epsfxsize=6cm
      \epsffile{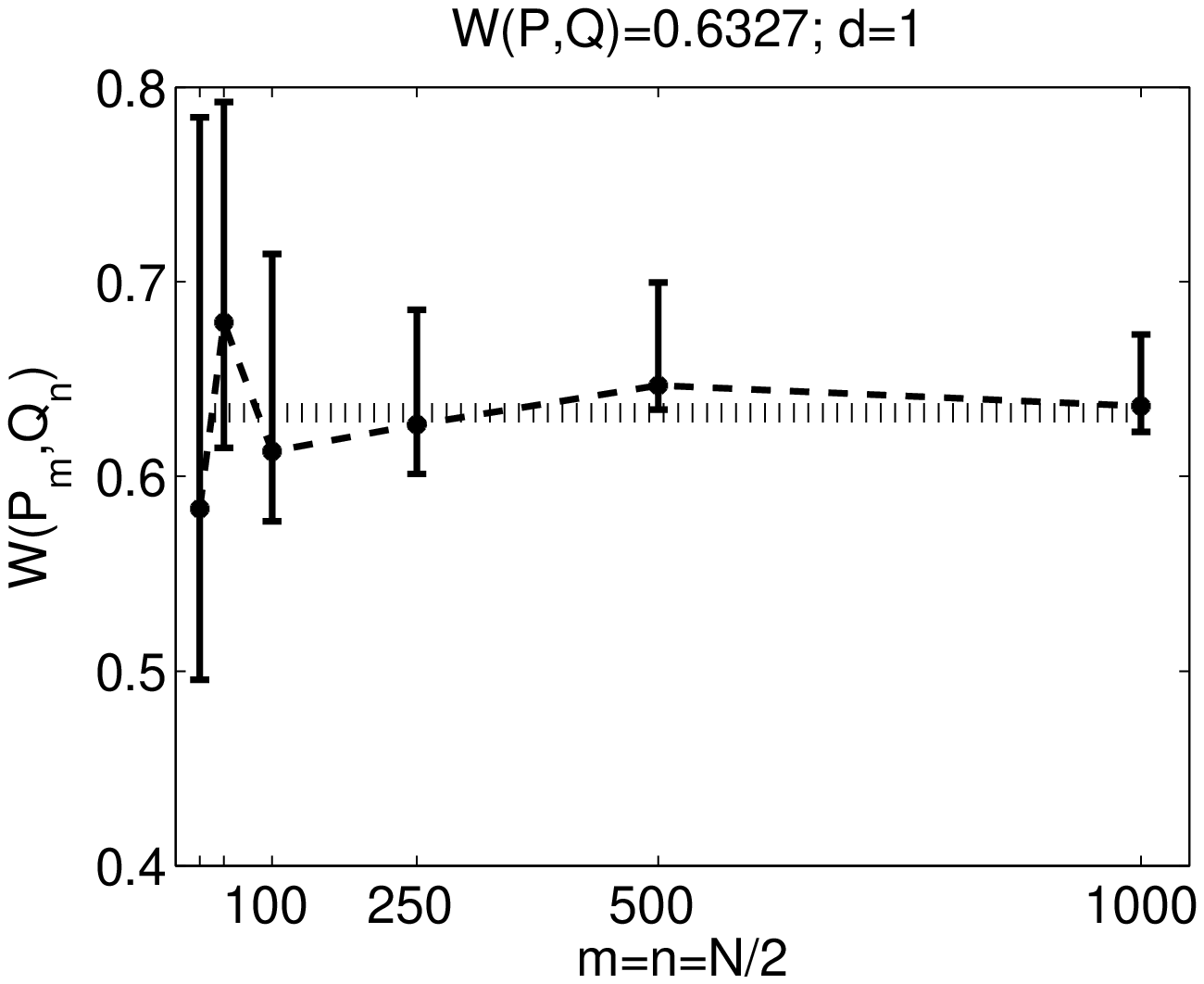}}\vspace{-2.5mm}
      {\small \center{(a$^\prime$)}}
    \end{minipage}\hspace{-3mm}
    \begin{minipage}{6cm}
      \center{\epsfxsize=6cm
      \epsffile{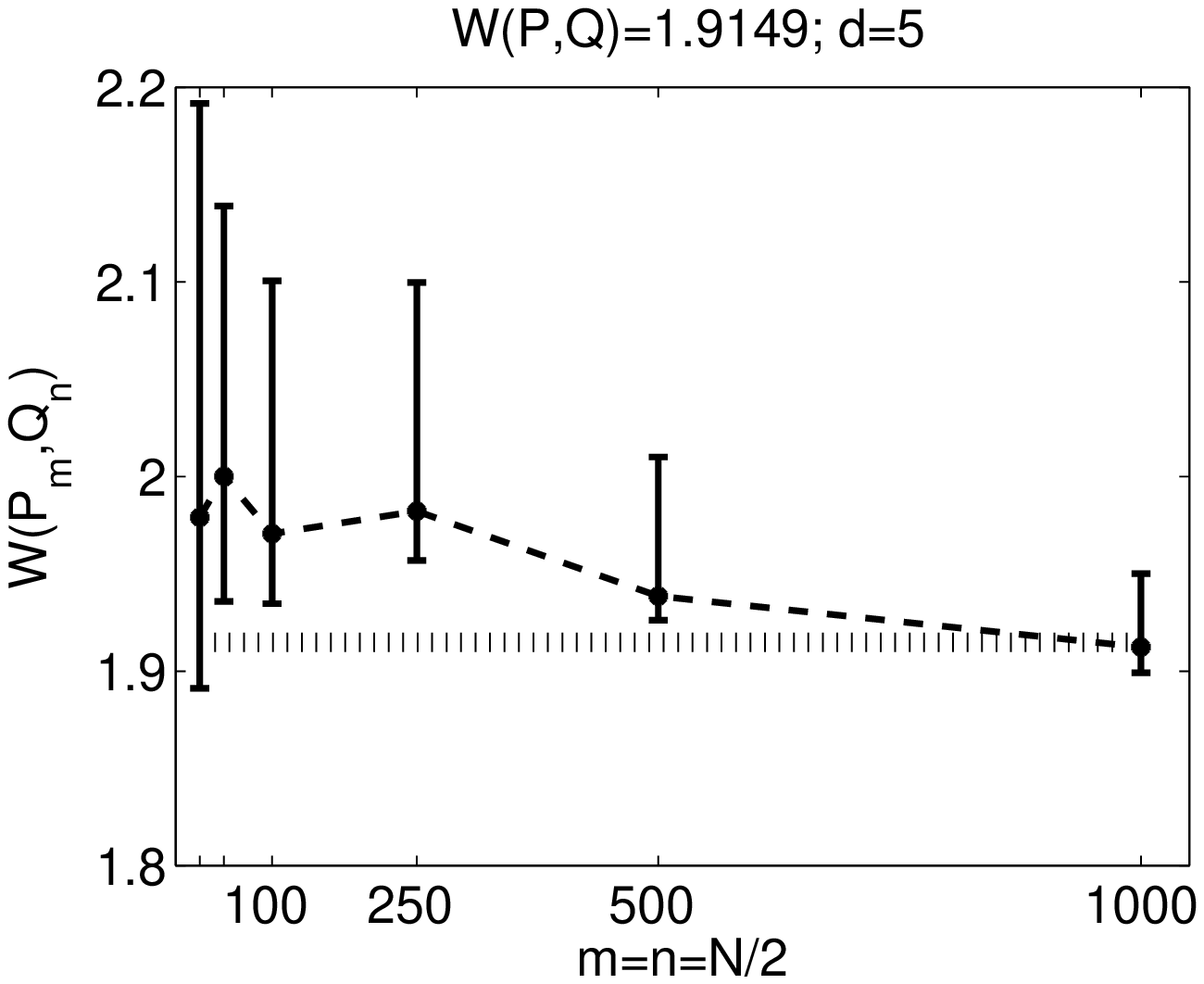}}\vspace{-2.5mm}
      {\small \center{(b$^\prime$)}}
    \end{minipage}\hspace{-3mm}
    \begin{minipage}{6cm}
      \center{\epsfxsize=6cm
      \epsffile{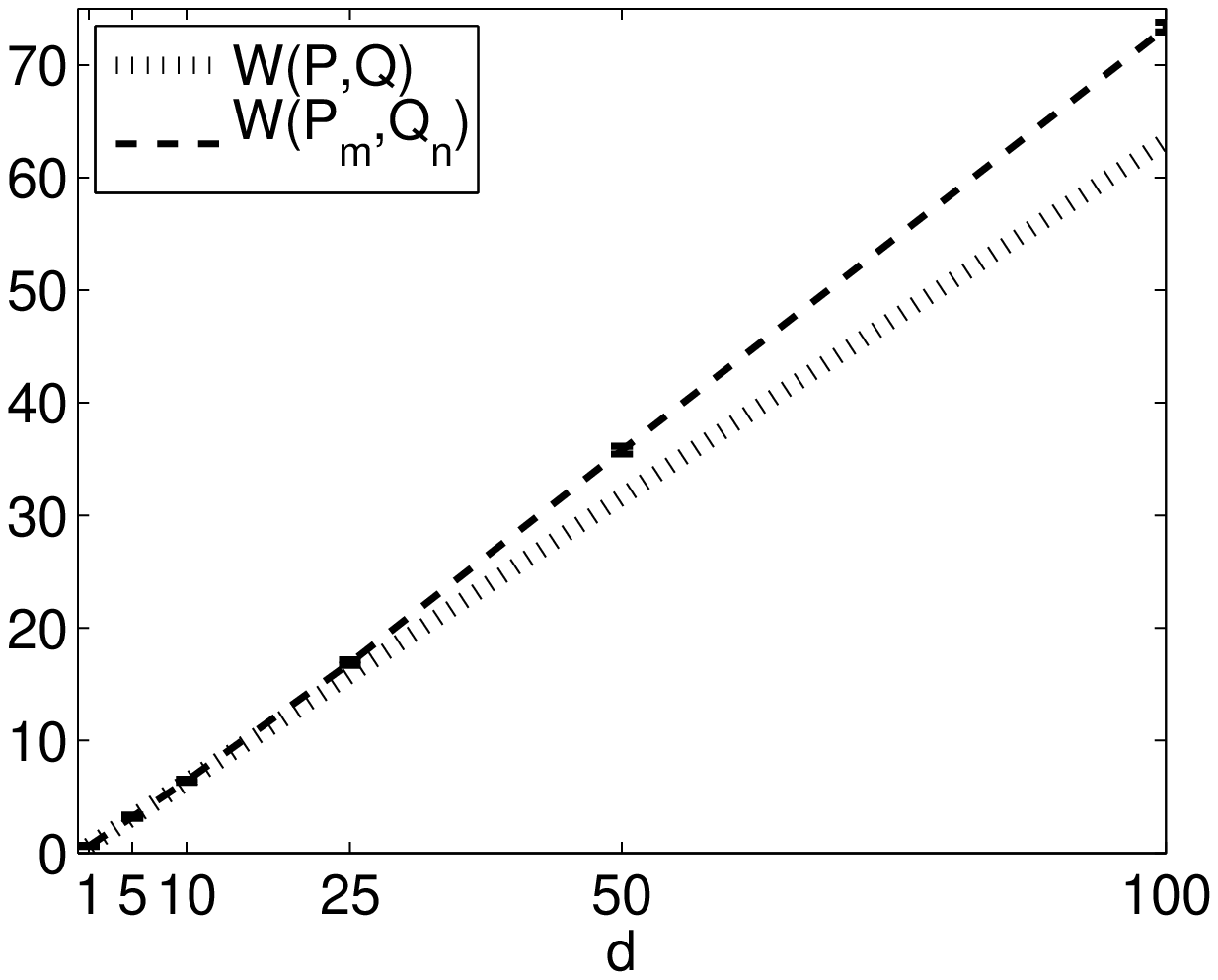}}\vspace{-2.5mm}
      {\small \center{(c$^\prime$)}}
    \end{minipage}\hspace{-3mm}
    \vspace{1mm}
  \end{tabular}
  \caption{(a-b) represent the empirical estimates of the Wasserstein distance (shown in thick dotted lines) between $\bb{P}=U[-\frac{1}{2},\frac{1}{2}]^d$ and $\bb{Q}=U[0,1]^d$ with $\rho(x,y)=\Vert x-y\Vert_1$, for increasing sample size $N$, where $d=1$ in (a) and $d=5$ in (b). Here $U[l_1,l_2]^d$ represents a uniform distribution on $[l_1,l_2]^d$ (see Example~\ref{exm:wass-1} for details). Similarly, (a$^\prime$-b$^\prime$) represent the empirical estimates of the Wasserstein distance (shown in thick dotted lines) between $\bb{P}$ and $\bb{Q}$, which are truncated exponential distributions on $\bb{R}^d_+$ (see Example~\ref{exm:wass-2} for details), for increasing sample size $N$. Here $d=1$ in (a$^\prime$) and $d=5$ in (b$^\prime$) with $\rho(x,y)=\Vert x-y\Vert_1$. The population values of the Wasserstein distance between $\bb{P}$ and $\bb{Q}$ are shown in thin dotted lines in (a-c, a$^\prime$-c$^\prime$). (c, c$^\prime$) represent the behavior of $W(\bb{P}_m,\bb{Q}_n)$ and $W(\bb{P},\bb{Q})$ for varying $d$ with a fixed sample size of $m=n=250$ (see Examples~\ref{exm:wass-1} and \ref{exm:wass-2} for details on the choice of $\bb{P}$ and $\bb{Q}$). Error bars are obtained by replicating the experiment $20$ times.}
\label{Fig:estimate}
\vspace{-4mm}
\end{figure*}
\begin{rem}
\emph{(i)} Note that the rate of convergence of $W$ and $\beta$ is dependent on the dimension, $d$, which means that in large dimensions, more samples are needed to obtain useful estimates of $W$ and $\beta$. Also note that the rates are independent of the metric, $\Vert\cdot\Vert_s,\,1\le s\le\infty$.\vspace{2mm}\\
\emph{(ii)} Note that when $M$ is a bounded, convex subset of $(\bb{R}^d,\Vert\cdot\Vert_s)$, faster rates are obtained than for the case where $M$ is just a bounded (but not convex) subset of $(\bb{R}^d,\Vert\cdot\Vert_s)$. \vspace{2mm}\\
\emph{(iii)} In the case of MMD, we have not made any assumptions on $M$ except it being a measurable space. This means in the case of $\bb{R}^d$, the rate is independent of $d$, which is a very useful property. The condition of the kernel being bounded is satisfied by a host of kernels, the examples of which include the Gaussian kernel, $k(x,y)=\exp(-\sigma\Vert x-y\Vert^2_2),\,\sigma>0$, Laplacian kernel, $k(x,y)=\exp(-\sigma\Vert x-y\Vert_1),\,\sigma>0$, inverse multiquadrics, $k(x,y)=(c^2+\Vert x-y\Vert^2_2)^{-t},\,c>0,\,t>d/2$, etc. on $\bb{R}^d$. See Wendland \cite{Wendland-05} for more examples. %AG: again, I do not see the point in re-proving a result that has been proved already. Just state the result without proof and describe the implications
%BK: i removed the proof.
 As mentioned before, the estimates for $R_m(\eu{F}_k;\{X^{(1)}_i\}^m_{i=1})$ can be directly obtained by using the entropy numbers of $\eu{F}_k$. See Cucker and Zhou~\cite[Chapter 5]{Cucker-07} for the estimates of entropy numbers for various $\eu{H}$.\vspace{2mm}
\end{rem}

The results derived so far in this section show that the estimators of the Wasserstein distance, Dudley metric and MMD exhibit good convergence behavior, irrespective of the distributions, unlike the case with $\phi$-divergence.
\subsection{Simulation results}\label{subsec:simulation}
So far, in Sections \ref{subsec:empirical} and \ref{subsec:consistency}, we have presented the empirical estimation of $W$, $\beta$ and $\gamma_k$ and their convergence analysis. Now, the question is how good are these estimators in practice? In this section, we demonstrate the performance of these estimators through simulations. 
\par As we have mentioned before, given $\bb{P}$ and $\bb{Q}$, it is usually difficult to exactly compute $W$, $\beta$ and $\gamma_k$. However, in order to test the performance of their estimators, in the following, we consider some examples where $W$, $\beta$ and $\gamma_k$ can be computed exactly.\vspace{2mm}
\subsubsection{Estimator of $W$}\label{subsubsec:wasserstein-results} 
%\par Let us consider $\bb{P}$ and $\bb{Q}$ for which $W(\bb{P},\bb{Q})$ can be exactly computed such that the usefulness of $W(\bb{P}_m,\bb{Q}_n)$ can be verified. As mentioned before (see footnote~\ref{footnote:vallander}), for $(M,\rho(x,y))=(\bb{R},|x-y|)$, $W$ can be equivalently written as 
%\begin{equation}\label{Eq:vallander}
%W(\bb{P},\bb{Q})=\int_\bb{R}\left|F_\bb{P}(x)-F_\bb{Q}(x)\right|\,dx,
%\end{equation}
%where $F_\bb{P}(x)=\bb{P}((-\infty,x])$~\cite{Vallander-73}. %It is easy to see that computing $W$ (see (\ref{Eq:primal-wasserstein})) when $(M,\rho(x,y))=(\bb{R}^d,\Vert x-y\Vert_1)$ involves $d$ computations of the form of (\ref{Eq:vallander}). 
\par For the ease of computation, let us consider $\bb{P}$ and $\bb{Q}$ (defined on the Borel $\sigma$-algebra of $\bb{R}^d$) as product measures, $\bb{P}=\otimes^d_{i=1}\bb{P}^{(i)}$ and $\bb{Q}=\otimes^d_{i=1}\bb{Q}^{(i)}$, where $\bb{P}^{(i)}$ and $\bb{Q}^{(i)}$ are defined on the Borel $\sigma$-algebra of $\bb{R}$. In this setting, when $\rho(x,y)=\Vert x-y\Vert_1$, it is easy to show that 
\begin{equation}\label{Eq:multi-wasserstein}
W(\bb{P},\bb{Q})=\sum^d_{i=1}W(\bb{P}^{(i)},\bb{Q}^{(i)}),
\end{equation}
where \begin{equation}\label{Eq:vallander}
W(\bb{P}^{(i)},\bb{Q}^{(i)})=\int_\bb{R}\left|F_{\bb{P}^{(i)}}(x)-F_{\bb{Q}^{(i)}}(x)\right|\,dx,
\end{equation}
and $F_{\bb{P}^{(i)}}(x)=\bb{P}^{(i)}((-\infty,x])$~\cite{Vallander-73} (see footnote~\ref{footnote:vallander}).
%$W(\bb{P}^{(i)},\bb{Q}^{(i)})$ is computed as in (\ref{Eq:vallander}). 
%Therefore, the computation of $W(\bb{P},\bb{Q})$ reduces to $d$ computations of the form of (\ref{Eq:vallander}). 
Now, in the following, we consider two examples where $W$ in (\ref{Eq:vallander}) can be computed in closed form. Note that we need $M$ to be a bounded subset of $\bb{R}^d$ such that the consistency of $W(\bb{P}_m,\bb{Q}_n)$ is guaranteed by Corollary~\ref{cor:rate}.\vspace{2mm}
\begin{figure*}[t]
  \centering
  \begin{tabular}{cccccc}\hspace{-5mm}
    \begin{minipage}{6cm}
      \center{\epsfxsize=6cm
      \epsffile{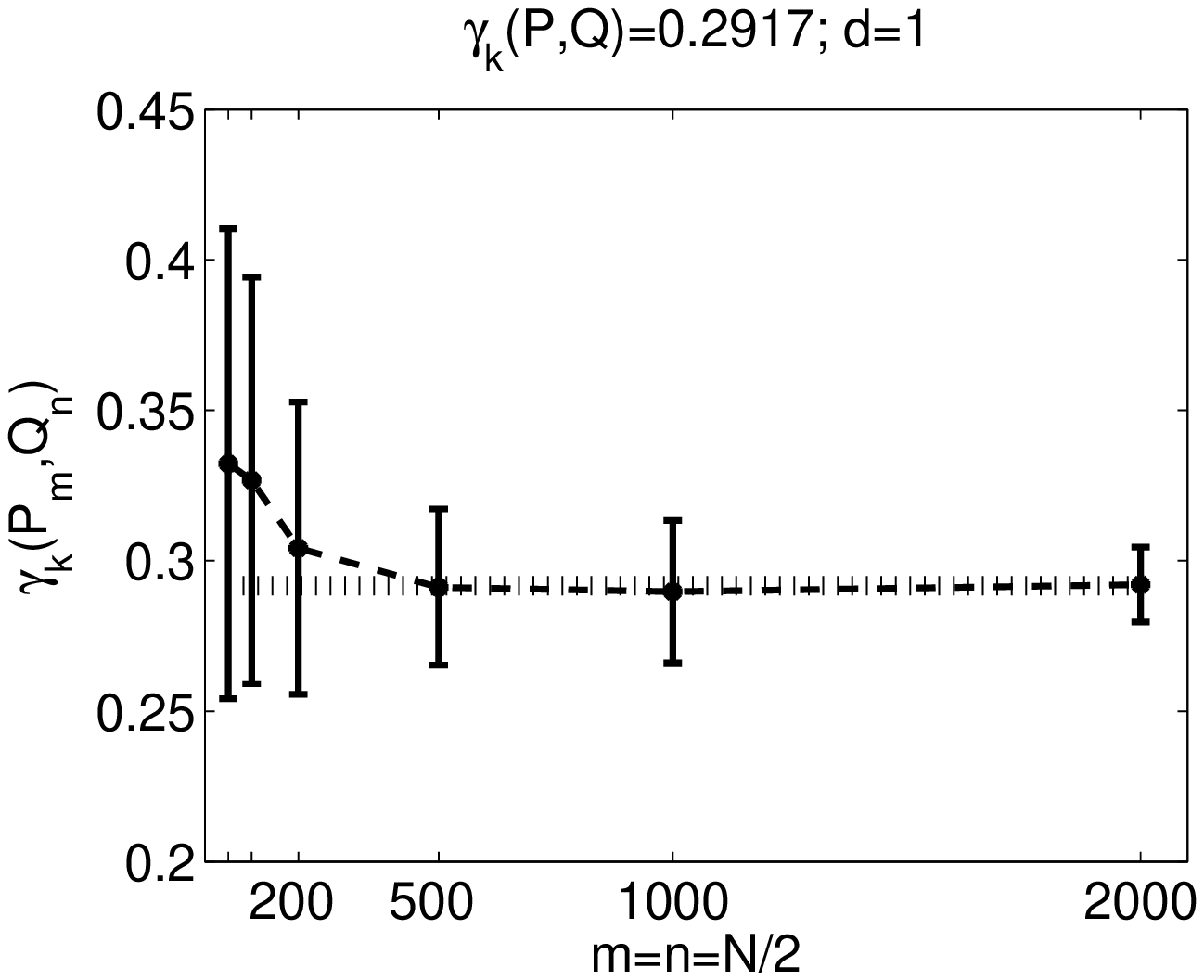}}\vspace{-2.5mm}
      {\small \center{(a)}}
    \end{minipage}\hspace{-3mm}
    \begin{minipage}{6cm}
      \center{\epsfxsize=6cm
      \epsffile{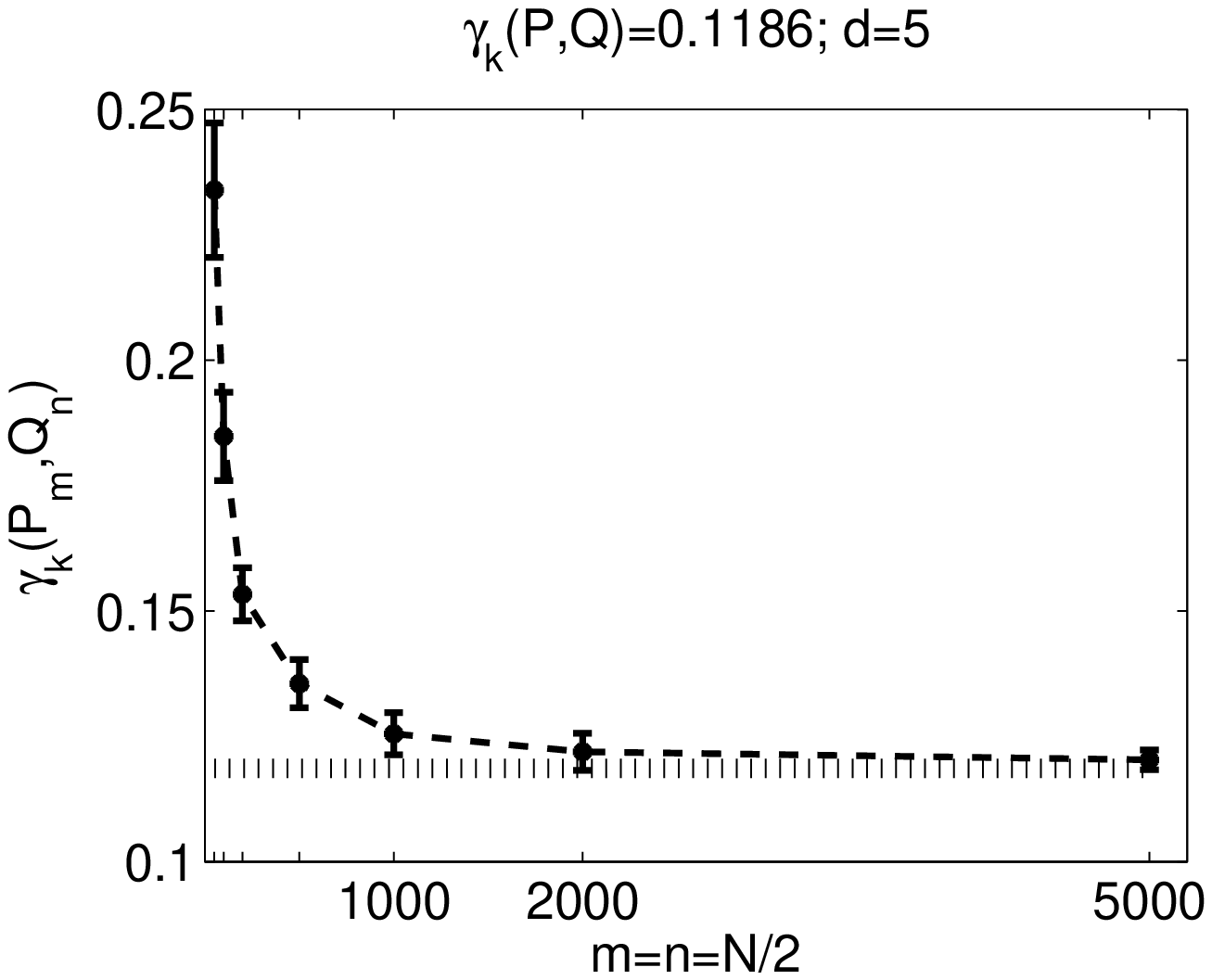}}\vspace{-2.5mm}
      {\small \center{(b)}}
    \end{minipage}\hspace{-3mm}
    \begin{minipage}{6cm}
      \center{\epsfxsize=6cm
      \epsffile{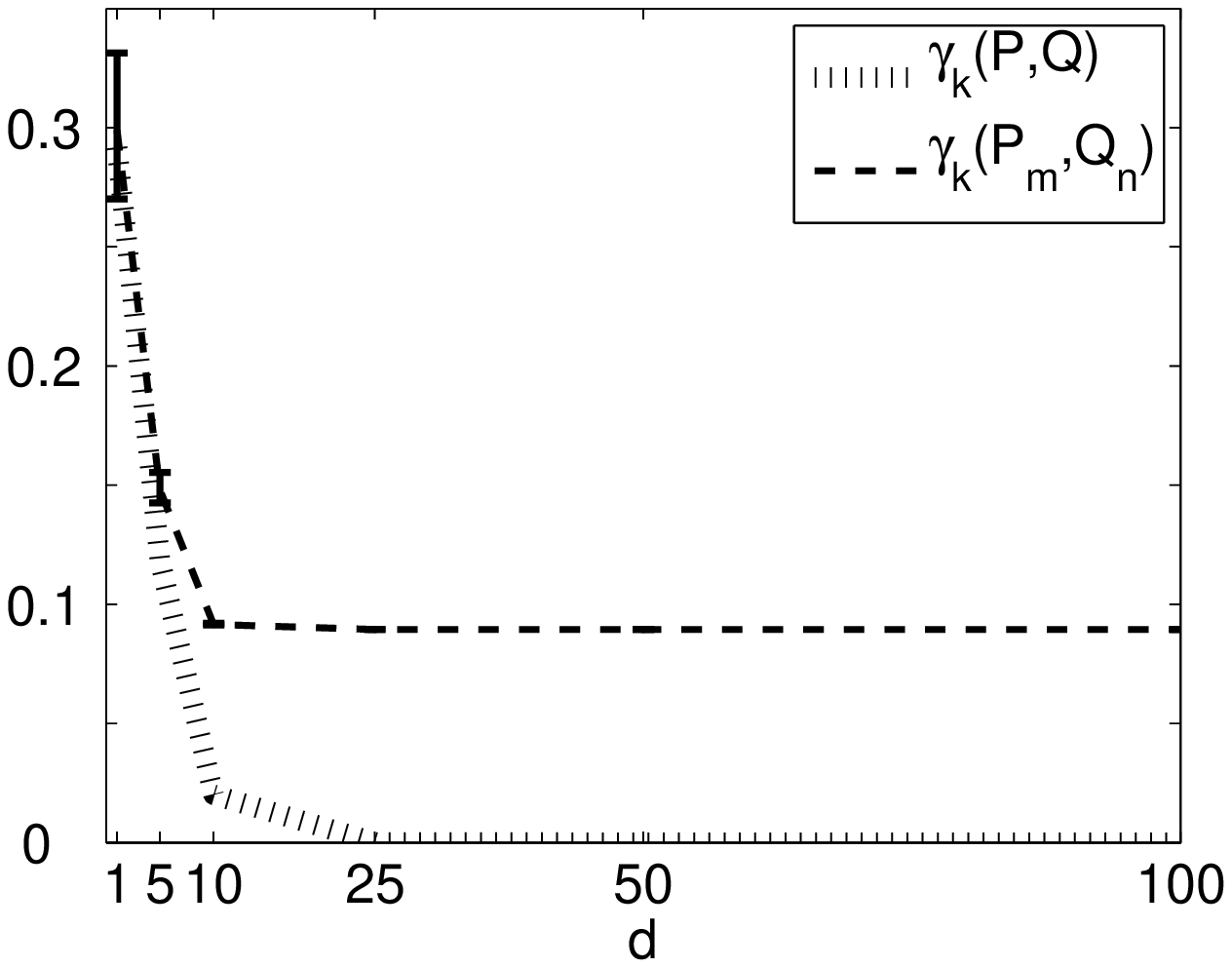}}\vspace{-2.5mm}
      {\small \center{(c)}}
    \end{minipage}
       \vspace{3.5mm}
  \end{tabular}
\begin{tabular}{cccccc}\hspace{-5mm}
    \begin{minipage}{6cm}
      \center{\epsfxsize=6cm
      \epsffile{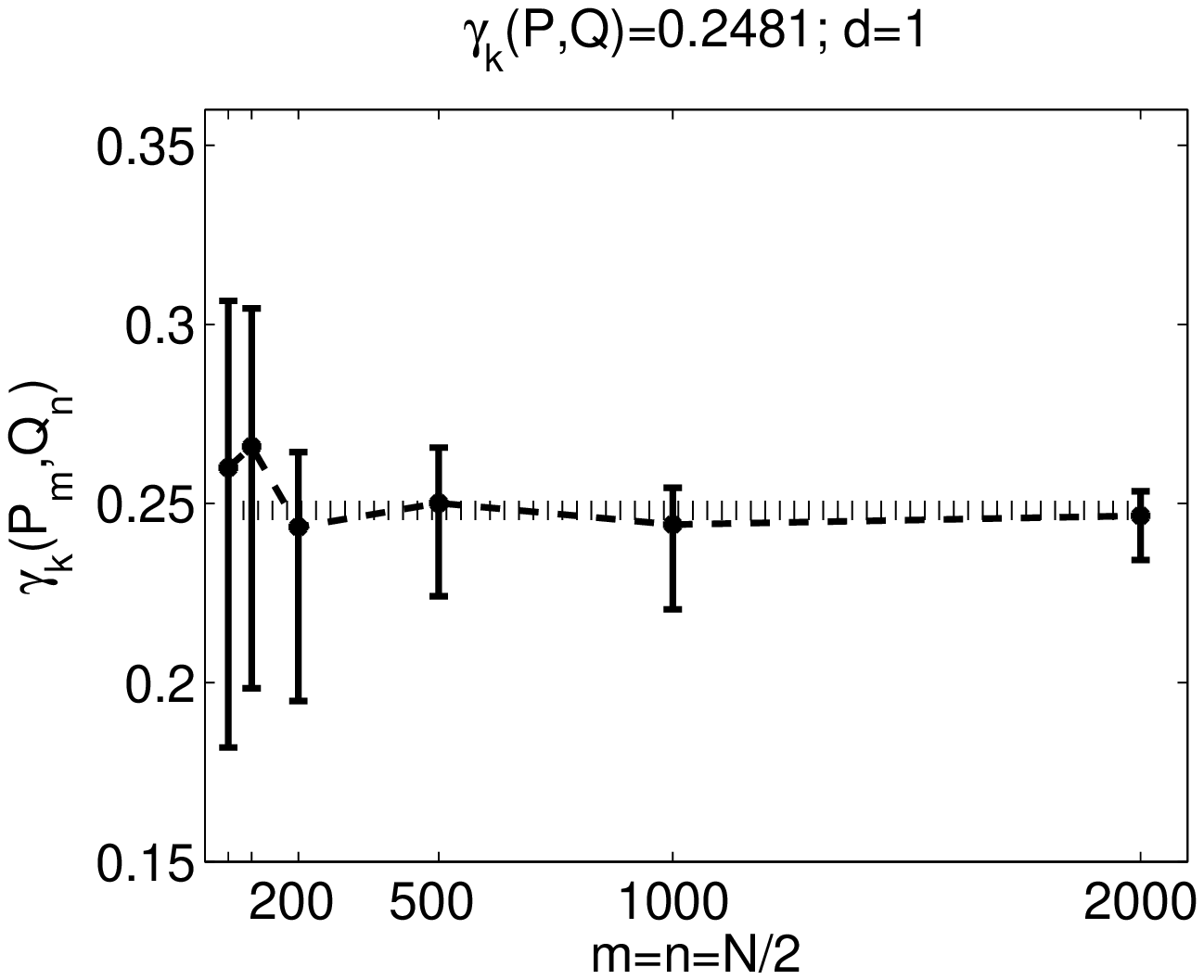}}\vspace{-2.5mm}
      {\small \center{(a$^\prime$)}}
    \end{minipage}\hspace{-3mm}
    \begin{minipage}{6cm}
      \center{\epsfxsize=6cm
      \epsffile{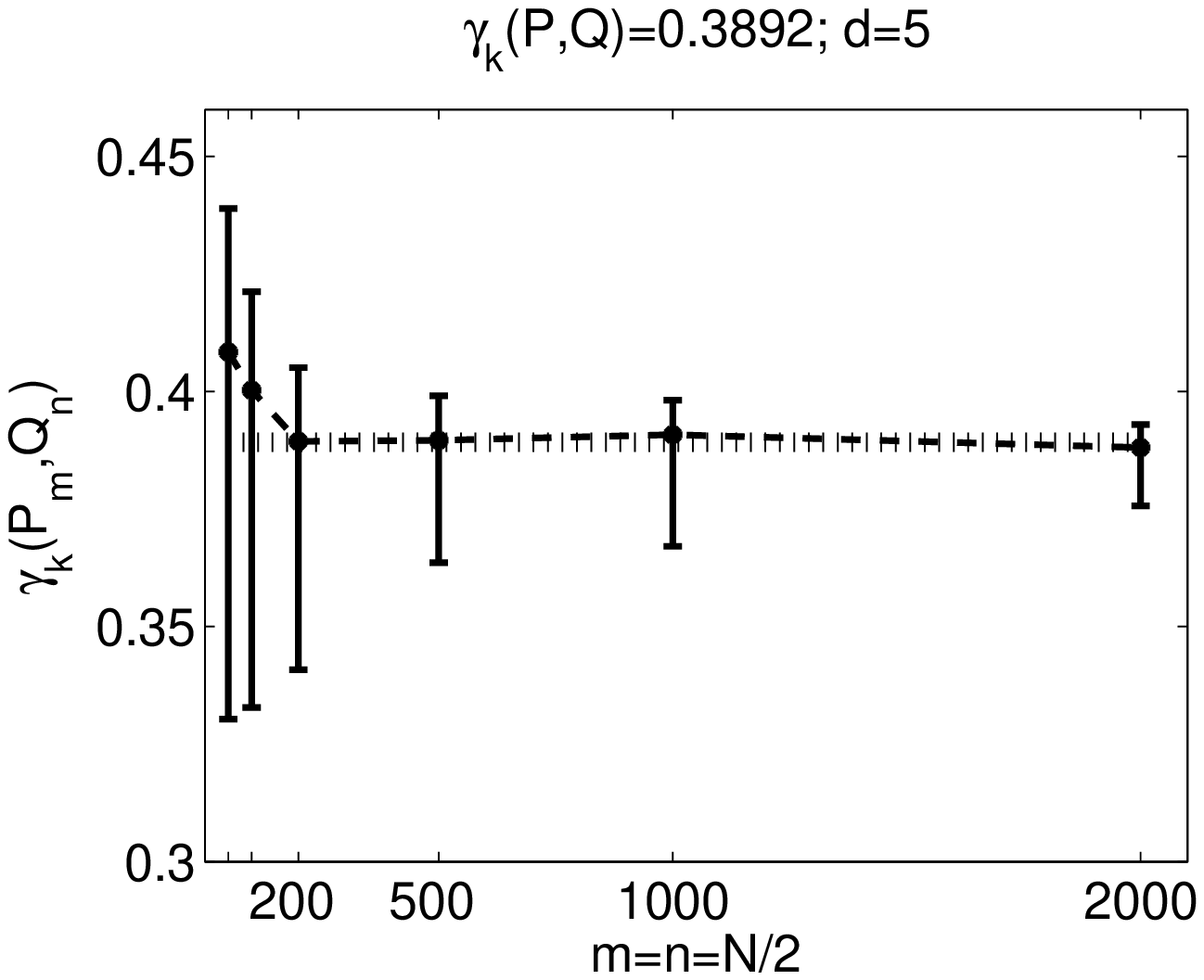}}\vspace{-2.5mm}
      {\small \center{(b$^\prime$)}}
    \end{minipage}\hspace{-3mm}
    \begin{minipage}{6cm}
      \center{\epsfxsize=6cm
      \epsffile{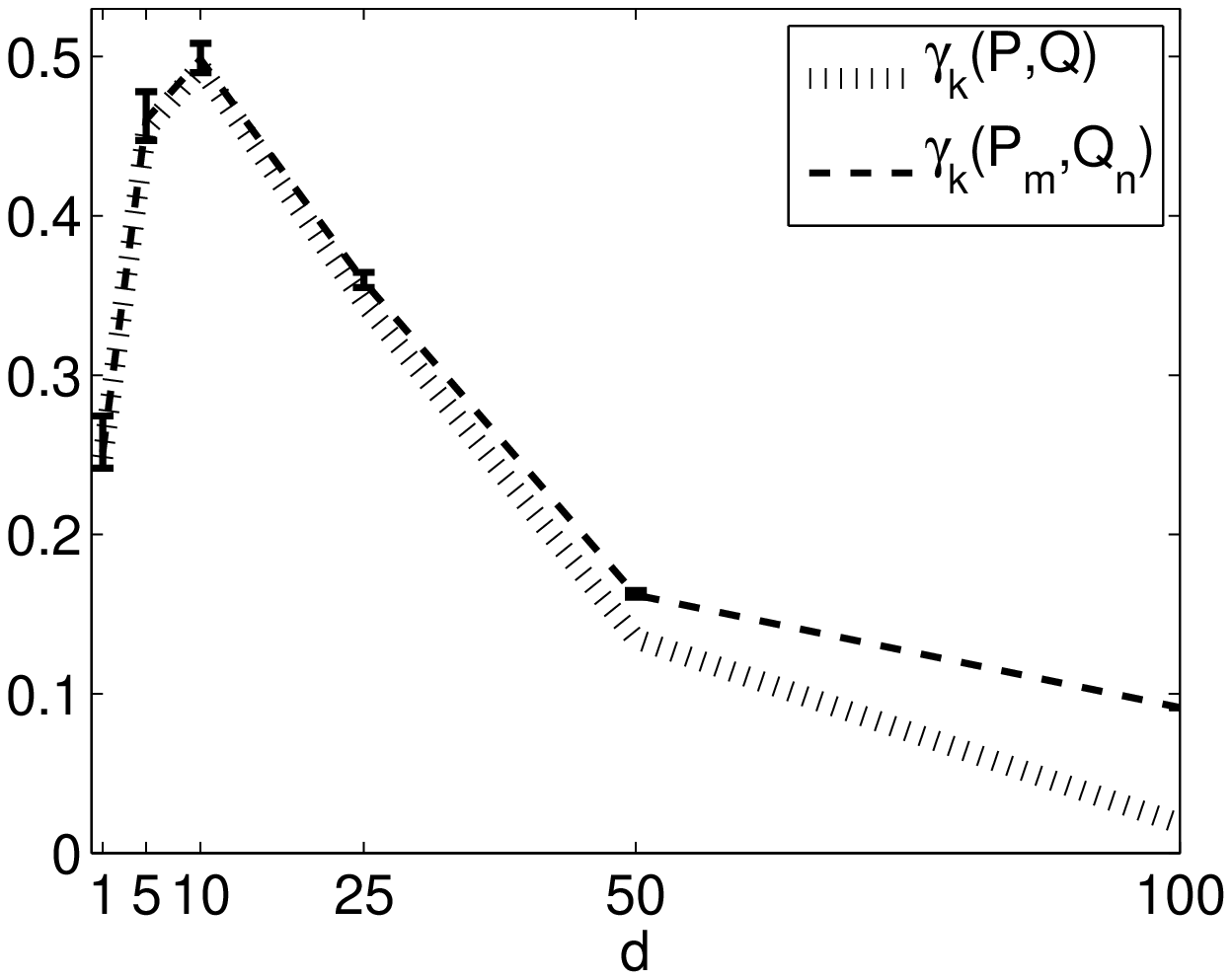}}\vspace{-2.5mm}
      {\small \center{(c$^\prime$)}}
    \end{minipage}\hspace{-3mm}
    \vspace{1mm}
  \end{tabular}
  \caption{(a-b) represent the empirical estimates of MMD (shown in thick dotted lines) between $\bb{P}=N(0,2I_d)$ and $\bb{Q}=N(1,2I_d)$ with $k(x,y)=\exp(-\frac{1}{2}\Vert x-y\Vert^2_2)$, for increasing sample size $N$, where $d=1$ in (a) and $d=5$ in (b) (see Example~\ref{exm:mmd-1} for details). Here $N(\mu,\sigma^2I_d)$ represents a normal distribution with mean vector $(\mu_1,\stackrel{d}{\ldots},\mu_d)$ and covariance matrix $\sigma^2I_d$. $I_d$ represents the $d\times d$ identity matrix. Similarly, (a$^\prime$-b$^\prime$) represent the empirical estimates of MMD (shown in thick dotted lines) between $\bb{P}$ and $\bb{Q}$, which are exponential distributions on $\bb{R}^d_+$ (see Example~\ref{exm:mmd-2} for details), for increasing sample size $N$. Here $d=1$ in (a$^\prime$) and $d=5$ in (b$^\prime$) with $k(x,y)=\exp(-\frac{1}{4}\Vert x-y\Vert_1)$. The population values of MMD are shown in thin dotted lines in (a-c, a$^\prime$-c$^\prime$). (c, c$^\prime$) represent the behavior of $\gamma_k(\bb{P}_m,\bb{Q}_n)$ and $\gamma_k(\bb{P},\bb{Q})$ for varying $d$ with a fixed sample size of $m=n=250$ (see Examples~\ref{exm:mmd-1} and \ref{exm:mmd-2} for details on the choice of $\bb{P}$ and $\bb{Q}$). Error bars are obtained by replicating the experiment $20$ times.}
\label{Fig:estimate-mmd}
\vspace{-4mm}
\end{figure*}
\begin{exm}\label{exm:wass-1}
Let $M=\times^d_{i=1}[a_i,s_i]$. Suppose $\bb{P}^{(i)}=U[a_i,b_i]$ and $\bb{Q}^{(i)}=U[r_i,s_i]$, which are uniform distributions on $[a_i,b_i]$ and $[r_i,s_i]$ respectively, where $-\infty<a_i\le r_i\le b_i\le s_i<\infty$. Then, it is easy to verify that $W(\bb{P}^{(i)},\bb{Q}^{(i)})=(s_i+r_i-a_i-b_i)/2$ and $W(\bb{P},\bb{Q})$ follows from (\ref{Eq:multi-wasserstein}).
\par Figures~\ref{Fig:estimate}(a) and \ref{Fig:estimate}(b) show the empirical estimates of $W$ (shown in thick dotted lines) for $d=1$ and $d=5$ respectively. Figure~\ref{Fig:estimate}(c) shows the behavior of $W(\bb{P}_m,\bb{Q}_n)$ and $W(\bb{P},\bb{Q})$ for various $d$ with a fixed sample size of $m=n=250$. Here, we chose $a_i=-\frac{1}{2}$, $b_i=\frac{1}{2}$, $r_i=0$ and $s_i=1$ for all $i=1,\ldots,d$ such that $W(\bb{P}^{(i)},\bb{Q}^{(i)})=\frac{1}{2},\,\forall i$ and $W(\bb{P},\bb{Q})=\frac{d}{2}$, shown in thin dotted lines in Figures~\ref{Fig:estimate}(a-c). Note that the present choice of $\bb{P}$ and $\bb{Q}$ would result in a KL-divergence of $+\infty$. \IEEEQED \vspace{2mm}
\end{exm}
\begin{exm}\label{exm:wass-2}
Let $M=\times^d_{i=1}[0,c_i]$. Suppose $\bb{P}^{(i)}$, $\bb{Q}^{(i)}$ have densities \begin{equation}
p_i(x)=\frac{d\bb{P}^{(i)}}{dx}=\frac{\lambda_ie^{-\lambda_ix}}{1-e^{-\lambda_ic_i}},\,\,q_i(x)=\frac{d\bb{Q}^{(i)}}{dx}=\frac{\mu_ie^{-\mu_ix}}{1-e^{-\mu_ic_i}}\nonumber
\end{equation}
respectively, where $\lambda_i>0,\,\mu_i>0$. Note that $\bb{P}^{(i)}$ and $\bb{Q}^{(i)}$ are exponential distributions supported on $[0,c_i]$ with rate parameters $\lambda_i$ and $\mu_i$. Then, it can be shown that \begin{equation}
W(\bb{P}^{(i)},\bb{Q}^{(i)})=\left|\frac{1}{\lambda_i}-\frac{1}{\mu_i}-\frac{c_i(e^{-\lambda_ic_i}-e^{-\mu_ic_i})}{(1-e^{-\lambda_ic_i})(1-e^{-\mu_ic_i})}\right|,\nonumber
\end{equation}
and $W(\bb{P},\bb{Q})$ follows from (\ref{Eq:multi-wasserstein}).
\par Figures~\ref{Fig:estimate}(a$^\prime$) and \ref{Fig:estimate}(b$^\prime$) show the empirical estimates of $W$ (shown in thick dotted lines) for $d=1$ and $d=5$ respectively. Let $\lambda=(\lambda_1,\stackrel{d}{\ldots},\lambda_d)$, $\mu=(\mu_1,\stackrel{d}{\ldots},\mu_d)$ and $c=(c_1,\stackrel{d}{\ldots},c_d)$. In Figure~\ref{Fig:estimate}(a$^\prime$), we chose $\lambda=(3)$, $\mu=(1)$ and $c=(5)$ which gives $W(\bb{P},\bb{Q})=0.6327$. %In Figure~\ref{Fig:estimate}(b$^\prime$), we chose $\lambda=(3,2)$, $\mu=(1,5)$ and $c=(5,6)$, which gives $W(\bb{P},\bb{Q})=0.9327$. 
In Figure~\ref{Fig:estimate}(b$^\prime$), we chose $\lambda=(3,2,1/2,2,7)$, $\mu=(1,5,5/2,1,8)$ and $c=(5,6,3,2,10)$, which gives $W(\bb{P},\bb{Q})=1.9149$. The population values $W(\bb{P},\bb{Q})$ are shown in thin dotted lines in Figures~\ref{Fig:estimate}(a$^\prime$) and \ref{Fig:estimate}(b$^\prime$). Figure~\ref{Fig:estimate}(c$^\prime$) shows $W(\bb{P}_m,\bb{Q}_n)$ and $W(\bb{P},\bb{Q})$ for various $d$ with a fixed sample size of $m=n=250$, $\lambda=(3,3,\stackrel{d}{\ldots},3)$, $\mu=(1,1,\stackrel{d}{\ldots},1)$ and $c=(5,5,\stackrel{d}{\ldots},5)$. \IEEEQED
\vspace{2mm}
\end{exm}
The empirical estimates in Figure~\ref{Fig:estimate} are obtained by drawing $N$ i.i.d. samples (with $m=n=N/2$) from $\bb{P}$ and $\bb{Q}$ and then solving the linear program in (\ref{Eq:lp}). It is easy to see from Figures~\ref{Fig:estimate}(a-b, a$^\prime$-b$^\prime$) that the estimate of $W(\bb{P},\bb{Q})$ improves with increasing sample size and that $W(\bb{P}_m,\bb{Q}_n)$ estimates $W(\bb{P},\bb{Q})$ correctly, which therefore demonstrates the efficacy of the estimator. Figures~\ref{Fig:estimate}(c) and \ref{Fig:estimate}(c$^\prime$) show the effect of dimensionality, $d$ of the data on the estimate of $W(\bb{P},\bb{Q})$. They show that at large $d$, the estimator has a large bias and more samples are needed to obtain better estimates. Error bars are obtained by replicating the experiment $20$ times.\vspace{2mm}
%So, for the easy computation of $W$ in (\ref{Eq:vallander}), let us consider $\bb{P}=U[-\frac{1}{2},\frac{1}{2}]^d$ and $\bb{Q}=U[0,1]^d$ where $U[l_1,l_2]^d$ represents a uniform distribution on $[l_1,l_2]^d$. Simple calculus would show that $W(\bb{P},\bb{Q})=\frac{d}{2}$. Note that the present choice of $\bb{P}$ and $\bb{Q}$ would result in a KL-divergence of $+\infty$. Now, we test the performance of the estimator $W(\bb{P}_m,\bb{Q}_n)$ by drawing i.i.d. samples (with $m=n$) from $\bb{P}$ and $\bb{Q}$ and then solving the linear program in (\ref{Eq:lp}). Figure~\ref{Fig:estimate}(a-c) shows the performance of $W(\bb{P}_m,\bb{Q}_n)$ for increasing sample sizes and different $d$ and it is clear that $W(\bb{P}_m,\bb{Q}_n)$ estimates $W(\bb{P},\bb{Q})$ correctly. Error bars are obtained by replicating the experiment $50$ times.
\subsubsection{Estimator of $\gamma_k$} We now consider the performance of $\gamma_k(\bb{P},\bb{Q})$. \cite{Gretton-06,Sriperumbudur-08} have shown that when $k$ is measurable and bounded,
%\begin{equation}
%\gamma_k(\bb{P},\bb{Q})=\left\Vert\int_M k(\cdot,x)\,d\bb{P}(x)-\int_M k(\cdot,x)\,d\bb{Q}(x)\right\Vert_\eu{H},\nonumber
%\end{equation}
%which is equivalent to
\begin{eqnarray}%\label{Eq:gammak}
\gamma_k(\bb{P},\bb{Q})&{}={}&\left\Vert\int_M k(\cdot,x)\,d\bb{P}(x)-\int_M k(\cdot,x)\,d\bb{Q}(x)\right\Vert_\eu{H}\nonumber
\end{eqnarray}
%\gamma_k(\bb{P},\bb{Q})
\begin{eqnarray}\label{Eq:gammak}
%&{}={}&\Big[\int k(x,y)\,d\bb{P}(x)\,d\bb{P}(y)\nonumber\\
%&{}{}&\quad +\int k(x,y)\,d\bb{Q}(x)\,d\bb{Q}(y)\nonumber\\
%&{}{}&\quad\quad -2\int k(x,y)\,d\bb{P}(x)\,d\bb{Q}(y)\Big]^{\frac{1}{2}}.
&{}={}&\Big[\int k(x,y)\,d\bb{P}(x)\,d\bb{P}(y)+\int k(x,y)\,d\bb{Q}(x)\,d\bb{Q}(y)\nonumber\\
&{}{}&\quad\quad -2\int k(x,y)\,d\bb{P}(x)\,d\bb{Q}(y)\Big]^{\frac{1}{2}}.
\end{eqnarray}
Note that, although $\gamma_k(\bb{P},\bb{Q})$ has a closed form in (\ref{Eq:gammak}), exact computation is not always possible for all choices of $k$, $\bb{P}$ and $\bb{Q}$. In such cases, one has to resort to numerical techniques to compute the integrals in (\ref{Eq:gammak}). In the following, we present two examples where we choose $\bb{P}$ and $\bb{Q}$ such that $\gamma_k(\bb{P},\bb{Q})$ can be computed exactly, which is then used to verify the performance of $\gamma_k(\bb{P}_m,\bb{Q}_n)$. Also note that for the consistency of $\gamma_k(\bb{P}_m,\bb{Q}_n)$, by Corollary~\ref{cor:wasserstein-dudley}, we just need the kernel, $k$ to be measurable and bounded and no assumptions on $M$ are required.\vspace{2mm}
\begin{exm}\label{exm:mmd-1}
Let $M=\bb{R}^d$, $\bb{P}=\otimes^d_{i=1}\bb{P}^{(i)}$ and $\bb{Q}=\otimes^d_{i=1}\bb{Q}^{(i)}$. Suppose $\bb{P}^{(i)}=N(\mu_i,\sigma^2_i)$ and $\bb{Q}^{(i)}=N(\lambda_i,\theta^2_i)$, where $N(\mu,\sigma^2)$ represents a Gaussian distribution with mean $\mu$ and variance $\sigma^2$. Let $k(x,y)=\exp(-\Vert x-y\Vert^2_2/2\tau^2)$. Clearly $k$ is measurable and bounded. With this choice of $k$, $\bb{P}$ and $\bb{Q}$, $\gamma_k$ in (\ref{Eq:gammak}) can be computed exactly as 
\begin{eqnarray}\label{Eq:gaussian-calculation}
\gamma^2_k(\bb{P},\bb{Q}){}&={}&\prod^d_{i=1}\frac{\tau}{\sqrt{2\sigma^2_i+\tau^2}}+\prod^d_{i=1}\frac{\tau}{\sqrt{2\theta^2_i+\tau^2}}\nonumber\\
{}&&{}\quad -2\prod^d_{i=1}\frac{\tau e^{-\frac{(\mu_i-\lambda_i)^2}{2(\sigma^2_i+\theta^2_i+\tau^2)}}}{\sqrt{\sigma^2_i+\theta^2_i+\tau^2}},
\end{eqnarray}
as the integrals in (\ref{Eq:gammak}) simply involve the convolution of Gaussian distributions.
\par Figures~\ref{Fig:estimate-mmd}(a-b) show the empirical estimates of $\gamma_k$ (shown in thick dotted lines) for $d=1$ and $d=5$ respectively. Figure~\ref{Fig:estimate-mmd}(c) shows the behavior of $\gamma_k(\bb{P}_m,\bb{Q}_n)$ and $\gamma_k(\bb{P},\bb{Q})$ for varying $d$ with a fixed sample size of $m=n=250$. Here we chose $\mu_i=0$, $\lambda_i=1$, $\sigma_i=\sqrt{2}$, $\theta_i=\sqrt{2}$ for all $i=1,\ldots,d$ and $\tau=1$. Using these values in (\ref{Eq:gaussian-calculation}), it is easy to check that $\gamma_k(\bb{P},\bb{Q})=5^{-d/4}(2-2e^{-d/10})^{1/2}$, which is shown in thin dotted lines in Figures~\ref{Fig:estimate-mmd}(a-c). We remark that an alternative estimator
of $\gamma_k$ exists which does not suffer from bias at small sample sizes: see \cite{Gretton-06}. \IEEEQED\vspace{2mm}
\end{exm}
\begin{exm}\label{exm:mmd-2}
Let $M=\bb{R}^d_+$, $\bb{P}=\otimes^d_{i=1}\bb{P}^{(i)}$ and $\bb{Q}=\otimes^d_{i=1}\bb{Q}^{(i)}$. Suppose $\bb{P}^{(i)}=\text{Exp}(1/\lambda_i)$ and $\bb{Q}^{(i)}=\text{Exp}(1/\mu_i)$, which are exponential distributions on $\bb{R}_+$ with rate parameters $\lambda_i>0$ and $\mu_i>0$ respectively. Suppose $k(x,y)=\exp(-\alpha\Vert x-y\Vert_1)$, $\alpha>0$, which is a Laplacian kernel on $\bb{R}^d$. Then, it is easy to verify that $\gamma_k(\bb{P},\bb{Q})$ in (\ref{Eq:gammak}) reduces to
\begin{eqnarray}\label{Eq:laplacian-calculation}
\gamma^2_k(\bb{P},\bb{Q})&{}={}&\prod^d_{i=1}\frac{\lambda_i}{\lambda_i+\alpha}+\prod^d_{i=1}\frac{\mu_i}{\mu_i+\alpha}\nonumber\\
&{}{}&\quad -2\prod^d_{i=1}\frac{\lambda_i\mu_i(\lambda_i+\mu_i+2\alpha)}{(\lambda_i+\alpha)(\mu_i+\alpha)(\lambda_i+\mu_i)}.\nonumber
\end{eqnarray}
\par Figures~\ref{Fig:estimate-mmd}(a$^\prime$-b$^\prime$) show the empirical estimates of $\gamma_k$ (shown in thick dotted lines) for $d=1$ and $d=5$ respectively. Figure~\ref{Fig:estimate-mmd}(c$^\prime$) shows the dependence of $\gamma_k(\bb{P}_m,\bb{Q}_n)$ and $\gamma_k(\bb{P},\bb{Q})$ on $d$ at a fixed sample size of $m=n=250$. Here, we chose $\{\lambda_i\}^d_{i=1}$ and $\{\mu_i\}^d_{i=1}$ as in Example~\ref{exm:wass-2} with $\alpha=\frac{1}{4}$, which gives $\gamma_k(\bb{P},\bb{Q})=0.2481$ for $d=1$ %, $0.2745$ for $d=2$ 
and $0.3892$ for $d=5$, shown in thin dotted lines in Figures~\ref{Fig:estimate-mmd}(a$^\prime$-c$^\prime$). \IEEEQED
\vspace{2mm}
\end{exm}
As in the case of $W$, the performance of $\gamma_k(\bb{P}_m,\bb{Q}_n)$ is verified by drawing $N$ i.i.d. samples (with $m=n=N/2$) from $\bb{P}$ and $\bb{Q}$ and computing $\gamma_k(\bb{P}_m,\bb{Q}_n)$ in (\ref{Eq:gamma-opt}). Figures~\ref{Fig:estimate-mmd}(a-b, a$^\prime$-b$^\prime$) show the performance of $\gamma_k(\bb{P}_m,\bb{Q}_n)$ for various sample sizes and some fixed $d$. It is easy to see that the quality of the estimate improves with increasing sample size and that $\gamma_k(\bb{P}_m,\bb{Q}_n)$ estimates $\gamma_k(\bb{P},\bb{Q})$ correctly. On the other hand, Figures~\ref{Fig:estimate-mmd}(c, c$^\prime$) demonstrate that $\gamma_k(\bb{P}_m,\bb{Q}_n)$ is biased at large $d$ and more samples are needed to obtain better estimates. As in the case of $W$, the error bars are obtained by replicating the experiment $20$ times.\vspace{2mm}
\subsubsection{Estimator of $\beta$}\label{subsubsec:beta}
\begin{figure}[t]
  \centering
  \begin{tabular}{cccccc}\hspace{-5mm}
    \begin{minipage}{6cm}
      \center{\epsfxsize=6cm
      \epsffile{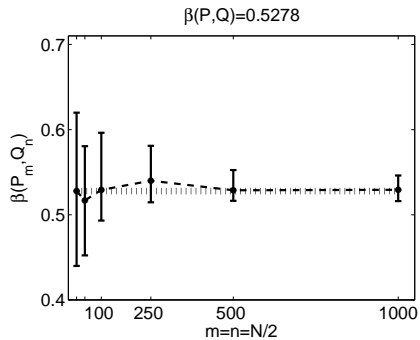}}\vspace{-2.5mm}
      \end{minipage}\hspace{-3mm}
    \vspace{4mm}
  \end{tabular}
  \caption{Empirical estimates of the Dudley metric (shown in a thick dotted line) between discrete distributions $\bb{P}$ and $\bb{Q}$ on $\bb{R}$ (see Example~\ref{exm:Dudley} for details), for increasing sample size $N$. The population value of the Dudley metric is shown in a thin dotted line. Error bars are obtained by replicating the experiment $20$ times.}
\label{Fig:estimate-dudley}
\vspace{-3mm}
\end{figure}

In the case of $W$ and $\gamma_k$, we have some closed form expression to start with (see (\ref{Eq:vallander}) and (\ref{Eq:gammak})), which can be solved by numerical methods. The resulting value is then used as the baseline to test the performance of the estimators of $W$ and $\gamma_k$. On the other hand, in the case of $\beta$, we are not aware of any such closed form expression to compute the baseline. However, it is possible to compute $\beta(\bb{P},\bb{Q})$ when $\bb{P}$ and $\bb{Q}$ are discrete distributions on $M$, i.e., $\bb{P}=\sum^r_{i=1}\lambda_i\delta_{X_i}$, $\bb{Q}=\sum^s_{i=1}\mu_i\delta_{Z_i}$, where $\sum^r_{i=1}\lambda_i=1$, $\sum^s_{i=1}\mu_i=1$, $\lambda_i\ge 0,\,\forall\,i$, $\mu_i\ge 0,\,\forall\,i$, and $X_i,Z_i\in M$. This is because, for this choice of $\bb{P}$ and $\bb{Q}$, we have
\begin{eqnarray}
\beta(\bb{P},\bb{Q})&{}={}&\sup\Big\{\sum^r_{i=1}\lambda_if(X_i)-\sum^s_{i=1}\mu_if(Z_i):\Vert f\Vert_{BL}\le 1\Big\}\nonumber\\
\label{Eq:beta-trivial}
&{}={}&\sup\Big\{\sum^{r+s}_{i=1}\theta_if(V_i):\Vert f\Vert_{BL}\le 1\Big\},
\end{eqnarray}
where $\theta=(\lambda_1,\ldots,\lambda_r,-\mu_1,\ldots,-\mu_s)$, $V=(X_1,\ldots,X_r,Z_1,\ldots,Z_s)$ with $\theta_i:=(\theta)_i$ and $V_i:=(V)_i$. Now, (\ref{Eq:beta-trivial}) is of the form of (\ref{Eq:ipm-empirical}) and so, by Theorem~\ref{thm:Dudley}, $\beta(\bb{P},\bb{Q})=\sum^{r+s}_{i=1}\theta_ia^\star_i$, where $\{a^\star_i\}$ solve the following linear program, 
\begin{eqnarray}\label{Eq:lp-dudley-estimate}
\max_{a_1,\ldots,a_{r+s},b,c} &{}{}&\,\,\, \sum^{r+s}_{i=1}\theta_ia_i\nonumber\\
\text{s.t.} &{}{}& \,\,\,-b\,\rho(V_i,V_j)\le a_i-a_j\le b\,\rho(V_i,V_j),\,\forall\,i,j\nonumber\\
&{}{}&\,\,\, -c\le a_i\le c,\,\forall\,i\nonumber\\
&{}{}&\,\,\, b+c\le 1.
\end{eqnarray}
Therefore, for these distributions, one can compute the baseline which can then be used to verify the performance of $\beta(\bb{P}_m,\bb{Q}_n)$. In the following, we consider a simple example to demonstrate the performance of $\beta(\bb{P}_m,\bb{Q}_n)$.\vspace{2mm}
\begin{exm}\label{exm:Dudley}
Let $M=\{0,1,2,3,4,5\}\subset\bb{R}$, $\lambda=(\frac{1}{3},\frac{1}{6},\frac{1}{8},\frac{1}{4},\frac{1}{8})$, $\mu=(\frac{1}{4},\frac{1}{4},\frac{1}{4},\frac{1}{4})$, $X=(0,1,2,3,4)$ and $Z=(2,3,4,5)$. With this choice, $\bb{P}$ and $\bb{Q}$ are defined as $\bb{P}=\sum^5_{i=1}\lambda_i\delta_{X_i}$ and $\bb{Q}=\sum^4_{i=1}\mu_i\delta_{Z_i}$. By solving (\ref{Eq:lp-dudley-estimate}) with $\rho(x,y)=|x-y|$, we get $\beta(\bb{P},\bb{Q})=0.5278$. Note that the KL-divergence between $\bb{P}$ and $\bb{Q}$ is $+\infty$.
\par Figure~\ref{Fig:estimate-dudley} shows the empirical estimates of $\beta(\bb{P},\bb{Q})$ (shown in a thick dotted line) which are computed by drawing $N$ i.i.d. samples (with $m=n=N/2$) from $\bb{P}$ and $\bb{Q}$ and solving the linear program in (\ref{Eq:lp-dudley}). It can be seen that $\beta(\bb{P}_m,\bb{Q}_n)$ estimates $\beta(\bb{P},\bb{Q})$ correctly. \IEEEQED
\vspace{2mm}
\end{exm}
Since we do not know how to compute $\beta(\bb{P},\bb{Q})$ for $\bb{P}$ and $\bb{Q}$ other than the ones we discussed here, we do not provide any other non-trivial examples to test the performance of $\beta(\bb{P}_m,\bb{Q}_n)$.
\subsection{Non-parametric estimation of total variation distance}\label{subsec:tv}
So far, the results in Section~\ref{subsec:empirical}--\ref{subsec:simulation} show that IPMs exhibit nice properties compared to that of $\phi$-divergences. As shown in Section~\ref{Sec:intersection}, since the total variation distance,
\begin{equation}\label{Eq:tv}
TV(\bb{P},\bb{Q}):=\sup\{\bb{P}f-\bb{Q}f:\Vert f\Vert_\infty\le 1\},
\end{equation}
is both an IPM and $\phi$-divergence, in this section, we consider its empirical estimation and the consistency analysis. Let $TV(\bb{P}_m,\bb{Q}_n)$ be an empirical estimator of $TV(\bb{P},\bb{Q})$. Using similar arguments as in Theorems~\ref{thm:Lipschitz} and \ref{thm:Dudley}, it can be shown that \begin{equation}\label{Eq:empirical-tv}
TV(\bb{P}_m,\bb{Q}_n)=\sum^N_{i=1}\widetilde{Y}_ia^\star_i,
\end{equation}
where $\{a^\star_i\}^N_{i=1}$ solve the following linear program,
\setlength{\arraycolsep}{0.0em}
\begin{eqnarray}
\max_{a_1,\ldots,a_N}&{}{}&\,\,\,\sum^N_{i=1}\widetilde{Y}_ia_i\nonumber\\
\text{s.t.}&{}{}&\,\,\,-1\le a_i\le 1,\,\forall\,i.
\end{eqnarray}
Now, the question is whether this estimator consistent. To answer this question, we consider an equivalent representation of $TV$ given as \begin{equation}
TV(\bb{P},\bb{Q})=2\sup_{A\in\eu{A}}\vert\bb{P}(A)-\bb{Q}(A)\vert,
\end{equation}
where the supremum is taken over all measurable subsets $A$ of $M$ \cite{Devroye-90}. Note that $\vert TV(\bb{P}_m,\bb{Q}_n)-TV(\bb{P},\bb{Q})\vert\le TV(\bb{P}_m,\bb{P})+TV(\bb{Q}_n,\bb{Q})$. It is easy to see that $TV(\bb{P}_m,\bb{P})\stackrel{a.s.}{\nrightarrow} 0$ as $m\rightarrow\infty$ for all $\bb{P}$ and therefore, the estimator in (\ref{Eq:empirical-tv}) is not strongly consistent. This is because if $\bb{P}$ is absolutely continuous, then $TV(\bb{P}_m,\bb{P})=2$, where we have considered the set $A$ that is the finite support of $\bb{P}_m$ such that $\bb{P}_m(A)=1$ and $\bb{P}(A)=0$. In fact, Devroye and Gy\"{o}rfi \cite{Devroye-90} have proved that for any empirical measure, $\widehat{\bb{P}}_m$ (a function depending on $\{X^{(1)}_i\}^m_{i=1}$ assigning a nonnegative number to any measurable set), there exists a distribution, $\bb{P}$ such that for all $m$,
\begin{equation}
\sup_{A\in\eu{A}}\vert\widehat{\bb{P}}_m(A)-\bb{P}(A)\vert>\frac{1}{4}\,\,\text{a.s.}
\end{equation}
This indicates that, for the strong consistency of distribution estimates in total variation, the set of probability measures has to be restricted. Barron \emph{et al.}~\cite{Barron-92} have studied the classes of distributions that can be estimated consistently in total variation. Therefore, for such distributions, the total variation distance between them can be estimated by an estimator that is strongly consistent.
\par The issue in the estimation of $TV(\bb{P},\bb{Q})$ is that the set $\eu{F}_{TV}:=\{f:\Vert f\Vert_\infty\le 1\}$ is too large to obtain meaningful results if no assumptions on distributions are made. On the other hand, one can choose a more manageable subset $\eu{F}$ of $\eu{F}_{TV}$ such that $\gamma_\eu{F}(\bb{P},\bb{Q})\le TV(\bb{P},\bb{Q}),\,\forall\,\bb{P},\bb{Q}\in\Scr{P}$ and $\gamma_\eu{F}(\bb{P}_m,\bb{Q}_n)$ is a consistent estimator of $\gamma_\eu{F}(\bb{P},\bb{Q})$. Examples of such choice of $\eu{F}$ include $\eu{F}_\beta$ and $\{\mathds{1}_{(-\infty,t]}:t\in\bb{R}^d\}$, where the former yields the Dudley metric while the latter results in the Kolmogorov distance. The empirical estimator of the Dudley metric and its consistency have been presented in Sections \ref{subsec:empirical} and \ref{subsec:consistency}. The empirical estimator of the Kolmogorov distance between $\bb{P}$ and $\bb{Q}$ is well studied and is strongly consistent, which simply follows from the famous Glivenko-Cantelli theorem \cite[Theorem 12.4]{Devroye-96}.
\par Since the total variation distance between $\bb{P}$ and $\bb{Q}$ cannot be estimated consistently for all $\bb{P},\bb{Q}\in\Scr{P}$, in the following, we present two lower bounds on $TV$, one involving $W$ and $\beta$ and the other involving $\gamma_k$, which can be estimated consistently.\vspace{2mm}
\begin{thm}[Lower bounds on $TV$]\label{thm:TVbound}
\begin{itemize}
\item[\emph{(i)}] For all $\bb{P}\ne \bb{Q}$, $\bb{P},\bb{Q}\in\Scr{P}$, we have
\begin{equation}\label{Eq:w-beta-tv}
TV(\bb{P},\bb{Q})\ge \frac{W(\bb{P},\bb{Q})\beta(\bb{P},\bb{Q})}{W(\bb{P},\bb{Q})-\beta(\bb{P},\bb{Q})}.
\end{equation}
\item[\emph{(ii)}] Suppose $C:=\sup_{x\in M} k(x,x)<\infty$. Then \begin{equation}\label{Eq:mmd-tv}
TV(\bb{P},\bb{Q})\ge\frac{\gamma_k(\bb{P},\bb{Q})}{\sqrt{C}}.
\end{equation}
\end{itemize}
\end{thm}
Before, we prove Theorem~\ref{thm:TVbound}, we present a simple lemma.\vspace{2mm}
\begin{lem}\label{lem:primal-dual}
Let $\theta:V\rightarrow\bb{R}$ and $\psi:V\rightarrow\bb{R}$ be convex functions on a real vector space $V$. Suppose 
\begin{equation}\label{Eq:primal}
a=\sup\{\theta(x):\psi(x)\le b\},
\end{equation}
where $\theta$ is not constant on $\{x:\psi(x)\le b\}$ and $a<\infty$.
Then 
\begin{equation}\label{Eq:dual}
b=\inf\{\psi(x):\theta(x)\ge a\}.
\end{equation}
\end{lem}
\begin{IEEEproof}
See Appendix~\ref{appendix-optim}.
\vspace{2mm}
\end{IEEEproof}
\begin{IEEEproof}[Proof of Theorem~\ref{thm:TVbound}]
\emph{(i)} Note that $\Vert f\Vert_L$, $\Vert f\Vert_{BL}$ and $\Vert f\Vert_\infty$ are convex functionals on the vector spaces $\text{Lip}(M,\rho)$, $BL(M,\rho)$ and $U(M):=\{f:M\rightarrow\bb{R}\,|\,\Vert f\Vert_\infty<\infty\}$ respectively. Similarly, $\bb{P}f-\bb{Q}f$ is a convex functional on $\text{Lip}(M,\rho)$, $BL(M,\rho)$ and $U(M)$. Since $\bb{P}\ne \bb{Q}$, $\bb{P}f-\bb{Q}f$ is not constant on $\eu{F}_W$, $\eu{F}_\beta$ and $\eu{F}_{TV}$. Therefore, by appropriately choosing $\psi$, $\theta$, $V$ and $b$ in Lemma~\ref{lem:primal-dual}, the following sequence of inequalities are obtained. Define $\beta:=\beta(\bb{P},\bb{Q})$, $W:=W(\bb{P},\bb{Q})$, $TV:=TV(\bb{P},\bb{Q})$.  
\setlength{\arraycolsep}{0.0em}
\begin{eqnarray}
1&{}={}&\inf\{\Vert f\Vert_{BL}:\bb{P}f-\bb{Q}f\ge \beta,\,f\in BL(M,\rho)\}\nonumber\\
&{}\ge{}&\inf\{\Vert f\Vert_{L}:\bb{P}f-\bb{Q}f\ge \beta,\,f\in BL(M,\rho)\}\nonumber\\
&{}{}&\quad+\inf\{\Vert f\Vert_\infty:\bb{P}f-\bb{Q}f\ge \beta,\,f\in BL(M,\rho)\}\nonumber\\
&{}={}&\frac{\beta}{W}\inf\{\Vert f\Vert_{L}:\bb{P}f-\bb{Q}f\ge W,\,f\in BL(M,\rho)\}\nonumber \\
%\end{eqnarray}
%\begin{eqnarray}
&{}{}&\quad +\frac{\beta}{TV}\inf\{\Vert f\Vert_\infty:\bb{P}f-\bb{Q}f\ge TV,\,f\in BL(M,\rho)\}\nonumber\\
&{}\ge{}&\frac{\beta}{W}\inf\{\Vert f\Vert_{L}:\bb{P}f-\bb{Q}f\ge W,\,f\in\text{Lip}(M,\rho)\}\nonumber\\
&{}{}&\quad+\frac{\beta}{TV}\inf\{\Vert f\Vert_\infty:\bb{P}f-\bb{Q}f\ge TV,\,f\in U(M)\}\nonumber\\
&{}={}&\frac{\beta}{W}+\frac{\beta}{TV},\nonumber
\end{eqnarray}
%\setlength{\arraycolsep}{0.0em}
%\begin{eqnarray}
%&{}\ge{}&\frac{\beta}{W}\inf\{\Vert f\Vert_{L}:\bb{P}f-\bb{Q}f\ge W,\,f\in\text{Lip}(M,\rho)\}\nonumber\\
%&{}{}&\quad+\frac{\beta}{TV}\inf\{\Vert f\Vert_\infty:\bb{P}f-\bb{Q}f\ge TV,\,f\in U(M)\}\nonumber\\
%&{}={}&\frac{\beta}{W}+\frac{\beta}{TV},\nonumber
%\end{eqnarray}
which gives (\ref{Eq:w-beta-tv}).\vspace{2mm}\\
\emph{(ii)} To prove (\ref{Eq:mmd-tv}), we use the coupling formulation for $TV$ \cite[p. 19]{Lindvall-92} given by
\begin{equation}\label{Eq:coupling-tv}
TV(\bb{P},\bb{Q})=2\inf_{\mu\in\eu{L}(\bb{P},\bb{Q})}\mu(X\ne Y),
\end{equation}
where $\eu{L}(\bb{P},\bb{Q})$ is the set of all measures on $M\times M$ with marginals $\bb{P}$ and $\bb{Q}$. Here, $X$ and $Y$ are distributed as $\bb{P}$ and $\bb{Q}$ respectively. Let $\lambda\in \eu{L}(\bb{P},\bb{Q})$ and $f\in\eu{H}$. Then 
\setlength{\arraycolsep}{0.0em}
\begin{eqnarray}
\left\vert\int_M f\,d(\bb{P}-\bb{Q})\right\vert&{}={}&\left\vert\int (f(x)-f(y))\,d\lambda(x,y)\right\vert\nonumber\\
&{}\le{}&\int\vert f(x)-f(y)\vert\,d\lambda(x,y)\nonumber\\
&{}\stackrel{(a)}{=}{}&\int\vert\langle f,k(\cdot,x)-k(\cdot,y)\rangle_\eu{H}\vert\,d\lambda(x,y)\nonumber\\
&{}\stackrel{(b)}{\le}{}&\Vert f\Vert_\eu{H}\int\Vert k(\cdot,x)-k(\cdot,y)\Vert_\eu{H}\,d\lambda(x,y),\nonumber
\end{eqnarray}
where we have used the reproducing property of $\eu{H}$ in $(a)$ and the Cauchy-Schwartz inequality in $(b)$. Taking the supremum over $f\in\eu{F}_k$ and the infimum over $\lambda\in\eu{L}(\bb{P},\bb{Q})$ gives
\begin{equation}\label{Eq:coupling-rkhs}
\gamma_k(\bb{P},\bb{Q})\le\inf_{\lambda\in\eu{L}(\bb{P},\bb{Q})}\int \Vert k(\cdot,x)-k(\cdot,y)\Vert_\eu{H}\,d\lambda(x,y).
\end{equation}
Consider 
\setlength{\arraycolsep}{0.0em}
\begin{eqnarray}\label{Eq:mmd-to-tv}
\Vert k(\cdot,x)-k(\cdot,y)\Vert_\eu{H}&{}\le{}&\mathds{1}_{x\ne y}\Vert k(\cdot,x)-k(\cdot,y)\Vert_\eu{H}\nonumber\\
&{}\le{}& \mathds{1}_{x\ne y}\left[\Vert k(\cdot,x)\Vert_\eu{H}+\Vert k(\cdot,y)\Vert_\eu{H}\right]\nonumber\\
&{}={}&\mathds{1}_{x\ne y}\left[\sqrt{k(x,x)}+\sqrt{k(y,y)}\right]\nonumber\\
&{}\le{}& 2\sqrt{C}\mathds{1}_{x\ne y}.
\end{eqnarray}
Using (\ref{Eq:mmd-to-tv}) in (\ref{Eq:coupling-rkhs}) yields (\ref{Eq:mmd-tv}).\vspace{2mm}
\end{IEEEproof}
\begin{rem}\label{rem:tv}
\emph{(i)} As mentioned before, a simple lower bound on $TV$ can be obtained as $TV(\bb{P},\bb{Q})\ge \beta(\bb{P},\bb{Q}),\,\forall\,\bb{P},\bb{Q}\in\Scr{P}$. It is easy to see that the bound in (\ref{Eq:w-beta-tv}) is tighter as $\frac{W(\bb{P},\bb{Q})\beta(\bb{P},\bb{Q})}{W(\bb{P},\bb{Q})-\beta(\bb{P},\bb{Q})}\ge\beta(\bb{P},\bb{Q})$ with equality if and only if $\bb{P}=\bb{Q}$.\vspace{2mm}\\
\emph{(ii)} From (\ref{Eq:w-beta-tv}), it is easy to see that $TV(\bb{P},\bb{Q})=0$ or $W(\bb{P},\bb{Q})=0$ implies $\beta(\bb{P},\bb{Q})=0$ while the converse is not true. This shows that the topology induced by $\beta$ on $\Scr{P}$ is coarser than the topology induced by either $W$ or $TV$.\vspace{2mm}\\
\emph{(iii)} The bounds in (\ref{Eq:w-beta-tv}) and (\ref{Eq:mmd-tv}) translate as lower bounds on the KL-divergence through Pinsker's inequality: $TV^2(\bb{P},\bb{Q})\le 2\,KL(\bb{P},\bb{Q}),\,\forall\,\bb{P},\bb{Q}\in\Scr{P}$. See Fedotov \emph{et al.} \cite{Fedotov-03} and references therein for more refined bounds between $TV$ and $KL$. Therefore, using these bounds, one can obtain a consistent estimate of a lower bound on $TV$ and $KL$. The bounds in (\ref{Eq:w-beta-tv}) and (\ref{Eq:mmd-tv}) also translate to lower bounds on other distance measures on $\Scr{P}$. See \cite{Gibbs-02} for a detailed discussion on the relation between various metrics.\vspace{2mm}
\end{rem}
To summarize, in this section, we have considered the empirical estimation of IPMs along with their convergence rate analysis. We have shown that IPMs such as the Wasserstein distance, Dudley metric and MMD are simpler to estimate than the KL-divergence. This is because the Wasserstein distance and Dudley metric are estimated by solving a linear program while estimating the KL-divergence involves solving a quadratic program \cite{Nguyen-08}. Even more, the estimator of MMD has a simple closed form expression. On the other hand, space partitioning schemes like in \cite{Wang-05}, to estimate the KL-divergence, become increasingly difficult to implement as the number of dimensions increases whereas an increased number of dimensions has only a mild effect on the complexity of estimating $W$, $\beta$ and $\gamma_k$. In addition, the estimators of IPMs, especially the Wasserstein distance, Dudley metric and MMD, exhibit good convergence behavior compared to KL-divergence estimators as the latter can have an arbitrarily slow rate of convergence depending on the probability distributions \cite{Wang-05,Nguyen-08}. With these advantages, we believe that IPMs can find applications in information theory, detection theory, image processing, machine learning, neuroscience and other areas. As an example, in the following section, we show how IPMs are related to binary classification. %AG: I do not believe this is a convincing justification for the classification. See notes.
%BK: the justification is presented in the following section. however, we cannot use the witness function formalism for wasserstein and dudley unlike in mmd. in mmd, taking the sign of the witness function gives u quite a bit of an idea. if u see, the witness functions of the wasserstein and dudley, it is not that clear as the function depends on the output of a linear program.
\section{Interpretability of IPMs: Relation to Binary Classification}\label{Sec:surrogate}
In this section, we provide different interpretations of IPMs by relating them to the problem of binary classification. %Many previous works, e.g., \cite{Buja-05,Liese-06,Nguyen-09} have studied the problem of relating the risk (expected loss) in binary classification problems to $\phi$-divergences 
%, i.e., determining the loss function related to a $\phi$-divergence and vice-versa 
%(see \cite[Section 1.3]{Reid-09} for a list of detailed references). %Since IPMs are essentially different from $\phi$-divergences (see Section~\ref{Sec:intersection}), we are interested to study the relation between IPMs and binary classification. 
%In this section, we show how IPMs, measuring the distance between class conditional distributions, appear naturally in binary classification problems. 
First, in Section~\ref{Sec:Lrisk_interpret}, we provide a novel interpretation for $\beta$, $W$, $TV$ and $\gamma_k$ (see Theorem~\ref{thm:functionclass}), as the optimal risk associated with an appropriate binary classification problem. Second, in Section~\ref{Sec:Lipschitz}, we relate $W$ and $\beta$ to the margin of the Lipschitz classifier \cite{Luxburg-04} and the bounded Lipschitz classifier respectively. The significance of this result is that the smoothness of Lipschitz and bounded Lipschitz classifiers is inversely related to the distance between the class-conditional distributions, computed using $W$ and $\beta$ respectively. Third, in Section~\ref{Sec:mmd}, we discuss the relation between $\gamma_k$ and the Parzen window classifier \cite{Scholkopf-02,ShaweTaylor-04} (also called the kernel classification rule~\cite[Chapter 10]{Devroye-96}).%, Fisher discriminant analysis~\cite[Section 4.3]{Devroye-96} and hard-margin support vector machine \cite{Cortes-95}.
\subsection{Interpretation of $\beta$, $W$, $TV$ and $\gamma_k$ as the optimal risk of a binary classification problem}\label{Sec:Lrisk_interpret}
Let us consider the binary classification problem with $X$ being a $M$-valued random variable, $Y$ being a $\{-1,+1\}$-valued random variable and the product space, $M\times \{-1,+1\}$, being endowed with a Borel probability measure $\mu$. A discriminant function, $f$ is a real valued measurable function on $M$, whose sign is used to make a classification decision. Given a loss function, $L:\{-1,+1\}\times\bb{R}\rightarrow\bb{R}$, the goal is to choose an $f$ that minimizes the risk associated with $L$, with the optimal $L$-risk being defined as, 
%probability of making the incorrect classification, also known as the \emph{Bayes risk}. Formally, the Bayes decision rule involves solving 
%\setlength{\arraycolsep}{0.0em}
%\begin{eqnarray}\label{Eq:0-1loss}
%\widetilde{f}&{}={}&\arg\inf_{f\in\eu{F}_\star}\mu(Y\ne \text{sign}(f(X)))\nonumber\\
%&{}={}&\arg\inf_{f\in\eu{F}_\star}\int_M 
%\llbracket y\ne\text{sign}(f(x))\rrbracket\,d\mu(x,y),
%\end{eqnarray}
%where $(a,b)\mapsto\llbracket a\ne b\rrbracket$ is the $0-1$ loss function and $\mu(Y\ne\text{sign}(\widetilde{f}(X)))$ is the Bayes risk. (\ref{Eq:0-1loss}) can be generalized by replacing the $0-1$ loss function with some loss function, $L:\{-1,+1\}\times\mathbb{R}\rightarrow\mathbb{R}$, for example the squared-loss function, $L(a,b)=(a-b)^2$. Given $L$, the associated optimal $L$-risk is defined as
%For computational reasons, usually $L$ is chosen to be convex. %non-convex, for computational reasons, a surrogate loss function that is convex is usually considered. %the problem in (\ref{Eq:0-1loss}) is NP-hard to solve because of the non-convexity of the loss function, usually a surrogate loss function that is convex is considered. 
%Given $\ell$, the associated optimal $\ell$-risk is defined as 
\setlength{\arraycolsep}{0.0em}
\begin{eqnarray}\label{Eq:lrisk}
R^L_{\eu{F}_\star}&{}={}&\inf_{f\in\eu{F}_\star}\int_ML(y,f(x))\,d\mu(x,y)\nonumber\\
&{}={}&\inf_{f\in\eu{F}_\star}\Big\{\varepsilon\int_M L_1(f(x))\,d\bb{P}(x)\nonumber\\
&{}{}&\quad\quad +\,(1-\varepsilon)\int_M L_{-1}(f(x))\,d\bb{Q}(x)\Big\},
\end{eqnarray} where
%\begin{eqnarray}\label{Eq:lrisk}
%R_l(g)=\mu l(Yg(X))&=&\int_M\left[\eta(x)l(g(x))+(1-\eta(x))l(-g(x))\right]\,d\bb{W}(x)\nonumber\\
%&=&\pi\int_M l(g(x))\,d\bb{P}(x)+(1-\pi)\int_M l(-g(x))\,d\bb{Q}(x).
%\end{eqnarray}
%Here 
$L_1(\alpha):=L(1,\alpha)$, $L_{-1}(\alpha):=L(-1,\alpha)$, %AG: do we really need all this new notation?
%BK: yes..kind of..because i use this L_1 and L_{-1} in few other places and it is simpler to use this than say L(1,\alpha) and L(-1,\alpha)
$\bb{P}(X):=\mu(X|Y=+1)$, $\bb{Q}(X):=\mu(X|Y=-1)$, $\varepsilon:=\mu(M,Y=+1)$. Here, $\bb{P}$ and $\bb{Q}$ represent the class-conditional distributions and $\varepsilon$ is the prior distribution of class $+1$.
%$\bb{W}:=\pi \bb{P}+(1-\pi)\bb{Q}$ and 
%$\eta(X):=\mu(Y=1|X)=\pi\frac{d\bb{P}}{d\bb{W}}$.
%AG: the following paragraph might be shortened. Or at least better put into context, rather than dropped into the text as an aside. 
%BK: I reduced the paragraph and tried to put the results in context.
\par By appropriately choosing $L$, Nguyen \emph{et al.} \cite{Nguyen-09} have shown an equivalence between $\phi$-divergences (between $\bb{P}$ and $\bb{Q}$) and $R^L_{\eu{F}_\star}$. In particular, they showed that 
%an equivalence between $\phi$-divergences (between $\bb{P}$ and $\bb{Q}$) and the optimal $L$-risk associated with a loss-function, $L$, that satisfies $L_1(\alpha)=L_{-1}(-\alpha)$. 
%They showed that 
for each loss function, $L$, there exists exactly one corresponding $\phi$-divergence such that the $R^L_{\eu{F}_\star}=-D_\phi(\bb{P},\bb{Q})$. 
%is equal to the negative $\phi$-divergence between $\bb{P}$ and $\bb{Q}$.
%\footnote{This result holds even if $L$ does not satisfy the property $L_1(\alpha)=L_{-1}(-\alpha),\,\forall\alpha\in\bb{R}$. However, such an assumption is made to completely analyze the relation between $L$ and $\phi$.} 
For example, the total-variation distance, Hellinger distance and $\chi^2$-divergence are shown to be related to the optimal $L$-risk where $L$ is the hinge loss ($L(y,\alpha)=\max(0,1-y\alpha)$), exponential loss ($L(y,\alpha)=\exp(-y\alpha)$) and logistic loss ($L(y,\alpha)=\log(1+\exp(-y\alpha))$) respectively. %\footnote{By choosing $L_1(\alpha)=-\frac{\alpha}{\varepsilon}$ and $L_{-1}(\alpha)=\frac{e^{\alpha-1}}{1-\varepsilon}$, it can be shown that the associated optimal $L$-risk is the negative of the KL-divergence between $\bb{P}$ and $\bb{Q}$.} 
In statistical machine learning, these losses are well-studied and are shown to result in various binary classification algorithms like support vector machines, Adaboost and logistic regression. See \cite{Evgeniou-00,Scholkopf-02} for details.%In Appendix, using the results in \cite{Nguyen-05}, we compute the loss function, $\ell$ associated with $\phi_\star$.
\par Similarly, by appropriately choosing $L$, we present and prove the following result that relates IPMs (between the class-conditional distributions) and the optimal $L$-risk of a binary classification problem.\vspace{2mm} %Before that we need a notion of the loss function, $\ell$ to \emph{classification-calibrated}. A loss function $\ell$ is \emph{classification-calibrated} if for any $a,b\ge 0$ and $a\ne b$:
%\begin{equation}
%\inf_{\{\alpha\in\mathbb{R}\,|\,\alpha(a-b)<0\}}\left[\ell(\alpha)a+\ell(-\alpha)b\right]>\inf_{\alpha\in\mathbb{R}}\left[\ell(\alpha)a+\ell(-\alpha)b\right].
%\end{equation}
%This definition ensures that the decision rule $g$ behaves equivalently to the Bayes decision rule in the binary classification setting. \cite{Bartlett-06} provides an equivalent condition for $\ell$ to be classification-calibrated: a convex function $\ell$ is classification-calibrated if and only if it is differentiable at $0$ and $\ell^\prime(0)<0$. Now, we are ready to present the result.
\begin{thm}[$\gamma_\eu{F}$ and associated $L$]\label{thm:functionclass}
Let $L_1(\alpha)=-\frac{\alpha}{\varepsilon}$ and $L_{-1}(\alpha)=\frac{\alpha}{1-\varepsilon}$. Let $\eu{F}\subset\eu{F}_\star$ be such that $f\in\eu{F}\Rightarrow -f\in\eu{F}$. Then, $\gamma_\eu{F}(\bb{P},\bb{Q})=-R^L_\eu{F}$.\vspace{2mm} 
\end{thm}
\begin{IEEEproof}
%\begin{itemize}
%\item[($\Rightarrow$)]
From (\ref{Eq:lrisk}), we have 
\begin{eqnarray}
&&\varepsilon\int_M L_1(f)\,d\bb{P}+(1-\varepsilon)\int_M L_{-1}(f)\,d\bb{Q}\nonumber\\
&&\quad=\int_M f\,d\bb{Q}-\int_M f\,d\bb{P}=\bb{Q}f-\bb{P}f.
\end{eqnarray}
%$\pi\int_M L_1(f)\,d\bb{P}+(1-\pi)\int_M L_{-1}(f)\,d\bb{Q}=\int_M f\,d\bb{Q}-\int_M f\,d\bb{P}=\bb{Q}f-\bb{P}f$. 
Therefore, 
\setlength{\arraycolsep}{0.0em}
\begin{eqnarray}
R^L_\eu{F}&{}={}&\inf_{f\in\eu{F}}(\bb{Q}f-\bb{P}f)=
-\sup_{f\in\eu{F}}(\bb{P}f-\bb{Q}f)\nonumber\\
&{}\stackrel{(a)}{=}{}&
-\sup_{f\in\eu{F}}|\bb{P}f-\bb{Q}f|=
-\gamma_{\eu{F}}(\bb{P},\bb{Q}),
\end{eqnarray} where $(a)$ follows from the fact that $\eu{F}$ is symmetric around zero, i.e., $f\in\eu{F}\Rightarrow -f\in\eu{F}$.\vspace{2mm}
%\item[($\Leftarrow$)] Suppose $\gamma_\eu{G}(\bb{P},\bb{Q})=-R^\ell_\eu{G}$. This means 
%\begin{equation}\label{Eq:calibrated}
%\forall\,\bb{P},\bb{Q}\in\Scr{P},\,-\inf_{g\in\eu{G}}\left[\pi\int_M \ell(g)\,d\bb{P}+(1-\pi)\int_M\ell(-g)\,d\bb{Q}\right]=\sup_{g\in\eu{G}}|\bb{P}g-\bb{Q}g|.
%\end{equation}
%Since it holds for all $\bb{P},\bb{Q}\in\Scr{P}$, it also holds for $\bb{P}=\bb{Q}$, which means \begin{equation}
%\forall\,\bb{P}\in\Scr{P},\,\inf_{g\in\eu{G}}\int_M\left[\pi \ell(g)+(1-\pi)\ell(-g)\right]\,d\bb{P}=0.\nonumber
%\end{equation} Therefore, $\forall\,\alpha\in\bb{R},\,\pi\ell(\alpha)=-(1-\pi)\ell(-\alpha)$, which also means, $\pi\ell(-\alpha)=-(1-\pi)\ell(\alpha)$ and so we get $\pi=\frac{1}{2}$ and $\ell(\alpha)=-\ell(-\alpha)$. Since $\ell$ is convex on $\bb{R}$, this means $\ell(-\alpha)$ is concave on $\bb{R}$ and therefore $\ell(\alpha)$ should be of the form $\ell(\alpha)=b\alpha+c,\,b\in\bb{R},\,c\in\bb{R}$. With $\ell(\alpha)=-\ell(-\alpha)$, we get $c=0$ and so $\ell(\alpha)=b\alpha$. Since $\ell$ is classification-calibrated, we have $\ell^\prime(0)=b<0$. Substituting for $\ell(\alpha)$ in (\ref{Eq:calibrated}), we have $-\inf_{g\in\eu{G}}\frac{b}{2}(\bb{P}g-\bb{Q}g)=\sup_{g\in\eu{G}}|\bb{P}g-\bb{Q}g|$. This means, $\forall\,\bb{P},\bb{Q}\in\Scr{P},\,-\frac{b}{2}\sup_{g\in\eu{G}}(\bb{P}g-\bb{Q}g)=\sup_{g\in\eu{G}}(\bb{P}g-\bb{Q}g)$, which implies $b=-2$.
%\qedhere
%\end{itemize}
\end{IEEEproof}
Theorem~\ref{thm:functionclass} shows that $\gamma_\eu{F}(\bb{P},\bb{Q})$ is the negative of the optimal $L$-risk that is associated with a binary classifier that classifies the class-conditional distributions $\bb{P}$ and $\bb{Q}$ using the loss function, $L$, in Theorem~\ref{thm:functionclass}, when the discriminant function is restricted to $\eu{F}$. Therefore, Theorem~\ref{thm:functionclass} provides a novel interpretation for the total variation distance, Dudley metric, Wasserstein distance and MMD, which can be understood as the optimal $L$-risk associated with binary classifiers where the discriminant function, $f$ is restricted to $\eu{F}_{TV}$, $\eu{F}_\beta$, $\eu{F}_W$ and $\eu{F}_k$ respectively. %Note that the loss function in Theorem~\ref{thm:functionclass} is peculiar as unlike in the usual setting (e.g., support vector machine, logistic regression, Adaboost), correctly classified points are rewarded based on how well they are classified. 
%It is easy to see from Theorem~\ref{thm:functionclass} that if $\eu{G}=\eu{F}_\star$, then we attain the result in Theorem~\ref{pro:loss} and $\tilde{R}_{\eu{F}_\star}=0\llbracket\bb{P}=\bb{Q}\rrbracket-\infty\llbracket \bb{P}\ne\bb{Q}\rrbracket$. Since $\eu{F}_\star$ is not an interesting function class, one can use the result in Proposition~\ref{pro:functionclass} to determine a classifier whose risk corresponds to $\gamma_{\eu{G}}(\bb{P},\bb{Q})$, i.e., $\tilde{g}_{\eu{G}}=\arg\inf_{g\in\eu{G}}R_l(g)=\arg\inf_{g\in\eu{G}} (\bb{Q}g-\bb{P}g)$. 
\par Suppose, we are given a finite number of samples $\{(X_i,Y_i)\}^N_{i=1}$, $X_i\in M$, $Y_i\in\{-1,+1\},\,\forall\,i$ drawn i.i.d. from $\mu$ and we would like to build a classifier, $f\in\eu{F}$ that minimizes the expected loss (with $L$ as in Theorem~\ref{thm:functionclass}) based on this finite number of samples. This is usually carried out by solving an empirical equivalent of (\ref{Eq:lrisk}), %i.e., replacing $\mu$ by $\mu_N:=\frac{1}{N}\sum^N_{i=1}\delta_{X_i}\delta_{Y_i}$, 
which reduces to (\ref{Eq:ipm-empirical}), i.e., $\gamma_\eu{F}(\bb{P}_m,\bb{Q}_n)=\sup\{|\sum^N_{i=1}\widetilde{Y_i}f(X_i)|:f\in\eu{F}\}$ by noting that $X^{(1)}_.:=X_i$ when $Y_i=1$, $X^{(2)}_.:=X_i$ when $Y_i=-1$, %$\varepsilon=\frac{m}{N}$ 
and $f\in\eu{F}\Rightarrow -f\in\eu{F}$. This means the sign of $f\in\eu{F}$ that solves (\ref{Eq:ipm-empirical}) is the classifier we are looking for. 
\subsection{Wasserstein distance and Dudley metric: Relation to Lipschitz and bounded Lipschitz classifiers}\label{Sec:Lipschitz}
The Lipschitz classifier is defined as the solution, $f_{\text{lip}}$ to the following program:
\setlength{\arraycolsep}{0.0em}
\begin{eqnarray}\label{Eq:lipclassifier}
\inf_{f\in \text{Lip}(M,\rho)}&{}{}&\,\,\,\Vert f\Vert_L\nonumber\\
\text{s.t.}&{}{}&\,\,\, Y_if(X_i)\ge 1,\,i=1,\ldots,N,
\end{eqnarray}
which is a large margin classifier with margin\footnote{The margin is a technical term used in statistical machine learning. See \cite{Scholkopf-02} for details.} $\frac{1}{\Vert f_{\text{lip}}\Vert_L}$. The program in (\ref{Eq:lipclassifier}) computes a \emph{smooth} function, $f$ that classifies the training sequence, $\{(X_i,Y_i)\}^N_{i=1}$ correctly (note that the constraints in (\ref{Eq:lipclassifier}) are such that $\text{sign}(f(X_i))=Y_i$, which means $f$ classifies the training sequence correctly, assuming the training sequence is separable). The smoothness is controlled by $\Vert f\Vert_L$ (the smaller the value of $\Vert f\Vert_L$, the smoother $f$ and vice-versa). See \cite{Luxburg-04} for a detailed study on the Lipschitz classifier. Replacing $\Vert f\Vert_{L}$ by $\Vert f\Vert_{BL}$ in (\ref{Eq:lipclassifier}) gives the bounded Lipschitz classifier, $f_{\text{BL}}$ which is the solution to the following program:
\setlength{\arraycolsep}{0.0em}
\begin{eqnarray}\label{Eq:dudleyclassifier}
\inf_{f\in BL(M,\rho)}&{}{}&\,\,\,\Vert f\Vert_{BL}\nonumber\\
\text{s.t.}&{}{}& \,\,\,Y_if(X_i)\ge 1,\,i=1,\ldots,N.
\end{eqnarray}
Note that replacing $\Vert f\Vert_L$ by $\Vert f\Vert_\eu{H}$ in (\ref{Eq:lipclassifier}), taking the infimum over $f\in\eu{H}$, yields the hard-margin support vector machine (SVM) \cite{Cortes-95}. We now show how the empirical estimates of $W$ and $\beta$ appear as upper bounds on the margins of the Lipschitz and bounded Lipschitz classifiers, respectively.\vspace{2mm}
\begin{thm}\label{thm:lip-nn}
The Wasserstein distance and Dudley metric are related to the margins of Lipschitz and bounded Lipschitz classifiers as
\setlength{\arraycolsep}{0.0em}
\begin{eqnarray}
\frac{1}{\Vert f_{\text{lip}}\Vert_L}&{}\le{}& \frac{W(\bb{P}_m,\bb{Q}_n)}{2},\label{Eq:lipschitz-class-thm}\\
\frac{1}{\Vert f_{\text{BL}}\Vert_{BL}}&{}\le{}& \frac{\beta(\bb{P}_m,\bb{Q}_n)}{2}.\label{Eq:dudley-class-thm-1}
\end{eqnarray}
%In addition, there exists $f^\star\in\text{Lip}(M,\rho)$, $f_\star\in BL(M,\rho)$ such that $\Vert f^\star\Vert_L=\Vert f_{\text{lip}}\Vert_L$, $\Vert f_\star\Vert_{BL}=\Vert f_{\text{BL}}\Vert_{BL}$ and $\text{sign}(f^\star)$, $\text{sign}(f_\star)$ are $1$-nearest neighbor (NN) classifiers.\footnote{The 1-nearest neighbor rule, $f_{\text{NN}}$ is defined as: $f_{\text{NN}}(x)=1$ if $\inf\{\rho(x,X_i):Y_i=1\}\le\inf\{\rho(x,X_i):Y_i=-1\}$ and $f_{\text{NN}}(x)=-1$ otherwise.}\vspace{2mm}
%Let $L^\star=\frac{2}{\rho(X^+,X^-)}$. For all $\alpha\in[0,1]$, the following function solves (\ref{Eq:lipclassifier}):
%\begin{equation}\label{Eq:g-alpha}
%g_\alpha(x):=\alpha\min_{i=1,\ldots,n}(Y_i+L^\star\rho(x,X_i))+(1-\alpha)\max_{i=1,\ldots,n}(Y_i-L^\star\rho(x,X_i)),
%\end{equation}
%and $\text{sign}(g_{\frac{1}{2}})$ is a $1$-NN classifier.
\end{thm}
\begin{IEEEproof}
Define $W_{mn}:=W(\bb{P}_m,\bb{Q}_n)$. By Lemma~\ref{lem:primal-dual}, we have
%Let $f_1$ be a solution to
%\begin{equation}\label{Eq:g1}
%W(\bb{P}_m,\bb{Q}_n)=\sup\left\{\sum^N_{i=1}\widetilde{Y}_if(X_i)\,:\,\Vert f\Vert_L\le 1\right\}.
%\end{equation} 
%We first show that %$g_1$ is also an optimizer of 
%\begin{eqnarray}\label{Eq:alternate}
%1&=&\inf\Big\{\Vert f\Vert_L\,:\,\sum^N_{i=1}\widetilde{Y}_if(X_i)\ge W(\bb{P}_m,\bb{Q}_n),\nonumber\\
%&&\qquad\,f\in\text{Lip}(M,\rho)\Big\}.
%\end{eqnarray}
\begin{equation}\label{Eq:alternate}
1=\inf\Big\{\Vert f\Vert_L:\sum^N_{i=1}\widetilde{Y}_if(X_i)\ge W_{mn},\,f\in\text{Lip}(M,\rho)\Big\},\nonumber
\end{equation}
which can be written as
\begin{equation}\label{Eq:alternate-1}
\frac{2}{W_{mn}}=\inf\Big\{\Vert f\Vert_L:\sum^N_{i=1}\widetilde{Y}_if(X_i)\ge 2,\,f\in\text{Lip}(M,\rho)\Big\}.\nonumber
\end{equation}
Note that $\{f\in \text{Lip}(M,\rho):Y_if(X_i)\ge 1,\,\forall\,i\}\subset
\{f\in\text{Lip}(M,\rho):\sum^N_{i=1}\widetilde{Y}_if(X_i)\ge 2\}$, and therefore %To show this, consider \begin{equation}
%F:=\Big\{f\in\text{Lip}(M,\rho)\,|\,\sum^N_{i=1}\widetilde{Y}_if(X_i)>W(\bb{P}_m,\bb{Q}_n)\Big\}.\nonumber
%\end{equation}
%For any $f\in F$, $\Vert f\Vert_L>1$ because if it were not the case, then $f_1$ is not a solution to (\ref{Eq:g1}). Let \begin{equation}
%H:=\{f\in\text{Lip}(M,\rho)\,|\,\sum^N_{i=1}\widetilde{Y}_if(X_i)=W(\bb{P}_m,\bb{Q}_n)\}.\nonumber
%\end{equation}
%Clearly any $f\in H$ is a solution to (\ref{Eq:g1}) and therefore %. It is easy to see that any solution to (\ref{Eq:g1}) also lies in $\{g:\Vert g\Vert_L=1\}$. Therefore, 
%$\Vert f\Vert_L=1,\,\forall\,f\in H$, which proves (\ref{Eq:alternate}). %So the solution to (\ref{Eq:alternate}) satisfies $\Vert g\Vert_L=1$ and $\sum^N_{i=1}\widetilde{Y}_ig(X_i)=W(\bb{P}_m,\bb{Q}_n)$, which is also satisfied by any solution to (\ref{Eq:g1}). Hence, $g_1$ is also an optimizer of (\ref{Eq:alternate}). 
%Since $m=n=\frac{N}{2}$, (\ref{Eq:alternate}) can be equivalently written as
%\setlength{\arraycolsep}{0.0em}
%\begin{eqnarray}
%1&{}={}&\inf\Big\{\Vert f\Vert_L:\sum^N_{i=1}Y_if(X_i)\ge \frac{NW_{mn}}{2},\,f\in\text{Lip}(M,\rho)\Big\}\nonumber\\
%&{}\le{}&\inf\Big\{\Vert f\Vert_L:Y_if(X_i)\ge \frac{W_{mn}}{2},\,\forall\,i,\, f\in\text{Lip}(M,\rho)\Big\},\nonumber
%\end{eqnarray}
%such that
\begin{equation}
\frac{2}{W_{mn}}\le\inf\Big\{\Vert f\Vert_L:Y_if(X_i)\ge 1,\,\forall\,i,\,f\in\text{Lip}(M,\rho)\Big\},\nonumber
\end{equation}
hence proving (\ref{Eq:lipschitz-class-thm}). Similar analysis for $\beta$ yields (\ref{Eq:dudley-class-thm-1}).
\vspace{2mm}
\end{IEEEproof}
%AG: the discussion below seems very interesting, and should be *explained* in the introduction, as well as being mentioned at the start of the section as motivation, etc etc.
%BK: I have mentioned and explained in the introduction and at the start of the section.
The significance of this result is as follows. (\ref{Eq:lipschitz-class-thm}) shows that $\Vert f_{\text{lip}}\Vert_L\ge \frac{2}{W(\bb{P}_m,\bb{Q}_n)}$, which means the smoothness of the classifier, $f_{\text{lip}}$, computed as $\Vert f_{\text{lip}}\Vert_L$ is bounded by the inverse of the Wasserstein distance between $\bb{P}_m$ and $\bb{Q}_n$. So, if the distance between the class-conditionals $\bb{P}$ and $\bb{Q}$ is ``small" (in terms of $W$), then the resulting Lipschitz classifier is less smooth, i.e., a ``complex" classifier is required to classify the distributions $\bb{P}$ and $\bb{Q}$. A similar explanation holds for the bounded Lipschitz classifier.
\subsection{Maximum mean discrepancy: Relation to Parzen window classifier and support vector machine}\label{Sec:mmd}
Consider the maximizer $f$, for the empirical estimator of MMD, in (\ref{Eq:optimum-g}). Computing $y=\text{sign}(f(x))$ gives
\begin{equation}\label{Eq:Parzen}
y=\left\{\begin{array}{c@{\quad}l}
+1,& \frac{1}{m}\sum_{Y_i=1}k(x,X_i)>\frac{1}{n}\sum_{Y_i=-1}k(x,X_i)\\
-1,& \frac{1}{m}\sum_{Y_i=1}k(x,X_i)\le \frac{1}{n}\sum_{Y_i=-1}k(x,X_i)\\
\end{array}\right.,
\end{equation}
which is exactly the classification function of a Parzen window classifier \cite{ShaweTaylor-04, Scholkopf-02}. %Note that here $m=n=\frac{N}{2}$. 
It is easy to see that (\ref{Eq:Parzen}) can be rewritten as
\begin{equation}\label{Eq:mean-classifier-1}
y=\text{sign}(\langle w, k(\cdot,x)\rangle_\eu{H}),
\end{equation} 
where $w=\mu^+-\mu^-$, $\mu^{+}:=\frac{1}{m}\sum_{Y_i=1}k(\cdot,X_i)$ and $\mu^{-}:=\frac{1}{n}\sum_{Y_i=-1}k(\cdot,X_i)$. $\mu^+$ and $\mu^-$ represent the class means associated with $X^+:=\{X_i:Y_i=1\}$ and $X^-:=\{X_i:Y_i=-1\}$ respectively.
\par The Parzen window classification rule in (\ref{Eq:mean-classifier-1}) can be interpreted as a \emph{mean classifier} in $\eu{H}$: $\langle w, k(\cdot,x)\rangle_\eu{H}$ represents a hyperplane in $\eu{H}$ passing through the origin with $w$ being its normal along the direction that joins the means, $\mu^{+}$ and $\mu^{-}$ in $\eu{H}$. From (\ref{Eq:gamma-opt}), we can see that $\gamma_k(\bb{P}_m,\bb{Q}_n)$ is the RKHS distance between the mean functions, $\mu^+$ and $\mu^-$. 
\par Suppose $\Vert \mu^+\Vert_\eu{H}=\Vert \mu^-\Vert_\eu{H}$, i.e., $\mu^+$ and $\mu^-$ are equidistant from the origin in $\eu{H}$. Then, the rule in (\ref{Eq:mean-classifier-1}) can be equivalently written as
\begin{equation}\label{Eq:rkhs-nn-1}
y=\text{sign}\left(\Vert k(\cdot,x)-\mu^-\Vert^2_\eu{H}-\Vert k(\cdot,x)-\mu^+\Vert^2_\eu{H}\right).
\end{equation}
(\ref{Eq:rkhs-nn-1}) provides another interpretation of the rule in (\ref{Eq:Parzen}), i.e., as a nearest-neighbor rule: assign to $x$ the label associated with the mean $\mu^+$ or $\mu^-$, depending on which mean function is closest to $k(\cdot,x)$ in $\eu{H}$.
%\par $\gamma_k(\bb{P}_m,\bb{Q}_n)$ can also be written as 
%\begin{equation}
%-\gamma_k(\bb{P}_m,\bb{Q}_n)=\inf_{f\in\eu{F}_k}-\sum^N_{i=1}\widetilde{Y}_if(X_i),
%\end{equation}
%which is in the form of empirical risk minimization with the regularizer determined by $\Vert f\Vert_\eu{H}$. This explicitly establishes the assumptions behind the Parzen window classifier and shows that the Parzen window classifier can be derived in the regularization framework.\footnote{Support vector machines, logistic regression and Adaboost can also be derived in the empirical risk minimization-regularization framework by choosing appropriate loss functions \cite{Scholkopf-02}. Choosing the loss function as in Theorem~\ref{thm:functionclass} with the regularizer being a unit ball in an RKHS yields the Parzen window classifier.} 
\par The classification rule in (\ref{Eq:Parzen}) differs from the ``classical" Parzen window classifier in two respects. (i) Usually, the kernel (called the smoothing kernel) in the Parzen window rule is translation invariant in $\mathbb{R}^d$. In our case, $M$ need not be $\mathbb{R}^d$ and $k$ need not be translation invariant. So, the rule in (\ref{Eq:Parzen}) can be seen as a generalization of the classical Parzen window rule. (ii) The kernel in (\ref{Eq:Parzen}) is positive definite unlike in the classical Parzen window rule where $k$ need not have to be so.
\par Recently, Reid and Williamson \cite[Section 8, Appendix E]{Reid-09} have related MMD to Fisher discriminant analysis \cite[Section 4.3]{Devroye-96} in $\eu{H}$ and SVM \cite{Cortes-95}. Our approach %AG: ``one approach''?? Or ``by contrast, our approach''....? Also, didn't Bob have a result at some stage where he was minimizing the margin? We could contrast with that.
%BK: I wrote it as our appraoch. i think constrating with bob's work will just make it explain quite a bit i believe.
 to relate MMD to SVM is along the lines of Theorem~\ref{thm:lip-nn}, where it is easy to see that the margin of an SVM, computed as $\frac{1}{\Vert f\Vert_\eu{H}}$, can be upper bounded by $\frac{\gamma_k(\bb{P}_m,\bb{Q}_n)}{2}$, which says that the smoothness of an SVM classifier is bounded by the inverse of the MMD between $\bb{P}$ and $\bb{Q}$.\\

To summarize, in this section, we have provided an intuitive understanding of IPMs by relating them to the binary classification problem. We showed that IPMs can be interpreted either in terms of the risk associated with an appropriate binary classifier or in terms of the smoothness of the classifier.
\section{Conclusion \& Discussion}
In this work, we presented integral probability metrics (IPMs) from a more practical perspective. We first proved that IPMs and $\phi$-divergences are essentially different: indeed, the total variation distance is the only ``non-trivial" $\phi$-divergence that is also an IPM. We then demonstrated consistency and convergence rates of the empirical estimators of IPMs, and showed that the empirical estimators of the Wasserstein distance, Dudley metric, and maximum mean discrepancy are strongly consistent and have a good convergence behavior. In addition, we showed these estimators to be very easy to compute, unlike for $\phi$-divergences. Finally, we found that IPMs naturally appear in a binary classification setting, first by relating them to the optimal $L$-risk of a binary classifier; and second, by relating the Wasserstein distance to the margin of a Lipschitz classifier, the Dudley metric to the margin of a bounded Lipschitz classifier, and the maximum mean discrepancy to the Parzen window classifier. With many IPMs having been used only as theoretical tools, we believe that this study highlights properties of IPMs that have not been explored before and would improve their practical applicability.
\par There are several interesting problems yet to be explored in connection with this work. The minimax rate for estimating $W$, $\beta$ and $\gamma_k$ has not been established, nor is it known whether the proposed estimators achieve this rate.
%AG: I do not see what is added by the following paragraph. I'd cut it.
%BK: removed it
%\par De Groot~\cite{DeGroot-62,DeGroot-70} introduced the concept of \emph{statistical information} that is widely studied in information theory and statistics. \cite{Osterreicher-93,Liese-06} have shown that every statistical information is a $\phi$-divergence and every $\phi$-divergence is a statistical information. Since an IPM is trivially a $\phi$-divergence (see Theorems~\ref{thm:phi-mmd} and \ref{thm:tv-mmd}), it can be related to statistical information (see Eq. (77) in \cite{Liese-06}).
It may also be possible to relate IPMs and Bregman divergences. On the most basic level, these two families do not intersect: Bregman divergences do not satisfy the triangle inequality, whereas IPMs do (which are pseudometrics on $\Scr{P}$). Recently, however, Chen \emph{et al.}~\cite{Chen-08a,Chen-08b} have studied ``square-root metrics" based on Bregman divergences. One could investigate conditions on $\eu{F}$ for which $\gamma_\eu{F}$ coincides with such a family. \par Similarly, in the case of $\phi$-divergences, some functions of $D_\phi$ are shown to be metrics on $\Scr{P}_\lambda$ (see Theorem~\ref{thm:tv-mmd} for the notation), for example, the square root of the variational distance, the square root of Hellinger's distance, the square root of the Jensen-Shannon divergence~\cite{Endres-03, Fuglede-03, Hein-05}, etc. Also, \"{O}sterreicher and Vajda \cite[Theorem 1]{Osterreicher-03} have shown that certain powers of $D_\phi$ are metrics on $\mathscr{P}_\lambda$. Therefore, one could investigate conditions on $\eu{F}$ for which $\gamma_\eu{F}$ equals such functions of $D_\phi$.
% if have a single appendix:
%\appendix[Proof of the Zonklar Equations]
% or
\appendices
\section{Proof of Lemma~\ref{lem:phipseudo}}\label{appendix-lem-phipseudo}
%\begin{IEEEproof}[Proof of Lemma~\ref{lem:phipseudo}] 
%\begin{itemize}
$(\Leftarrow)\,\,$ If $\phi$ is of the form in (\ref{Eq:discern-triangle}), then by Lemma~\ref{lem:tv}, we have $D_\phi(\bb{P},\bb{Q})=\frac{\beta-\alpha}{2}\int_M|p-q|\,d\lambda$, which is a metric on $\Scr{P}_\lambda$ if $\beta>\alpha$ and therefore is a pseudometric on $\mathscr{P}_\lambda$. If $\beta=\alpha$, $D_\phi(\bb{P},\bb{Q})=0$ for all $\bb{P},\bb{Q}\in\Scr{P}_\lambda$ and therefore is a pseudometric on $\Scr{P}_\lambda$.\vspace{2mm}\\
$(\Rightarrow)\,\,$ If $D_\phi$ is a pseudometric on $\mathscr{P}_\lambda$, then it satisfies the triangle inequality and ($\bb{P}=\bb{Q}\Rightarrow D_\phi(\bb{P},\bb{Q})=0$) and therefore by \cite[Theorem 2]{Khosravifard-07}, $\phi$ is of the form in (\ref{Eq:discern-triangle}).
\section{Proof of Theorem~\ref{Thm:consistency}}\label{appendix-strong}
Consider $|\gamma_\eu{F}(\bb{P}_m,\bb{Q}_n)-\gamma_\eu{F}(\bb{P},\bb{Q})|=
\Big|\sup_{f\in\eu{F}}|\bb{P}_mf-\bb{Q}_nf|-\sup_{f\in\eu{F}}|\bb{P}f-\bb{Q}f|\Big|
\le\sup_{f\in\eu{F}}||\bb{P}_mf-\bb{Q}_nf|-|\bb{P}f-\bb{Q}f||
\le\sup_{f\in\eu{F}}|\bb{P}_mf-\bb{Q}_nf-\bb{P}f+\bb{Q}f|
\le\sup_{f\in\eu{F}}[|\bb{P}_mf-\bb{P}f|+|\bb{Q}_nf-\bb{Q}f|]
\le \sup_{f\in\eu{F}}|\bb{P}_mf-\bb{P}f| + \sup_{f\in\eu{F}}|\bb{Q}_nf-\bb{Q}f|$. Therefore, by Theorem~\ref{Thm:vandegeer} (see Appendix~\ref{appendix-supp}), $\sup_{f\in\eu{F}}|\bb{P}_mf-\bb{P}f|\stackrel{a.s.}{\longrightarrow} 0$, $\sup_{f\in\eu{F}}|\bb{Q}_nf-\bb{Q}f|\stackrel{a.s.}{\longrightarrow} 0$ and the result follows.
\section{Proof of Theorem~\ref{Thm:rate}}\label{appendix-rate}
From the proof of Theorem~\ref{Thm:consistency}, we have  
$|\gamma_\eu{F}(\bb{P}_m,\bb{Q}_n)-\gamma_\eu{F}(\bb{P},\bb{Q})|\le
\sup_{f\in\eu{F}}|\bb{P}_mf-\bb{P}f| + \sup_{f\in\eu{F}}|\bb{Q}_nf-\bb{Q}f|$. We now bound the terms $\sup_{f\in\eu{F}}|\bb{P}_mf-\bb{P}f|$ and $\sup_{f\in\eu{F}}|\bb{Q}_nf-\bb{Q}f|$, which are the fundamental quantities that appear in empirical process theory. The proof strategy begins in a manner similar to \cite[Appendix A.2]{Gretton-07}, but with an additional step which will be flagged below.
\par Note that $\sup_{f\in\eu{F}}|\bb{P}_mf-\bb{P}f|$ satisfies (\ref{Eq:boundediff}) (see Appendix~\ref{appendix-supp}) with $c_i=\frac{2\nu}{m}$. Therefore, by McDiarmid's inequality in (\ref{Eq:mcdiarmid}) (see Appendix~\ref{appendix-supp}), we have that with probability at least $1-\frac{\delta}{4}$, the following holds:
\begin{eqnarray}\label{Eq:delta-1}
\lefteqn{\sup_{f\in\eu{F}}|\bb{P}_mf-\bb{P}f|\le\bb{E}\sup_{f\in\eu{F}}|\bb{P}_mf-\bb{P}f|+\sqrt{\frac{2\nu^2}{m}\log\frac{4}{\delta}}}\hspace{1cm}\nonumber\\
&&\stackrel{(a)}{\le }2\bb{E}\sup_{f\in\eu{F}}\Big|\frac{1}{m}\sum^m_{i=1}\sigma_if(X^{(1)}_i)\Big|+\sqrt{\frac{2\nu^2}{m}\log\frac{4}{\delta}},
\end{eqnarray}
where $(a)$ follows from bounding $\bb{E}\sup_{f\in\eu{F}}|\bb{P}_mf-\bb{P}f|$ by using the symmetrization inequality in (\ref{Eq:symmetrization}) (see Appendix~\ref{appendix-supp}). Note that the expectation in the second line of (\ref{Eq:delta-1}) is taken jointly over $\{\sigma_i\}^m_{i=1}$ and $\{X^{(1)}_i\}^m_{i=1}$. $\bb{E}\sup_{f\in\eu{F}}\Big|\frac{1}{m}\sum^m_{i=1}\sigma_if(X^{(1)}_i)\Big|$ can be written as $\bb{E}\bb{E}_\sigma\sup_{f\in\eu{F}}\Big|\frac{1}{m}\sum^m_{i=1}\sigma_if(X^{(1)}_i)\Big|$, where the inner expectation, which we denote as $\bb{E}_\sigma$, is taken with respect to $\{\sigma_i\}^m_{i=1}$ conditioned on $\{X^{(1)}_i\}^m_{i=1}$ and the outer expectation is taken with respect to $\{X^{(1)}_i\}^m_{i=1}$. 
Since $\bb{E}_\sigma\sup_{f\in\eu{F}}|\frac{1}{m}\sum^{m}_{i=1}\sigma_i f(X^{(1)}_i)|$ satisfies (\ref{Eq:boundediff}) (see Appendix~\ref{appendix-supp}) with $c_i=\frac{2\nu}{m}$, by McDiarmid's inequality in (\ref{Eq:mcdiarmid}) (see Appendix~\ref{appendix-supp}), with probability at least $1-\frac{\delta}{4}$, we have 
\setlength{\arraycolsep}{0.0em}
\begin{eqnarray}\label{Eq:rademacher-1}
\bb{E}\sup_{f\in\eu{F}}\left|\frac{1}{m}\sum^{m}_{i=1}\sigma_i f(X^{(1)}_i)\right|&{}\le{}& \bb{E}_\sigma\sup_{f\in\eu{F}}\left|\frac{1}{m}\sum^{m}_{i=1}\sigma_i f(X^{(1)}_i)\right|\nonumber\\
&{}{}&+\sqrt{\frac{2\nu^2}{m}\log\frac{4}{\delta}}.
\end{eqnarray}
Tying (\ref{Eq:delta-1}) and (\ref{Eq:rademacher-1}), we have that with probability at least $1-\frac{\delta}{2}$, the following holds:
\begin{equation}\label{Eq:p}
\sup_{f\in\eu{F}}|\bb{P}_mf-\bb{P}f|\le 2R_{m}(\eu{F};\{X^{(1)}_i\}^m_{i=1})+\sqrt{\frac{18\nu^2}{m}\log\frac{4}{\delta}}.
\end{equation}
Performing similar analysis for $\sup_{f\in\eu{F}}| \bb{Q}_nf-\bb{Q}f|$, we have that with probability at least $1-\frac{\delta}{2}$,
\begin{equation}\label{Eq:q}
\sup_{f\in\eu{F}}|\bb{Q}_nf-\bb{Q}f|\le 2R_{n}(\eu{F};\{X^{(2)}_i\}^n_{i=1})+\sqrt{\frac{18\nu^2}{n}\log\frac{4}{\delta}}.
\end{equation}
The result follows by adding (\ref{Eq:p}) and (\ref{Eq:q}). Note that the second
application of McDiarmid was not needed in \cite[Appendix A.2]{Gretton-07}, since in that case a simplification
was possible due to $\eu{F}$ being restricted to  RKHSs.
\section{Proof of Lemma~\ref{lem:primal-dual}}\label{appendix-optim}
Note that $A:=\{x:\psi(x)\le b\}$ is a convex subset of $V$. Since $\theta$ is not constant on $A$, by Theorem~\ref{Thm:rockafeller} (see Appendix~\ref{appendix-supp}), $\theta$ attains its supremum on the boundary of $A$. Therefore, any solution, $x_*$ to (\ref{Eq:primal}) satisfies $\theta(x_*)=a$ and $\psi(x_*)=b$. Let $G:=\{x:\theta(x)>a\}$. For any $x\in G$, $\psi(x)>b$. If this were not the case, then $x_*$ is not a solution to (\ref{Eq:primal}). Let $H:=\{x:\theta(x)=a\}$. Clearly, $x_*\in H$ and so there exists an $x\in H$ for which $\psi(x)=b$. Suppose $\inf\{\psi(x):x\in H\}=c<b$, which means for some $x^*\in H$, $x^*\in A$. From (\ref{Eq:primal}), this implies $\theta$ attains its supremum relative to $A$ at some point of relative interior of $A$. By Theorem~\ref{Thm:rockafeller}, this implies $\theta$ is constant on $A$ leading to a contradiction. Therefore, $\inf\{\psi(x):x\in H\}=b$ and the result in (\ref{Eq:dual}) follows.
\section{Supplementary Results}\label{appendix-supp}  % for no appendix heading
In this section, we collect results that are used to prove results in Section \ref{Sec:consistency}.% and \ref{Sec:surrogate}.
\vspace{2mm}\par\noindent We quote the following result on Lipschitz extensions from \cite{Luxburg-04} (see also \cite{McShane-34,Whitney-34}).\vspace{2mm}
\begin{lem}[Lipschitz extension]\label{lem:lip}
Given a function $f$ defined on a finite subset $x_1,\ldots,x_n$ of $M$, there exists a function $\widetilde{f}$ which coincides with $f$ on $x_1,\ldots,x_n$, is defined on the whole space $M$, and has the same Lipschitz constant as $f$. Additionally, it is possible to explicitly construct $\widetilde{f}$ in the form
\setlength{\arraycolsep}{0.0em}
\begin{eqnarray}
\widetilde{f}(x)&{}={}&\alpha\min_{i=1,\ldots,n}(f(x_i)+L(f)\rho(x,x_i))\nonumber\\
&{}{}& +(1-\alpha)\max_{i=1,\ldots,n}(f(x_i)-L(f)\rho(x,x_i)),
\end{eqnarray}
for any $\alpha\in[0,1]$, with $L(f)=\max_{x_i \ne x_j}\frac{f(x_i)-f(x_j)}{\rho(x_i,x_j)}$.\vspace{2mm}
\end{lem}
The following result on bounded Lipschitz extensions is quoted from \cite[Proposition 11.2.3]{Dudley-02}.\vspace{2mm}
\begin{lem}[Bounded Lipschitz extension]\label{lem:lip-bounded}
If $A\subset M$ and $f\in BL(A,\rho)$, then $f$ can be extended to a function $h\in BL(M,\rho)$ with $h=f$ on $A$ and $\Vert h\Vert_{BL}=\Vert f\Vert_{BL}$.
Additionally, it is possible to explicitly construct $h$ as
\begin{equation}\label{Eq:lem-dudley}
h=\max\left(-\Vert f\Vert_\infty,\min\left(g,\Vert f\Vert_\infty\right)\right),
\end{equation}
where $g$ is a function on $M$ such that $g=f$ on $A$ and $\Vert g\Vert_L=\Vert f\Vert_L$.\vspace{2mm}
\end{lem}
The following result is quoted from \cite[Theorem 3.7]{vandeGeer-00}.\vspace{2mm}
\begin{thm}\label{Thm:vandegeer}
Let $F(x)=\sup_{f\in\eu{F}}|f(x)|$ be the envelope function for $\eu{F}$. Assume that $\int F\,d\bb{P}<\infty$, and suppose moreover that for any $\varepsilon>0$, $\frac{1}{m}\mathcal{H}(\varepsilon,\eu{F},L_1(\bb{P}_m))\stackrel{\bb{P}}{\longrightarrow} 0$. Then $\sup_{f\in\eu{F}}(\bb{P}_mf-\bb{P}f)\stackrel{a.s.}{\longrightarrow} 0$.\vspace{2mm}
\end{thm}
\begin{thm}[\cite{McDiarmid-89}  McDiarmid's Inequality]\label{thm:mcdiarmid}
Let $X_1,\ldots,$ $X_n,X^\prime_1,\ldots,X^\prime_n$ be independent random variables taking values in a set $M$, and assume that $f:M^n\rightarrow\bb{R}$ satisfies
\begin{equation}\label{Eq:boundediff}
|f(x_1,\ldots,x_n)-f(x_1,\ldots,x_{i-1},x^\prime_i,x_{i+1},\ldots,x_n)|\le c_i,
\end{equation}
$\forall\,x_1,\ldots,x_n,x^\prime_1,\ldots,x^\prime_n\in M$. Then for every $\epsilon>0$,
\begin{equation}\label{Eq:mcdiarmid}
\text{Pr}\left(f(X_1,\ldots,X_n)-\bb{E}f(X_1,\ldots,X_n)\ge \epsilon\right)\le e^{\frac{-2\epsilon^2}{\sum^n_{i=1}c^2_i}}.
\end{equation}
\end{thm}
\begin{lem}[\cite{Vaart-96} Symmetrization]
Let $\sigma_1,\ldots,\sigma_N$ be i.i.d. Rademacher random variables. Then,
\begin{equation}\label{Eq:symmetrization}
\bb{E}\sup_{f\in\eu{F}}\left|\bb{E}f-\frac{1}{N}\sum^N_{i=1}f(x_i)\right|\le 2\bb{E}\sup_{f\in\eu{F}}\left|\frac{1}{N}\sum^N_{i=1}\sigma_if(x_i)\right|.
\end{equation}
\end{lem}
The following result is quoted from \cite[Theorem 32.1]{Rockafeller-70}.\vspace{2mm}
\begin{thm}\label{Thm:rockafeller}
Let $f$ be a convex function, and let $C$ be a convex set contained in the domain of $f$. If $f$ attains its supremum relative to $C$ at some point of relative interior of $C$, then $f$ is actually constant throughout $C$.
\end{thm}
% do not use \section anymore after \appendix, only \section*
% is possibly needed

% use appendices with more than one appendix
% then use \section to start each appendix
% you must declare a \section before using any
% \subsection or using \label (\appendices by itself
% starts a section numbered zero.)
%

%\appendices
%\section{Proof of the First Zonklar Equation}
%Appendix one text goes here.

% you can choose not to have a title for an appendix
% if you want by leaving the argument blank
%\section{}
%Appendix two text goes here.

% use section* for acknowledgement
\section*{Acknowledgments}
B. K. S. wishes to acknowledge the support from the Max Planck Institute (MPI) for Biological Cybernetics, National Science Foundation (grant DMS-MSPA 0625409), the Fair Isaac Corporation and the University of California MICRO program. Part of this work was done while he was visiting MPI. A. G. was supported by grants DARPA IPTO FA8750-09-1-0141, ONR MURI N000140710747, and ARO MURI W911NF0810242. The authors thank Robert Williamson and Mark Reid for helpful conversations.
% Can use something like this to put references on a page
% by themselves when using endfloat and the captionsoff option.
\ifCLASSOPTIONcaptionsoff
  \newpage
\fi

% trigger a \newpage just before the given reference
% number - used to balance the columns on the last page
% adjust value as needed - may need to be readjusted if
% the document is modified later
\IEEEtriggeratref{48}
\end{document}